\documentclass[onecolumn,sort&compress,numbers]{els-mrw} % For numbered references Style 

\usepackage{amsmath,amssymb,amsfonts,amsthm,makeidx,graphicx}
\usepackage{txfonts,bbold}
\usepackage{helvet}
\usepackage{CJKutf8}
\newcommand{\ket}[1]{|#1\rangle} % Ket
\begin{document}

%%%%%%%%%%%%%%%%%%%%%%%%%%%%%%%%%%%%%%%%%%%%%%%%%%%%%%%%%%%%%%%%
%% The following items are mandatory: 
%% - title
%% - author names
%% - affiliation details
%% - abstract
%% - keywords

%% Precise, concise, and informative description of the focus of this work. Avoid abbreviations and formulae in the title
\chapter{Light Meson Resonances}\label{chap1}

%% All author names and affiliations, and email address for the corresponding author
\author[1]{Jos\'e Ram\'on Pel\'aez}%
%\author[2]{Second Author}%
%\author[1,2]{Third Author}%

\address[1]{\orgname{Universidad Complutense}, \orgdiv{Departamento de F\'{\i}sica Te\'orica and IPARCOS}, \orgaddress{Plaza de las Ciencias 1, 28040 Madrid- SPAIN}}
%\address[2]{\orgname{Name of Institution}, \orgdiv{Division or Department}, \orgaddress{Address of Institution}}

\articletag{Chapter Article tagline: update of previous edition, reprint.}

\maketitle

%%%%%%%%%%%%%%%%%%%%%%%%%%%%%%%%%%%%%%%%%%%%%%%%%%%%%%%%%%%%%%%%
%% The following item is mandatory: 
%% 100-150 word summary of the chapter
\begin{abstract}[Abstract]
	We provide a modern pedagogical introduction to light mesons. We discuss their masses and widths, their classification into multiplets of isospin and flavor SU(3), as well as other quantum numbers, mixing schemes, observable states, and their disposition in Regge trajectories. We briefly introduce the relevant concepts of Quantum Chromodynamics, including confinement, chiral symmetry, and the relationship between poles and resonances. We discuss their interpretation in terms of quark-model ordinary quark-antiquark states or non-ordinary states like exotics, multiquarks, or glueballs. Finally, we comment on the open issues that are still under discussion.
\end{abstract}

%% 5-10 words that embody the key topics in the chapter. What terms would someone put into a search engine if they were looking for a chapter like this?
\begin{keywords}
 	light-mesons\sep light-quarks\sep resonances \sep quark-model \sep QCD
\end{keywords}

%%%%%%%%%%%%%%%%%%%%%%%%%%%%%%%%%%%%%%%%%%%%%%%%%%%%%%%%%%%%%%%%
%% The following item is optional: 
%% - Single figure visually illustrating the key topic/method/outcome described in the chapter
\begin{figure}
	\centering
	\includegraphics[width=\textwidth]{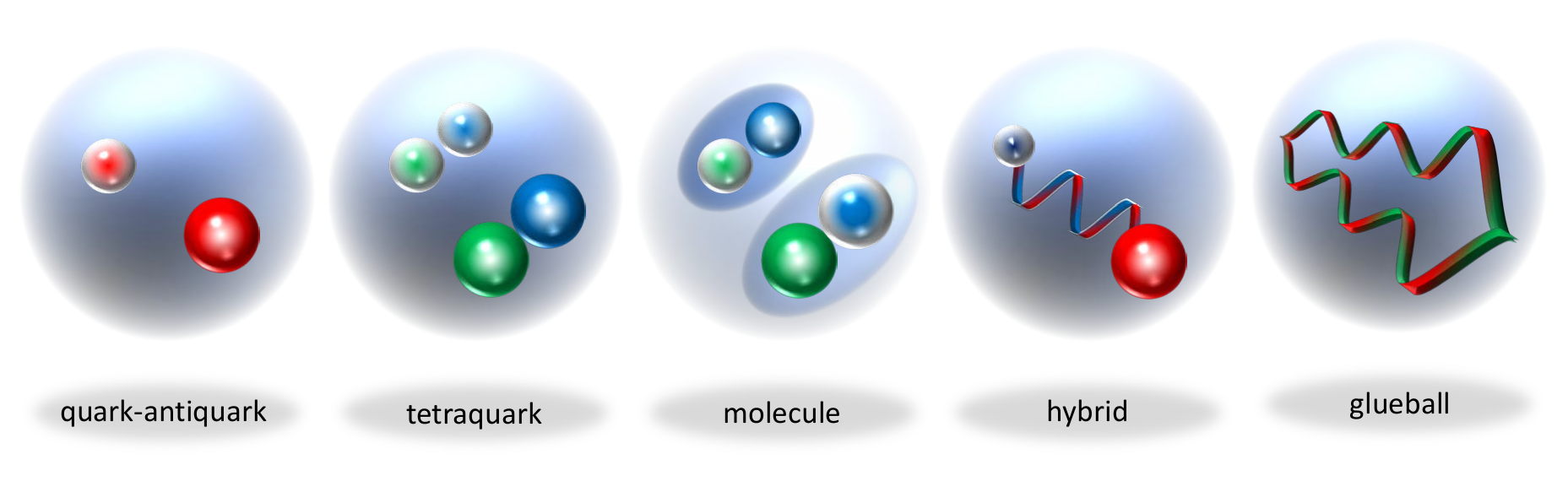}
	\caption{Pictorial illustration of possible meson compositions according to their extended quark-model valence constituents. The quark-antiquark component is dominant for ``ordinary mesons''---the most common. However, some light mesons require a significant or even dominant contribution from other ``non-ordinary'' configurations of which we show the simplest ones.
    This is an oversimplification; in real life, where all these components mix, the situation is much richer, as discussed in this review.
    }
	\label{fig:typesofmesons}
\end{figure}

%%%%%%%%%%%%%%%%%%%%%%%%%%%%%%%%%%%%%%%%%%%%%%%%%%%%%%%%%%%%%%%%
%% The following item is optional: 
%% - System of abbreviations/terms/symbols used in the specific field of study/community. List and define
\begin{glossary}[Nomenclature]
	\begin{tabular}{@{}lp{34pc}@{}}
        BW  & Breit-Wigner (formula).\\
		ChPT & Chiral Perturbation Theory.\\
        fm & femtometer or fermi $=10^{-15}$ meters. \\
        LO, NLO, NNLO & Leading Order (of an expansion), Next to Leading Order, Next to Next to Leading Order.\\
        L$\sigma$M & Linear sigma model.\\
        MeV, GeV & Mega-electronvolt, Giga-electronvolt.\\
        $N_c$ & Number of colors in QCD. \\
        NGB & Nambu-Goldstone Boson.\\
        NNG & Nishijima-Nakano-Gell-Mann (relation). \\
        NJL & Nambu-Jona-Lasinio (model).\\
        ns & nano-second $=10^{-9}$ seconds.\\
        QCD & Quantum Chromodynamics.\\
		QED & Quantum Electrodynamics.\\
        QM & Quark Model.\\
        RPP & Review of Particle Physics. \\
        SU(2) & 2-dimensional Special Unitary group ($2\times2$ complex unitary matrices with unit determinant). \\
        SU(3) & 3-dimensional Special Unitary group ($3\times3$ complex unitary matrices with unit determinant). \\
        SU(3)$_F$ & Flavor-symmetry group generated by the third component of isospin and strangeness. \\
        SU(3)$_c$ & Color-symmetry gauge group of QCD. \\
        UChPT & Unitarized Chiral Perturbation Theory.
	\end{tabular}
\end{glossary}

%%%%%%%%%%%%%%%%%%%%%%%%%%%%%%%%%%%%%%%%%%%%%%%%%%%%%%%%%%%%%%%%
%% The following item is mandatory: 
%% List of the key points and topics a reader can expect to learn from this chapter 
\section{Objectives}
\begin{itemize}
	\item To recognize what light mesons are, their relevance on their own, for other hadronic processes, and the strong interactions.
    \item To realize that mesons are unstable states, or resonances, whose lifetimes span several orders of magnitude, and that they are rigorously defined in terms of poles in the complex plane.
	\item To understand the classification of mesons according to their quantum numbers, like spin, parity, charge conjugation,  isospin, and flavor.
    \item To comprehend flavor symmetry and the classification of mesons into SU(3)$_F$ flavor multiplets.
	\item To learn very basic concepts of Quantum Chromodynamics (QCD) and the Quark Model (QM) of relevance for meson physics.
    \item To interpret mesons in terms of ordinary quark-antiquark states and or other non-ordinary configurations within the QM. 	
    \item To understand the Nambu-Goldstone boson nature of the lightest mesons and their relation to spontaneous chiral symmetry breaking of QCD. To gain insight into Chiral Perturbation Theory, its implications, and its unitarized version.
    \item To appreciate the implications of the meson classification into Regge trajectories.
    \item To understand that the quark model, the Regge approach, and other models are oversimplifications, helpful for classification purposes or understanding concepts, but that it is QCD that governs meson dynamics.   
\end{itemize}

%%%%%%%%%%%%%%%%%%%%%%%%%%%%%%%%%%%%%%%%%%%%%%%%%%%%%%%%%%%%%%%%
%% The following items are mandatory: 
%% - Section: Introduction 
%% - further sections
%% - Section: Conclusion
\section{Introduction}\label{intro}

Mesons are defined with just two words: bosonic hadrons. 
They are hadrons because they are subject to strong interactions besides the electromagnetic, weak, and gravitational forces. This means that they are primarily composed of quarks and gluons. They are bosons because they have integer spin, whereas hadrons with half-integer spin are called baryons.

In many popular science books and reviews, mesons are often defined in a naive manner as bound states of a quark and an antiquark. This configuration is represented pictorially in the leftmost picture in Fig.~\ref{fig:typesofmesons}. Although this oversimplification may be suitable for many mesons, from a modern perspective, it is also very inadequate for quite a few of the mesons to be discussed below, which, naively, are closer to tetraquarks, meson molecules, glueballs, hybrids, or mixtures of these compositions. These ``non-ordinary" configurations are also represented pictorially in Fig.\ref{fig:typesofmesons}.

Meson masses range from $\simeq140\,$MeV$/c^2$ 
$\simeq 2.5\times 10^{-28}\,$kg for the pions, which are the lightest, up to several thousands of MeV$/c^2$. Note that in this context, using kilograms or even grams is very inconvenient, and we use units of MeV$/c^2\simeq1.8\times 10^{-30}\,$kg, or GeV$/c^2=1000\,$MeV$/c^2\simeq 1.8\times 10^{-27}\,$kg. However, from now on, we will adopt the system of natural units, which is the default in this context, where $c=1$ and $ \hbar=1$, and masses are given in MeV or GeV.

The typical hadron radius is $\lesssim 1\,$fm (1 femtometer or fermi is 1fm$=10^{-15}\,$m). Except for the proton, the neutron, and sometimes the deuteron, nuclei are not usually referred to as hadrons since they are interpreted as composed of baryons, and their sizes typically range from about 2 to 12 fm.

All mesons are unstable and have very short lifetimes. Among them, the longest-lived are the few mesons that only decay via electroweak interactions. The largest mean life, $\tau_{K_L}\sim 50\,$ns=$5\cdot10^{-8}\,$s corresponds to the so called ``long-kaon" $K_L$, followed by the charged pion $\tau_{\pi^\pm}\sim26\,$ns=$2.6\cdot10^{-8}\,$s, whereas that of the neutral pion, or $\pi^0$, it is just $\tau_{\pi^0}\sim8.4\cdot10^{-17}\,$s$\simeq 10^{-7}\,$ns  and for the $\eta$ it is $\tau_\eta\sim5\cdot10^{-19}\,$s=$5\cdot10^{-10}\,$ns.
From the point of view of strong interactions, those few mesons would be stable and are sometimes referred to as quasi-stable.
In contrast, the mean life of most other mesons, which decay via strong interactions, is typically less than $10^{-14}$ns.  

In this review, unless otherwise stated, we will take the numerical values of meson properties like mass, decay times, etc, as well as their uncertainties, from the Review of Particle Physics (RPP) in its 2024 edition and 2025 update \cite{ParticleDataGroup:1986kuw}, which we consider a conservative compilation of the present state of the art. 

For our subsequent discussion, we will briefly introduce some key concepts.

\subsection{Resonances: Mass, width, and associated poles}

Relatively long-lived charged mesons, such as $\pi^\pm$ or $K^\pm$, can leave tracks in detectors. However, mesons with very short lifetimes do not reach the detectors, but disintegrate in flight, decaying into states containing the longest-lived mesons, i.e., pions and kaons, as well as photons and leptons.
By measuring the energy distribution of such decays, it is possible to observe resonant peaks or other salient structures in the cross sections, which are indicative of the presence of such short-lived mesons, which are therefore called ``resonances".

For such meson resonances, the mean life is not a very convenient quantity, and their decays are then characterized by their ``width'', denoted by $\Gamma$, which in natural units is the inverse of their mean life. In the simplest cases, this width is indeed the width of the observed resonant peaks. Typical light-meson widths range from tens to hundreds of MeV (to be compared with 1.31 keV of the $\eta$ discussed above). For example, the archetypal $\rho(770)$ meson has a width $\Gamma\simeq147\,$MeV, much more convenient to use than its lifetime of $\simeq4.7\cdot10^{-24}\,$s=$4.7\cdot10^{-15}\,$ns. This meson is simply referred to as ``the rho-resonance." 

Let us provide a succinct and intuitive introduction to the formalism of resonances. 
First, let us recall that the mass $m$ of a stable meson corresponds to its energy at rest. Its quantum wave function $\psi(0)$ at time $t=0$, if properly normalized, satisfies $P(0)=\vert \psi(0)\vert^2=1$, where $P(0)$ is the probability of finding that meson in space. The time evolution is given by $\psi(t)=e^{-imt}\,\psi(0)$, which implies that $P(t)=\vert \psi(t)\vert^2=1$, and the probability of finding the meson remains the same as it should be for a stable state. However, if a state disintegrates, the probability of finding it is known to decay exponentially as $P(t)=P(0)\,e^{-t/\tau}$, where $\tau$ is the mean lifetime. Such a behavior can be described mathematically if the mass of the unstable state were complex, i.e. $m\to m-i\Gamma/2$, since
then $\psi(t)=e^{-i(m-i\Gamma/2)t}\,\psi(0)=e^{-imt}e^{-\Gamma/2}\,\psi(0)$ and therefore
$P(t)=\vert \psi(t)\vert^2=P(0)\,e^{-\Gamma t}$. The mean life would be $\tau=1/\Gamma$. Note that $\Gamma$ has natural units of mass. The preceding argument is one of the reasons why, mathematically, it is often convenient to study complex energy values.

When these unstable states propagate, their propagators will now have a pole at this complex mass, or, in relativistic formalisms, in its complex mass squared $s_p=(m-i\Gamma/2)^2$.  Scattering or production amplitudes where these states can be produced will inherit those poles. Note that now we use the conventional Mandelstam $s$-variable, which is the total energy squared in the center-of-mass frame of the process.

In regions where these poles are close to the real axis, i.e., when the width is small and no other singularities are nearby (like threshold or other poles), the
amplitude behaves as $\sim g(s)/(s_p-s)$ with $g(s)$ an analytic function that can be expanded in the real axis around $s_p$, assuming $\Gamma/m\ll1$.  If the resonance couples to a decay channel $a$, this leads to the familiar Breit-Wigner (BW) formula:
\begin{equation}
 {\cal A}^{\rm BW}_a(s)=\frac{N_a(s)}{m_{\rm BW}^2-s-im_{\rm BW}\Gamma(s)},
\end{equation}
where the Breit-Wigner masses are related to the ``pole masses" as $m_{\rm BW}\simeq m$, $\Gamma_{\rm BW}=\Gamma(m_{\rm BW}^2)\sim \Gamma$.
$N_a(s)$ is a function that contains couplings, normalization, kinematics, and production factors of channel $a$, which have known expressions in simple cases like two-body decays or scattering. For details, we recommend the ``Resonances" note in the RPP \cite{ParticleDataGroup:2024cfk}.

For our purposes here, it is enough to remark that the modulus of the BW formula displays a peak shape at $\sqrt{s}=m_{BW}$, whose width is controlled by $\Gamma$: the larger $\Gamma$, the wider the peak and the shorter the mean life of the resonance. Thus, by looking for these peaks, one may discover resonances. Very often, this is expressed in terms of partial-wave amplitudes with definite angular momentum and in combinations of definite isospin. These partial waves are obtained by studying the angular dependence of the decay products and allow for the identification of the BW mass, width, as well as the spin $J$ of the resonance, together with other quantum numbers.

However, not all meson resonances are narrow, i.e., $\Gamma/m\ll1$, and their poles lie far from other analytic structures.  Despite the resonance pole being at the same position in all the amplitudes where it appears, when the narrowness and isolation requirements are not met, resonance peaks may appear distorted when viewed from the real axis, i.e., as seen in data.
For instance, interference between nearby resonances may deform the peaks to the point of becoming dips---as for the $f_0(980)$ in $\pi\pi$ scattering---, or leave almost no visible peak or dip---as for the $f_0(500)$ and $K^*_0(700)$. Note that these three examples are scalars.  In particular, the distortion can vary significantly across different processes, apparently leading to other masses and widths, to the point where the same resonance may be interpreted as distinct resonances. We will see that some situations like this are still unresolved for light mesons.  Furthermore, some analytic structures unrelated to resonances, such as thresholds or triangle singularities near thresholds (see \cite{Guo:2019twa} for a review), can give rise to substantial distortions or even apparent peaks. When the data are not precise enough or analyzed with too simple models, these distortions could be misinterpreted as resonances. Hence, although peaks that are clearly isolated from other structures commonly correspond to resonances, one must be aware that ``not all resonances are peaks and not all peaks are resonances".

In summary, the BW approximation, variations of it, or simple resonant models without the correct analytic properties do not always hold. Unfortunately, they have been used, and continue to be used, even when they should not, which has led to considerable confusion for some resonances. What to do then?

The model-independent and mathematically rigorous way of defining a resonance is through its associated pole in amplitudes. Unfortunately, the analytic continuation of functions known on the real axis to the complex plane is a challenging mathematical problem. Over the past few decades, the hadron physics community has made a relentless effort to determine resonance poles and their parameters as rigorously as possible.
The most rigorous approaches involve the use of dispersive methods, which are a consequence of causality, together with constraints from unitarity and crossing symmetry.
Dispersion theory is cast mathematically in terms of integral dispersion relations, which are the application of Cauchy's Integral formula to physical amplitudes. These are complicated equations that, as will be discussed below, have been implemented in a few cases so far, but have been essential in determining the existence of the lightest scalar mesons and a single hybrid candidate. Similarly, somewhat less rigorous results have been obtained by imposing unitarity and the two-body correct analytic structure by techniques generically called unitarization or coupled-channel approaches, which can be extended to more systems. For details on this subject, we refer the reader to the ``Theory of resonances" \cite{Mai:2025wjb} or ``Coupled-channel formalism" \cite{Oller:2025leg} reviews of this Encyclopedia, as well as the reviews \cite{Pelaez:2015qba,Oller:2019opk,Yao:2020bxx,Pelaez:2021dak}.

For this reason, the RPP lists a ``$ T$-Matrix pole", $\sqrt{s_0}=m-i\Gamma/2$, whenever it is available, in addition to a BW mass and width. When nothing is indicated, it is just the latter. It is easy to see that when the BW formula is applicable, the BW and pole parameters are similar. When they are not, the BW is not appropriate, but those values are still listed because, unfortunately, some analyses still use them. In this mini-review, we will always use the ``$T$-Matrix pole" whenever both are available.

\subsection{Meson quantum numbers and multiplets} 

Apart from their spin $J$, other quantum numbers can characterize a hadron or a family of hadrons. For instance, hadrons can be electrically neutral or possess an electric charge that is a multiple of the proton charge (or the opposite of the electron charge).  But there are more quantum numbers of interest that we review below.

\subsubsection{Parity and charge conjugation}
Two other multiplicative numbers also characterize hadrons: parity
and charge conjugation, which are conserved by the strong interactions. Parity refers to their behavior under total reflection. If their state stays the same, they have 
positive parity $P=+1=+$, but if it changes sign, they have negative parity $P=-1=-$. 
Charge conjugation, or $C$-parity, is the transformation that changes the charges of a state.  Only mesons that are neutral with respect to all charges can transform into themselves under $C$.  Depending on whether they stay the same or pick up a sign, we say these neutral states have $C=+1=+$ or $C=-1=-$, respectively. 
It is customary to use the notation $J^{P}$ to indicate
the spin and parity of a hadron, and for mesons with defined $C$ to write $J^{PC}$.

\subsubsection{Isospin}
Very often, families of hadrons can be found with the same $J^P$ and almost the same mass (up to a few MeV), differing in their charges, but behaving very similarly under strong interactions.
This suggests that the members of these families form a single entity from the point of view of the strong force, which presents a degeneracy only lifted by weaker interactions, such as electromagnetism. 
Such degeneracy was initially suggested in 1932 by Heisenberg \cite{Heisenberg:1932dw} when treating protons (p) and neutrons (n) equally in his model for their binding within nuclei. In 1937, Wigner observed that if their complex states are disposed in a doublet $(p,n)$, their strong dynamics are invariant under the action of the group of complex SU(2) matrices (also called two-dimensional complex rotations). Because of its similarity to the spin formulation, he called this symmetry "isobaric spin"; nowadays, it is referred to as isospin.
Contrary to the usual angular momentum, isospin transformations are unrelated to space-time, 
and thus are called ``internal". 
Being an approximate symmetry, it only gives rise to the approximate conservation of the three charges associated with the three group generators, which form an isospin vector $\vec{I} $, of which only its modulus $\vec{I}^2$ and one component, usually chosen to be the third one $I_3$, can be determined simultaneously. 

As with the familiar spin, any representation of this group can be completely decomposed into a (direct) sum of irreducible representations, i.e., a family of a few states that transform among themselves under SU(2) symmetry transformations. Each of these sets, labeled by its isospin $I$, spans a vector-space generated by a basis of states $\ket{I I_3}$, 
with $-I\leq I_3\leq I$, which are simultaneous eigenvectors of 
$\vec I^2$ and $I_3$. Namely: $\vec I^2\ket{I, I_3}=I(I+1)\ket{I, I_3}$ and $ I_3\ket{I, I_3}=I_3\ket{I, I_3}$. Note that for a given $I$, the basis is formed by $2I+1$ states and is referred to as a multiplet. 
There are three kinds of meson isospin multiplets that we will find repeatedly:
\begin{itemize}
	\item {\bf Isosinglets:} or isospin singlets. Composed of a single state with $I=I_3=0$, which under isospin transforms into itself. The $\eta$, $f_0(500)$, $f_0(980)$, $\phi(1020)$,  and many other states, are isosinglets. 
	\item {\bf Isodoublets:} or isospin doublets. Composed of two states with $I=1/2$ and $I_3=-1/2$ or $1/2$, which transform between themselves under isospin rotations. The $K^+,K^0$, the $K^-,\bar K^0$, and other similar pairs are examples of isospin doublets.
    \item {\bf Isotriplets:} or isospin triplets. Composed of three states with
    $I=1$ and $ I_3=-1, 0, 1$, which transform among themselves under isospin.
    Examples of isotriplets are the $\pi^+,\pi^0$ and $\pi^-$, or the $\rho^+(770), \rho^0(770)$ and $\rho^-(770)$. 
\end{itemize}
Note that no mesons with higher isospin have been found.

\subsubsection{Flavors, SU(3)$_F$ and meson multiplets} 

Besides $I_3$, mesons have three other compatible additive integer quantum numbers, or flavors, called strangeness $\mathsf{S}$, charm $\mathsf{C}$, and bottomness (or beauty) $\mathsf{B}$, conserved under strong interactions. 
Mesons with non-vanishing charm have masses bigger than 1.85 GeV 
and those with non-zero bottomness are heavier than 5 GeV.
There is an additional additive baryon number ${\cal B}$, which is zero for mesons, a positive integer for baryons, and negative for antibaryons.

The ``light mesons'' that the title of this review refers to are unflavored ($\mathsf{S=C=B=0}$) or strange mesons ($\mathsf{S=\pm1, B=C=0}$), whose masses are less than $\sim 2.5\,$GeV.

\begin{figure}
%\centering
    \includegraphics[width=\textwidth]{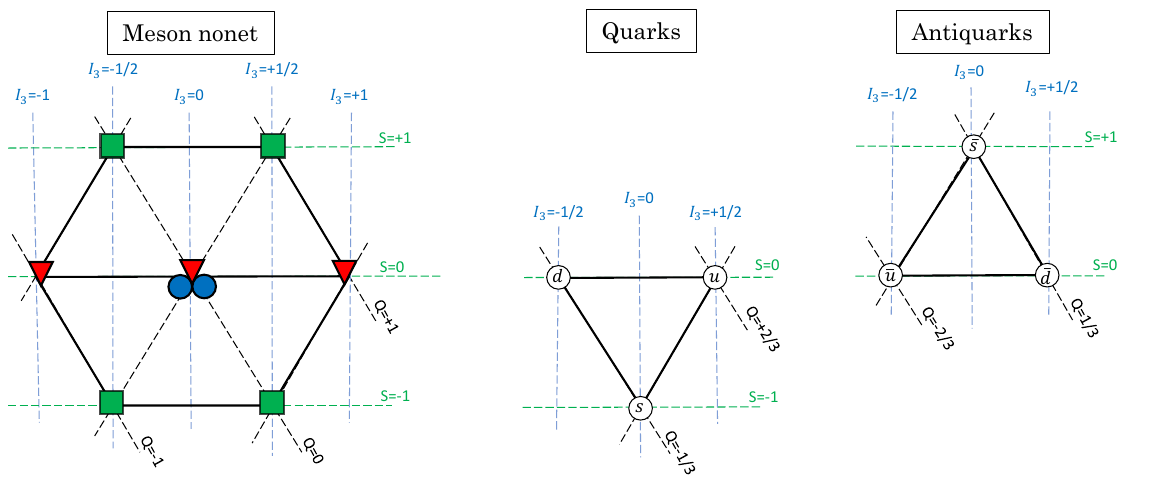}
	\caption{ Weight diagrams of SU(3)$_F$ representations of interest for light-meson spectroscopy and the quark model. Note the diagonals that contain particles with the same electric charge following the NNG formula $\mathsf{Q}=I_3+({\cal B}+\mathsf{S})/2$, where the baryon number ${\cal B}=0,1/3,-1/3$ for mesons, quarks, and antiquarks, respectively.
    \textbf{Left:} Typical SU(3)$_F$ meson nonet made of one SU(3)$_F$ singlet (blue circle) and an octet, i.e. $\bf 8\otimes 1$. The octet contains an isosinglet (blue circle), an isotriplet (red triangles), both with $\mathsf{S}=0$, and two strange isodoublets (green squares), with $\mathsf{S}=1$ and $\mathsf{S}=-1$, which are charge-conjugates of one another.	
    \textbf{Center:} Quarks belong to the SU(3)$_F$ fundamental representation, {\bf 3}. The $u,d$ quarks form an isodoublet with $\mathsf{S}=0$, and the strange quark $s$ forms an isosinglet with $\mathsf{S}=-1$. \textbf{Right:} Antiquarks belong to the conjugate fundamental representation $\bf \bar 3$.
      }
    \label{fig:SU3irreps}
\end{figure}

Another recurrent regularity can be observed in nonets of light mesons with the same $J^P$, illustrated in the left panel of Fig.\ref{fig:SU3irreps}.  Note that $I_3$ grows horizontally from left to right, whereas strangeness grows from bottom to top. Each nonet is made of:
\begin{itemize}
    \item One meson {\bf octet}, made in turn of:
    \begin{itemize} 
        \item[-] One unflavored {\bf isosinglet} (blue circle), with $I=0,\mathsf{S}=0$.
        \item[-] One unflavored {\bf isotriplet} (three red triangles), with $I=1,\mathsf{S}=0$. Very often, the isosinglet and isotriplet masses differ by tens of MeV.
        \item[-] {\bf Two isodoublets} with $I=1/2$  (four green squares). One with strangeness $\mathsf{S}=-1$ and the other $\mathsf{S}=+1$, which are each other's charge conjugate. The mass of the four members of these flavored isodoublets differs from the isotriplet mass by hundreds of MeV.
    \end{itemize} 
    \item One unflavored meson {\bf singlet} (a second blue circle), which is also an isosinglet.  The extra singlet mass differs by a few hundred MeV from the isotriplet mass.
\end{itemize}
 These light mesons satisfy the Nishijima-Nakano-Gell-Mann (NNG) formula for the electric charge \cite{Nakano:1953zz,Nishijima:1955gxk,Gell-Mann:1956iqa}, namely $\mathsf{Q}=I_3+({\cal B}+\mathsf{S})/2$, with baryon number ${\cal B}=0$. Discounting the kinematic effects of their masses, the members of these nonets, particularly those within the octet, behave as degenerate states of a single entity up to roughly a 20\% accuracy concerning strong interactions. 

Gell-Mann and Ne'eman first observed this symmetry \cite{Gell-Mann:1961omu,Neeman:1961jhl,Gell-Mann:1964ook}, and recognized the octet and its isomultiplet structure as an 8-dimensional representation of an SU(3)$_F$ group, denoted by {\bf 8}. Note that the SU(3) group has eight generators. For the recurrent appearance of the number 8, this scheme was dubbed ``The Eightfold Way" by Gell-Mann \cite{Gell-Mann:1961omu, Gell-Mann:1964ook}. Additional singlets, abbreviated as {\bf 1}, correspond to the trivial one-dimensional representations of that group, in which the state remains the same under SU(3)$_F$ transformations.  Octets and decuplets (\textbf{10}), which are other SU(3)$_F$ representations, are also frequently observed for baryons. 

The critical observation is that the SU(3)$_F$ multiplet structure of hadrons can be obtained from the combinations of smaller, irreducible multiplets which are not observed as such in the hadron spectrum. The discovery of the Eightfold Way soon suggested the existence of entities more elementary than hadrons, which we discuss right after introducing some general notation.

\begin{BoxA}[sec5:box1]{Mathematical interlude: SU(3) representation theory}

It is convenient to introduce some mathematical concepts and notation briefly. In Group Theory, a representation of dimension $n$ means that each group element is mapped
with a one-to-one correspondence into a (unitary) matrix acting over a complex vector space of dimension $n$. The transformation under the group of any vector within that space falls back into that space again.  Very naively, a representation is called irreducible if it does not contain representations of smaller dimensions. By conjugating all the matrices of a given representation, one obtains the conjugate representation, which is not always equivalent to the original one.

The plots in Fig.\ref{fig:SU3irreps} are called SU(3) weight diagrams of each representation.  Each state, or weight, represents an element of the basis of the vector space upon which the group representation is acting. Its coordinates correspond to its $I_3$ and $\mathsf{S}$ eigenvalues. Note that only two group generators can be diagonalized simultaneously, and that is why two numbers are enough to characterize the weights of the elements of the basis. All representations of the SU(3) group can be generated from different combinations of just two irreducible representations of dimension three, called fundamental, or ${\bf 3}$, and anti-fundamental (or conjugated), ${\bf \bar 3}$, whose weight diagrams we show in the center and right panels of Fig.\ref{fig:SU3irreps}, respectively. 

Mathematically, the operation to combine two representations is called the tensor product, denoted by $\otimes$. The dimension of the resulting representation is the product of the dimensions of the original representations. Its weight diagram is obtained by adding, in all combinations, the weight vectors of the first
representation to the weight vectors of the second (or vice versa). For example, a simple way to obtain the weight diagram of the nonet is from the ${\bf 3\otimes \bar{3}}$ tensor product. However, the nonet can be decomposed into irreducible representations as ${\bf 3\otimes \bar{3}=8\oplus1}$, i.e., the octet and the singlet. The ``direct sum" symbol $\oplus$ means that the representation has been decomposed into two independent irreducible representations. Note that there are other ways of obtaining nonets and octets.

Another example of relevance is ${\bf 3 \otimes 3=\bar 3\oplus 6}$, which means that the tensor product of two fundamental representations gives rise to a representation that is the direct sum of a fundamental representation ${\bf 3}$ and an anti-fundamental representation ${\bf \bar3}$. Actually, the antisymmetric combinations inside ${\bf 3 \otimes 3}$ are the ones that form the ${\bf \bar 3}$ representation. For us here, this means that two quarks in a conveniently antisymmetrized state, frequently called a diquark, can behave as an antiquark under SU(3) transformations. Conversely, ${\bf \bar 3 \otimes \bar 3=3\oplus \bar 6}$, and two antiquarks in an antisymmetric state, called an antidiquark, can behave as a quark.
\end{BoxA}

\subsubsection{Naming scheme}
In 1986, the Particle Data Group, in their RPP \cite{ParticleDataGroup:1986kuw}, adopted a naming scheme for mesons. As a general rule, it assigns a letter following the quantum numbers $I,P,C$ of each meson and a subscript $J$, followed by their approximate mass (rounded to MeV). 
As we have already seen, nonets come with two isosinglets, which have the same quantum numbers. Thus, when identified as members of a nonet, these two isosinglets share the same letter but are distinguished with a prime. This is the case of the $\eta,\eta'$, $f_J, f'_J$, $h_J,h_J'$.
However, for historical reasons, there are several exceptions to the general rule.
In particular, the two isosinglet vector mesons, $1^{--}$, are not differentiated by a prime, but are called $\omega_J$ and $\phi_J$ instead. 
In addition, the subscript $J$ and sometimes the mass are omitted for mesons in the two lowest nonets, $0^{-+}$ and $1^{--}$, since they were the first to be discovered, are the most common in all reactions, and their properties are very well known. 

When referring to individual members within each isotriplet or isodoublet, their charge is added as a superscript. 
Strange mesons have two additional rules.
 Isodoublets with strangeness $\mathsf{S}=-1$ are generically denoted with a bar over their letter $K$, which is omitted for $K^-$ but kept for $\bar K^0$. In addition, when the spin-parity belongs to the so-called ``natural" series, with $P=(-1)^{J}$, i.e.,
$J^P=0^+,1^-,2^+,...$, a superscript star, $^*$, is added.

In Table~\ref{tab:ordmesonnames}, we provide a list of light-meson names following this scheme. Note that names are assigned without any reference to the composition of the mesons. In constituent models, the constituents of these mesons possess a total angular momentum $L$ and spin $S$. For convenience in our discussion below, we also present these numbers in the table for the case of an ordinary $q\bar q$-meson assignment in the naive quark model that we will introduce later.

\begin{table}%[b]
%	\TBL{
\centering
\caption{\label{tab:ordmesonnames} Examples of the meson naming scheme. Recall that $C$ is only defined for the unflavored neutral members of the multiplet. For higher $J$, names follow the same pattern, but with $J$ as a subindex. Note that the names are assigned without any reference to their inner composition. Nevertheless, in this table we also provide the $L$ and $S$ assignments corresponding to the possible quantum numbers of ordinary $q\bar q$ mesons up to $L=3$ in the quark model, together with the spectroscopic notation. The later provides the combination $n^{2S+1}L_J$, where $n=1,2,...$ is the radial number and the angular momentum $L=0,1,2,3,...$ is given in spectroscopic notation $L=S,P,D,F...$, respectively. 
The same $J^{PC}$ can be obtained with the same $q \bar q$ valence composition but with different QM configurations. This can be achieved by changing the radial quantum number $n$, i.e.,  radial excitations, or with different combinations of $L$ and $S$. For example, we show two possible QM configurations for the $1^{--}$ and $2^{++}$ states.}
\medskip
{\normalsize
\begin{tabular}{|c|c|c|c|c|c|c|c|c|c|c|c|c|c|c|}
\hline
$J^{PC}$ & $0^{-+}$ & $1^{--}$ & $1^{+-}$ & $0^{++}$ & $1^{++}$ & $2^{++}$ & $2^{-+}$ & $1^{--}$ & $2^{--}$ & $3^{--}$  & $3^{+-}$ & $2^{++}$ & $3^{++}$ & $4^{++}$ \\
\hline
$I$=1 & $\pi$ & $\rho$ & $b_1$ & $a_0$ & $a_1$ & $a_2$ & $\pi_2$ & $\rho_1$ & $\rho_2$ & $\rho_3$  & $b_3$ & $a_2$ & $a_3$ & $a_4$ \\
$I$=0 & {\footnotesize $\eta$, $\eta'$} & {\footnotesize$\omega$, $\phi$} & {\footnotesize$h_1$, $h'_1$} & {\footnotesize$f_0$, $f'_0$} & {\footnotesize$f_1$, $f'_1$} & {\footnotesize$f_2$, $f'_2$} & {\footnotesize$\eta_2$, $\eta'_2$} & {\footnotesize$\omega_1$, $\phi_1$} & {\footnotesize$\omega_2$, $\phi_2$} & {\footnotesize$\omega_3$, $\phi_3$} & {\footnotesize$h_3$, $h'_3$} & {\footnotesize$f_2$, $f'_2$} & {\footnotesize$f_3$, $f'_3$} & {\footnotesize$f_4$, $f'_4$ }\\
$I$=1/2 & $K$ & $K^*$ & $K_1$ & $K^{*0}$ & $K_1$ & $K^*_2$ & $K_2$ & $K^*_1$ & $K_2$ & $K^*_3$ & $K_3$ & $K^*_2$ & $K_3$ & $K^*_4$\\
\hline\hline
$L$ & 0 & 0 & 1 & 1 & 1 & 1 & 2 & 2 & 2 & 2 &  3 & 3 & 3 & 3 \\
$S$ & 0 & 1 & 0 & 1 & 1 & 1 & 0 & 1 & 1 & 1 &  0 & 1 & 1 & 1 \\
$n^{2S+1}L_J$ & $n^{1}S_0$ & $n^{3}S_1$ & $n^{1}P_1$ & $n^{3}P_0$ & $n^{3}P_1$ & $n^{3}P_2$ & $n^{1}D_2$ & $n^{3}D_1$ & $n^{3}D_2$ & $n^{3}D_3$ &  $n^{1}F_3$ & $n^{3}F_2$ & $n^{3}F_3$ & $n^{3}F_4$ \\
\hline
\end{tabular}
}
%}
\end{table}

\subsubsection{Introducing quarks}
Indeed, the main features of the octet and other representations that appear in light-hadron spectroscopy can be
nicely explained in terms of more ``elementary'' fermions with $J^P=1/2^-$.
Gell-Mann \cite{Gell-Mann:1964ewy} and Zweig \cite{Zweig:1964ruk} first realized this and, respectively, named these particles ``quarks" and ``aces", but this last name did not prevail. Quarks appear in the three-dimensional or fundamental representation, called $\bf 3$. How they are identified with the representation weights is shown in the center panel of Fig.~\ref {fig:SU3irreps}. The unflavored ``up'' and ``down'' $u,d$ quarks form an isodoublet with $\mathsf{S}=0, I=1/2, I_3=1/2$ and $\mathsf{S}=0, I=1/2,I_3=-1/2$, respectively. In contrast, the strange quark is an isosinglet with $\mathsf{S}=-1,I=I_3=0$ (Note that, for historical reasons, the strange quark $s$ has negative strangeness). They all have baryon number ${\cal B}=1/3$. Their antiparticles, belonging to the conjugate representation $\bf \bar 3$, are called ``antiquarks'' and have opposite quantum numbers to their corresponding quarks. Quarks and antiquarks also satisfy the NNG formula $Q=I_3+({\cal B}+S)/2$. In particular, $u,d,s$ have $\mathsf{Q}=2/3,-1/3,-1/3$, respectively, whereas their respective antiquarks have opposite charges.

Most properties of everyday matter can be explained with the $u$ and $d$ quantum numbers alone, and they are often referred to as ``normal'' quarks $n=u,d$. Strange hadrons were initially found in cosmic rays. 
In the 1970s, with the advent of high-energy colliders, it became possible to access more massive hadrons, whose dynamics required the introduction of additional quarks and flavors, which are conserved under the strong interaction.
In particular, besides the up, down, and strange quarks, there are charm, bottom (or beauty), and top quarks, abbreviated by $ c$, $b$, and $t$, respectively. The $(c,s)$ and $(t,b)$
pairs replicate the isodoublet structure of $(u,d)$. 
Only the $s$ quark carries strangeness flavor, $\mathsf{S}=-1$, only the $c$ quark carries charm $\mathsf{C}=1$, only the $b$ quark carries bottomness $\mathsf{B}=-1$, and only the $t$ quark carries ``topness" $\mathsf{T}=1$. The same happens with their antiquarks, which carry the respective opposite flavors. 
The generalized NNG formula thus reads $\mathsf{Q}=I_3+({\cal B}+\mathsf{S+C+B+T})/2$.
Topness has not been observed at the hadron level because top quarks decay very rapidly, probably before forming hadrons (Nevertheless, toponium may have been observed recently\cite{CMS:2025kzt}, but its topness is still zero).

The first three quarks $u,d,s$ are known as ``light quarks", whereas $c,b$ and $t$ are called ``heavy quarks". Heavy quarks play a very minor role (only through higher-order quantum corrections) in the dynamics of the light mesons to which this review is dedicated. 

\subsection{Quantum Chromodynamics in a nutshell} 
The fundamental force that dominates the 
dynamics of quarks and is responsible for the formation of hadrons is called Quantum Chromodynamics, or QCD. 
Here, we provide a concise introduction to the fundamental concepts necessary to understand the physics of mesons. Detailed accounts of QCD can be found in other chapters of this Encyclopedia or the recent review \cite{Gross:2022hyw}, which celebrates its 50th birthday. For our purposes here, let us recall that QCD affects quarks, which are spin-1/2 fermions that, besides the flavor quantum numbers discussed before, possess one of the three types of so-called ``color'' charges \cite{Fritzsch:1973pi}.  Antiquarks possess the corresponding three opposite ``anti-colors".  As with electric charges, two opposite colors combine to form a color-neutral state, or a color singlet. However, there are many other ways of obtaining neutral states. In particular, a frequent neutral configuration can be obtained with three quarks, each with a different color. This configuration vaguely reminds us of our familiar experience with light, where the colors red, green, and blue combine to produce white. Indeed, this is why we use ``color" and ``Chromodynamics" in the context of the strong force, although they have nothing to do with the colors of objects in our everyday life.  QCD is invariant under yet another SU(3)$_c$ symmetry acting on the three-dimensional vector formed by the three color components of the quark field.   This symmetry is exact, and the $c$ subindex stands for color to differentiate it from the approximate SU(3)$_F$ flavor symmetry. In addition, it is a gauge or local symmetry, which means that the interaction Lagrangian is invariant under SU(3)$_c$ transformations that can differ from different locations in space-time. In the same way that photons, which are the gauge bosons of electromagnetism, 
carry the electromagnetic interaction, there are eight carriers of the color interactions, called gluons. They are spin-1 gauge bosons, each bringing one of the eight possible combinations of a color and a different anti-color. 
Quarks and gluons are considered fundamental particles because there is no evidence to suggest they have a size or structure. 

Unlike photons in quantum electrodynamics (QED), gluons are color-charged and interact directly with each other, resulting in distinct QCD and QED behaviors. In particular, the color interaction becomes so strong and attractive in the color-singlet channel that it confines the color charges within color-neutral composite states; these are the hadrons we have been discussing all along. 
There is not yet an analytic proof of color confinement in  QCD. Still, we cannot observe quarks and gluons directly; instead, we observe hadrons. 
Only at asymptotically small distances well below the hadron-size scale (or, equivalently, energies well above the GeV scale) can QCD be treated perturbatively in powers of its coupling constant, using quark and gluon degrees of freedom \cite{Politzer:1973fx,Gross:1973id}. 
Therefore, hadrons become the relevant degrees of freedom of the strong interaction between the ``asymptotic'' QCD regime and the nuclear realm.

In the hadron realm, QCD is not perturbative in terms of powers of its coupling constant, and calculations have to be tackled with non-perturbative techniques. Here, we provide brief definitions for three of them, which are of interest for our discussions below.
The first is lattice QCD, where space-time is discretized on a lattice, and the quark and gluon fields are represented by variables defined at the lattice sites and along the links between sites, respectively. This approach, very rigorous in theory, is computationally very intensive in practice. Hence, two approximations are usual: to neglect quark dynamics (quark loops), called quenched-lattice-QCD, and/or to have large quark masses, which then yield nonphysically large hadron masses. We refer the reader to the chapter on lattice and hadron spectroscopy in this Encyclopedia \cite{Prelovsek:2025gmd}.
A second, and very popular, non-perturbative approach is the use of Effective Field Theories. In this case, one first identifies the relevant degrees of freedom in a particular regime defined by a small parameter. Then one constructs the most general Lagrangian consistent with the QCD symmetries. This Lagrangian gives rise to an effective quantum field theory, which provides rigorous results as an expansion in that parameter, but is only valid when the parameter is small. 
We will explain the low-energy expansion of QCD, known as Chiral Perturbation Theory (ChPT), in more detail below.
A third approach is the $1/N_c$ expansion around $N_c=3$, where $N_c$ is the number of colors, which is very rigorous, but the expansion parameter is not very small, and the uncertainties are significant. It should not be confused with the $N_c\to\infty$ limit, which is of mathematical interest, but has limited phenomenological applications.
For detailed introductions to these topics, we refer to textbooks on QCD and the recent review \cite{Gross:2022hyw}.

Although hadrons are color-neutral, since they are made of quarks and gluons, they still interact strongly due to residual QCD effects. In contrast, the other particles in the Standard Model, all believed to be elementary, colorless particles, do not directly experience the strong color interaction; instead, they experience it feebly through quantum corrections (i.e., loops) involving quarks, gluons, or hadrons.

Note that hadrons inherit their baryon number, isospin, flavor, and electric charge from their quark and antiquark content since gluons are flavorless and electrically neutral. 
In particular, if the $u$ and $d$ masses were the same, the QCD Lagrangian would be invariant under isospin SU(2) transformations of the $(u,d)$ doublet. In fact, the $u$ and $d$ quarks owe their names to being the ``up'' and ``down'' components of this doublet when written as a vertical or column vector. 
In real life, the typical hadron physics scale of hundreds or thousands of MeV is much larger than the difference between the $u$ and $d$ so-called ``current masses'' \footnote{Since quarks are not observed freely, their ``current mass'' has a subtle definition. We have adopted the current-mass estimates of the RPP 2024 \cite{ParticleDataGroup:2024cfk},
which are provided in the mass-independent $\overline{\rm MS}$ renormalization scheme, at a renormalization scale of $\mu=2\,$GeV.},
which are the ones used in QCD calculations. For this reason, although the isospin symmetry is not exact, it is an excellent approximation.
Similarly, if the $u,d$ and $s$ current masses were equal, QCD would be symmetric under SU(3)$_F$ transformations of the $(u,d,s)$ vector of fields.
The strange-quark current-mass $\sim 93 {\rm MeV}$ is much bigger than those of the $u$ and $d$ quarks, but, still, it is less than 20\% of the mass of the lightest strange hadron, and thus SU(3)$_F$ is also an approximate symmetry of the strong interactions, although not as good as isospin symmetry.

Since quarks and antiquarks are spin-1/2 fermions, a meson, which is a colorless configuration with integer spin, can be made, for instance, of any number of bound quark-antiquark pairs with opposite color and any flavor. 
Moreover, binding these pairs inside a hadron requires gluons, which precisely receive their name because they ``glue'' (confine) the quarks and themselves inside the hadron. We have already seen that two quarks, conveniently anti-symmetrized, can behave as an antiquark under SU(3), and vice versa. Hence, it is also possible to obtain a color-singlet configuration with three quarks of different colors together with any number of quark-antiquark pairs, but that configuration corresponds to a baryon. 
In any case, it is the whole combination of quarks, antiquarks, and gluons, all bound and interacting among themselves through QCD, that constitutes a hadron.

Nevertheless, the simplest mathematical way of obtaining the quantum numbers of a color singlet is to consider
just one quark-antiquark pair, and that is why, too often, mesons are naively and erroneously defined as quark-antiquark bound states, i.e., only as depicted in the left panel of Fig.\ref{fig:typesofmesons}. Similarly, baryons are naively defined as made of three quarks. Still, this observation has given rise to the very popular Quark Model, which is not derived from QCD. 
However, since QCD becomes too strong below 1-2 GeV for the usual perturbative techniques, this drastic simplification has become very useful.
In particular, although not complete or extremely accurate, a very simple model for quarks has proven phenomenologically very successful in classifying hadrons and describing approximately many features of the hadron spectrum and dynamics. We introduce it next.

\section{The Quark Model as an organizational scheme: Types of mesons from their constituent partons}

Within a relatively simple ``Quark Model" (QM), it is indeed possible to describe a great deal of hadron spectroscopy.  A detailed account of the QM in its most naive or more elaborated versions (bag model, chiral-QM, etc ) can be found in other chapters of this Encyclopedia \cite{Entem:2025bqt} or in the recent reviews \cite{Gross:2022hyw} and \cite{Nefediev:2025vmo}, the latter focuses particularly on the chiral QM. Here, we will provide a very concise summary focused on mesons, their classification, and their potential interpretation as either ordinary or extraordinary configurations. We will discuss only the most qualitative aspects of the naive QM for classification purposes. Dynamical predictions of the QM, chiral-QM, or QCD-inspired quark models lie beyond our classification scope, and we refer the reader to the specialized literature (see references in \cite{Gross:2022hyw,Entem:2025bqt,Nefediev:2025vmo}).

Intuitively, the naive QM follows from the idea that within a hadron, there is always an indefinite number of quark-antiquark pairs and gluons, generically called ``partons", that form a neutral ``sea". It is neutral in the sense that the sea does not contribute to the hadron quantum numbers, but it is responsible for the strong binding dynamics and the generation of mass at the 100 MeV or GeV scale.  In addition, over the neutral sea population, there are always a few ``extra" partons, responsible for the hadron quantum numbers. In analogy with atomic physics, where unpaired electrons are responsible for the quantum numbers of the ionization state, these extra partons are called ``valence" quarks, antiquarks, or gluons. It is then convenient to think of a model where the hadron as composed only of ``constituent partons", which are effective degrees of freedom that carry quantum numbers as valence quarks or gluons would do, but that incorporate all the ``sea", and strong color-interaction non-perturbative effects into a ``constituent" mass $M\sim 300\,$ MeV to be added to their current mass.

One should keep in mind that the QM is not derived from QCD, although it shares its symmetries, particularly the approximate isospin and flavor symmetries. It is a basic scheme to classify hadrons. Nevertheless, by incorporating additional specific dynamics, it can also explain semi-quantitatively certain dynamical regularities, such as hadron masses, magnetic form factors, and decay rates, among others.

The original QM dealt only with the lightest quarks $u$, $d$, and $s$, which provide a basis of states for an approximate SU(3)$_F$ flavor symmetry, and the ``Eightfold way". This classification scheme can be easily extended to include $c$ and $b$. However, $N=4,5$ flavor SU(N) symmetries are not as good approximations as SU(3)$_F$ and, definitely, not as good as the SU(2) isospin. Here we deal with ``light mesons", which from the QM perspective 
means that they are only made of constituent $u$, $d$, $s$ quarks, antiquarks, and/or gluons.

Since constituent partons inherit their current quantum numbers, they have $\cal B$=+1/3 for each quark and -1/3 for each antiquark, which have spin 1/2, and $\cal B$=0 for gluons, with spin 1. Thus, since mesons have zero net baryon number and integer spin, if they contain constituent quarks, they must appear in quark-antiquark pairs. Otherwise, the hadron is a baryon. 

Thus, the first meson configuration that comes to mind is a color-neutral quark-antiquark pair, denoted $q^a\bar q^b$. Here, $a$ and $b$ stand for different flavors, and a sum over colors is implied to render the meson colorless. These generic 
$q\bar{q}$ states are usually called ``ordinary mesons", because they were the only ones considered in detail in the first QM proposals \cite{Gell-Mann:1964ewy, Zweig:1964ruk}. As we have already commented, at the popular level and quite often in specialized literature, mesons are naively defined simply as hadrons composed of a valence quark and an antiquark.
However, already in the original QM proposals, Gell-Mann made explicitly clear in  \cite{Gell-Mann:1964ewy} that ``{\it ...mesons are made out of ($q \bar q$), ($q  q \bar q\bar q$), etc.}'',  and Zweig wrote in a footnote of \cite{Zweig:1964ruk} that ``{\it ...mesons could be formed from $\bar A A$,$\bar A \bar A A A$, etc.}"  (recall Zweig called quarks ``aces" A).

Mesons in non-$(q\bar q)$ configurations, which may also include valence gluons, are nowadays called ``non-ordinary" states. When they have $J^{PC}$, charge or flavor quantum numbers impossible to obtain in $\bar q q$ configurations, they are also called ``exotic" mesons. If a $J^{PC}$ combination is not attainable by a $q\bar q$ configuration, it is called ``spin-exotic", whereas if the charge or isospin are
impossible for $q\bar q$ states, it is called ``flavor-exotic".
However, note that there can also be non-ordinary mesons that are not exotic, sometimes called ``crypto-exotics". Such states differ from ordinary mesons in various properties, including mass hierarchies, unexpected total or partial widths, and other more subtle characteristics, such as their Regge behavior or $N_c$ dependence. 
Unfortunately, the distinction between non-ordinary and exotic is often blurred in the literature. To add more confusion, some states (such as $X(3872)$) are usually called exotic, not because of their exotic quantum numbers, but because of other unusual properties.

There is indeed compelling evidence that at least the five different configurations
represented in Fig.1, play a significant role in the interpretation of light mesons in terms of their constituent partons. We discuss them next, following the order from left to right in Fig.\ref{fig:typesofmesons}. However, after briefly describing these simple configurations and their main characteristics, we will see that in real life, many of them appear mixed, which makes their identification a highly complex task.

\begin{figure}
%\centering
    \includegraphics[width=\textwidth]{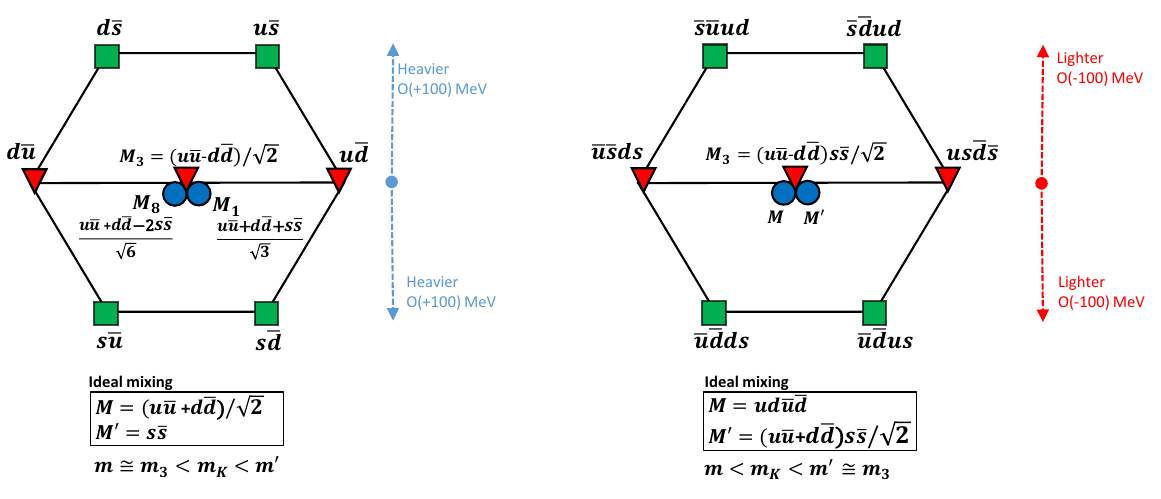}
	\caption{Interpretation of light-meson nonets: a) as ordinary $q\bar q$ mesons (hybrids as well, since valence gluons do not contribute to flavor and isospin) and b) as tetraquarks/molecules. Note the different mass hierarchy expectations.  In the ordinary hierarchy, the strange $I=1/2$ doublets (squares in green) contain one more strange quark than the $I=1$ triplet (triangles in red) and are expected to be $O(100\,{\rm MeV})$ heavier. In the tetraquark/molecule interpretation, it is just the other way around. The hierarchy is said to be inverted. Note that since the two isosinglets of the nonet have the same quantum numbers, they can mix. For these singlets, we provide as an example the relatively frequent ideal-mixing scenario, its valence composition, and its expected hierarchy. Note that in these two scenarios, the heaviest member of the nonet is one of the two isoscalars, a situation which is observed very frequently.
    }
	\label{fig:multiplets}
\end{figure}

    \subparagraph{\bf Ordinary mesons: Quarkonium}
    They are predominantly built as a $q\bar{q}$ constituent configuration, known as quarkonium, which does not allow for mesons with $I > 1$.  We insist that, unfortunately, this is often mistaken for the definition of a meson. The nonet composition in terms of quarkonia with different flavors is shown in the leftmost panel of Fig.\ref{fig:multiplets}. Observe that the strange mesons within the nonet are expected to be $O(100 {\rm MeV})$ {\it heavier} than the isotriplet, because they contain one $s$ or $\bar s$. The masses of the isosinglets depend on how they mix. In the so-called ``ideal" mixing scheme, also illustrated in Fig.\ref{fig:multiplets}, one of them is predominantly composed of just one $s\bar{s}$ pair, and it is expected to be the heaviest in the nonet. We will see below that this mixing scheme is an excellent approximation in the $1^{--}$ vector nonets, particularly in the lowest one, which contains the $\rho(770)$ and might be considered the archetypic ordinary-meson nonet, as well as in the $1^+$, $2^{++}$ and $3^{--}$ nonets.

 For these ordinary mesons, constituent normal quarks have a mass of $\sim 350\,$MeV
 whereas the strange constituent mass is $\sim500\,$MeV. The spin of the valence $q\bar q$ pair, $S$, can be 0 or 1. In addition, they have orbital angular momentum $L$, with each unit of $L$ adding $\simeq$400-600 MeV of mass. The total spin $J$ of the ordinary meson is then $\vert L+S\vert\leq J\leq L+S$,  
parity is given by $P=-(-1)^L$, and $C$-parity by $C=(-1)^{L+S}$.
Since $C$-parity is only defined for neutral mesons, it is generalized to non-strange mesons, defining $G$-parity as $G=(-1)^{I+L+S}$. It is an instrumental number to determine that some interactions and decays may not be allowed by $G$-conservation, but we will mostly ignore it in this introductory review.
With these constraints, not all $J^{PC}$ combinations can be obtained by  $q\bar q$ configurations. 
Note that states with $S=0$ have opposite $P=-C$, giving rise to meson states with $(J=L={\rm even})^{-+}$ or $(J=L={\rm odd})^{+-}$.
In contrast, states with $S=1$ have $P=C$, and provide mesons with $J^{++}$ and $J^{--}$ for all $J>0$ as well as with $0^{++}$. Consequently, mesons with $0^{--}, (J={\rm even})^{+-}$ or $(J={\rm odd})^{-+}$ cannot be quarkonia. In Table~\ref{tab:ordmesonnames}, we have listed all possible $J^{PC}$ quantum numbers attainable by ordinary mesons, up to $L=3$ and, therefore, the combinations $J^{PC}=0^{--}, 0^{+-},1^{-+},2^{+-},3^{-+},4^{+-}$, are absent.

As in atomic physics, the quark model radial excitations are labeled by the ``principal" or ``radial" quantum number $n\geq1$, with $n=1$ for ground states. Thus, $q\bar q$ states are named $J^{PC}$ for their measurable quantum numbers, but also $(n^{2S+1}L_J)$ for the quantum numbers in the QM description, although for $L$ the usual spectroscopic notation $L=S,P,D,F,...$ is used instead of $L=0,1,2,3,...$ In Table~\ref{tab:ordmesonnames} we provide the spectroscopic notation for the states of interest in this review, up to $L=3$. Quite roughly, and as happened with $L$, increasing $n$ by one unit increases the mass of the state by 400-600 MeV.

    \subparagraph{\bf Tetraquarks} 
    Tetraquarks are mesons whose valence constituents are two quarks and two antiquarks, i.e. $qq\bar q\bar q$. Following the chapter on ``Hadronic molecules and multiquark states" in this Encyclopedia \cite{Hanhart:2025bun}, we will describe two limiting cases, namely, ``Compact tetraquarks" and ``Meson molecules" --- where the quark-antiquark pairs bind first into two mesons, bound again by meson-scale strong interactions. However, tetraquark states may also lie somewhere in between these two extreme cases. Although this distinction has become more frequent in recent times, in the literature, authors may refer to tetraquarks generically, without specifying the type, or use the term "tetraquark" meaning a compact tetraquark.

    \subparagraph{\bf Compact Tetraquarks} 
    Compact tetraquarks are made of a diquark and an antidiquark, as illustrated in the second drawing from the left in Fig.\ref{fig:typesofmesons}. Recall that combining two fundamental representations of SU(3), we find ${\bf 3\otimes{3}= 6\oplus \bar{3}}$. The ${\bf \bar{3}}$, which is the antisymmetric configuration, behaves as the conjugate fundamental representation. In physics terms, a diquark, which is a two-quark antisymmetric configuration, transforms under SU(3) as an antiquark. And vice versa, for the antidiquark transforming and behaving as a quark. This works both for color and flavor. Thus, one can get a color singlet by combining a quark and a diquark, which would be a baryon. But binding a diquark and an antidiquark inside a colorless meson, one can also get a flavor singlet or a flavor octet, as we show in the right panel of Fig.\ref{fig:multiplets}. Note that the tetraquark mass hierarchy is inverted compared to the ordinary meson case, since we expect the isodoublets with non-zero strangeness (green squares) to be $O(100\, {\rm MeV})$ {\it lighter} than the isotriplet (red triangles). This is because the isodoublets now contain an $s\bar s$ valence pair, whereas the isodoublets only contain one $s$ or one $\bar s$. Concerning the two unflavored isosinglets, many mixing schemes are possible. The ``ideal mixing" scenario, provided in Fig.\ref{fig:multiplets}, corresponds to the case where only one of the two contains a valence $s\bar s$  pair, whereas the other is made only of valence normal quarks. In such a case, the former has a mass comparable to the triplet, and the latter is the lightest in the nonet. We will see below that the lightest scalar nonet $J^{P}=0^+$ follows this pattern. One frequent caveat raised for tetraquark interpretations is that the tensor product ${\bf (3\otimes 3)\otimes(\bar 3\otimes\bar 3)=1\oplus8\oplus36\oplus18\oplus\overline{18}}$, thus a plethora of multiplets different than octets and singlets could in principle exist but are not observed. Of course, the explanation could be that they are not seen because they are more massive and/or short-lived.

    \subparagraph{\bf Meson molecules}  Strictly speaking, because their valence content is $qq\bar q\bar q$, these states are tetraquarks. Depending on the context, molecules are also included when referring to tetraquarks in general.
    
    However, in molecular configurations, rather than grouping first into a diquark and an antidiquark and then binding together, the two quark-antiquark pairs bind first into two mesons and then form a bound state from residual strong interactions at the meson scale.
    Their mass hierarchy is expected to be similar to compact tetraquarks.
    We illustrate this meson-molecule configuration in the central panel of Fig.\ref{fig:typesofmesons}. On a first approximation, one would expect the binding of these two mesons to be weaker than in the tetraquark case, so that these states would find it easier to dissociate and decay. At first sight, the name ``molecule" suggests that these states have a binding energy that is small compared to the masses of the constituent mesons; therefore, one would expect these states to appear near two-body thresholds to which they couple strongly but have a limited phase space for decay, leading to narrow resonances. Very naively, molecules are expected to be less compact than compact tetraquarks (hence the name of the latter). Nevertheless, molecules can also mix with other compositions, and it is even possible that they behave as a ``meson cloud" that would give rise to large widths, larger radius, etc. In the presence of heavy quarks one such configuration is sometimes called hadroquarkonia \cite{Hanhart:2025bun}.
    We will see some of these features in the lightest scalar nonet $J^{P}=0^+$. For a recent review on hadron molecules, we refer the reader to \cite{Guo:2017jvc} and the ``Hadronic molecules and multiquark states" in this Encyclopedia \cite{Hanhart:2025bun}.

    \subparagraph{\bf Hybrids} These are states that contain valence gluons besides valence quarks and antiquarks. The simplest configuration would be $q\bar q g$, as illustrated in the second panel from the right in Fig.\ref{fig:typesofmesons}.
   Flavor structures are now similar to those of ordinary mesons, such as singlets and octets, but allow for ``spin-exotic" quantum numbers, like $1^{-+}$, in addition to the ordinary $J^{PC}$ configurations. Early flux-tube models \cite{Isgur:1984bm} suggested masses around 1.9 GeV whereas bag models preferred $\sim1.4\,$GeV and predicted four nonets \cite{Chanowitz:1982qj,Barnes:1982tx}, including an exotic one with $1^{-+}$. Lattice calculations \cite{Lacock:1996ny,Dudek:2011bn,Dudek:2013yja} suggest a mass of about 1.9 GeV, although they are quenched or have non-physically large quark masses; therefore, they could be lighter.
   We will see that the $\pi_1(1600)$ is a strong hybrid candidate, consistent with unquenched lattice calculations \cite{Woss:2020ayi}, and possibly the $\eta_1(1885)$ could be its partner. For reviews on hybrids, see \cite{Meyer:2015eta,Gross:2022hyw,Chen:2022asf}.
    
    \subparagraph{\bf Glueballs} As we have seen, one of the main features of QCD as a non-abelian gauge theory is that the gluons carry color charges and can interact among themselves.
    It is therefore possible that they could bind to form a meson by themselves, without any constituent quark or antiquark, as 
    represented pictorially in the rightmost panel of Fig.\ref{fig:typesofmesons}. Glueballs cannot have electric charge, isospin, or flavor.
    Thus, they appear as singlets. In principle, they could have either ordinary or spin-exotic numbers. Their decay to meson pairs like pions, or states containing photons, is typically more suppressed than for other mesons. 
    
   Generally, the lightest glueball is expected to be a scalar $J^{PC}=0^{++}$ between 1.5 and 2 GeV; the next one is a heavier tensor with $2^{++}$, and a pseudoscalar $1^{-+}$ of similar mass. 
   This pattern has been a stable prediction across various approaches since the 1990s (see the time evolution of predictions in \cite{Chen:2022asf}).
In particular, lattice-QCD calculations  \cite{Bali:1993fb, Morningstar:1999rf,Chen:2005mg,Gui:2019dtm,Athenodorou:2020ani,Gregory:2012hu}, although some of them quenched, find the
    lightest scalar glueball mass in the 1550-1800 MeV range, the tensor around 2.3 GeV, and a slightly heavier pseudoscalar.   However, the identification of these states with the observed resonances is hindered by the difficulty in implementing dynamical quarks as well as their light masses in lattice calculations. 
    Other phenomenological approaches suggest that the lightest glueball is a scalar, with a mass of around 1700-1900 MeV, and the next two are a few hundred MeV above. 
    Using QCD sum rules \cite{Chen:2021bck}, the tensor comes next. In contrast, the second lightest is the pseudoscalar glueball when using Hamiltonian QCD in Coulomb gauge \cite{Szczepaniak:2003mr}, or non-perturbative solutions of relativistic bound-state equations \cite{Huber:2020ngt}.

Apart from interactions with pairs of mesons, gluon-rich processes are obviously relevant for glueball searches. These include $p\bar p$ annihilation as well as radiative $J/\psi$ decays since, being a $c\bar c$ meson, it produces a photon and, primarily, two gluons. The first process was studied in Crystal Barrel a few decades ago, but it is expected to be further studied shortly in PANDA \cite{PANDA:2021ozp}, at GSI in Germany. BESIII, located in Beijing, is particularly well-suited for studying $J/\psi$ decays and has been providing exciting new results in recent years (see \cite{Jin:2021vct}). Central production via Pomeron exchange (see the Regge section below) is another glue-rich process that was studied at WA102 at CERN and is now being studied at STAR at BNL.

    \subparagraph{\bf Baryonia} 
   Also called baryon molecules or di-baryons. It is not forbidden that $N \bar N$ may resonate or form quasi-bound states
    around 2 GeV, which would look like light-meson resonances. However, these are very difficult to identify, and there are no clear indications that they have been observed, although we will comment on possible candidates. We refer to \cite{Guo:2017jvc} for further details.

    \subparagraph{\bf Mixing} 
    Unfortunately, in real life, all of the above configurations and states, with the same quantum numbers, can and do mix, often distorting the mass hierarchies within each multiplet and making it difficult to identify the valence content of the mesons. Small admixtures violating isospin or flavor have also been observed.  The possibility of mixing grows as the overlap between two mesons increases, i.e., their masses are closer than the size of their widths.  We will see that very often two or more mesons assigned to different multiplets have widths so large that they clearly overlap with each other. We are going to find several types of mixing: 
    
    \begin{itemize}
    
        \item {\bf Isosinglet mixing within a nonet: ideal mixing}
        
     Let us explain in some detail the mixing of isosinglets within the same nonet, because we will find it repeatedly below. It combines the octet-isoscalar state $M_8$ and the singlet $M_1$ to form the two isoscalars $M$ and $M'$. For scalars, these coincide with $f$ and $f'$ respectively. However, for historical reasons, for pseudoscalars $M$ is identified with $\eta'$ and $M'$ with $\eta$. In addition, recall that for vector mesons the notation is $\omega$, identified with $M$, and $\phi$, identified with $M'$. 
     It is usual to keep the same meson names for their QM wave functions and the same quark names for their QM wave functions. Then the mixing is written as:
    \begin{equation}
         M=M_8 \sin\theta+M_1\cos\theta,\quad M'=M_8 \cos\theta-M_1\sin\theta,
    \end{equation}
    where $\theta$ is called the mixing angle. 
   
    The so-called ``ideal-mixing" scenario is the one in which $M$ does not contain strange valence quarks or antiquarks. Different angles yield distinct mass hierarchies and partial decay widths. For ordinary mesons, the ideal-mixing valence-composition of $M$, $M'$ reads:
    \begin{equation}
        M_8=(u\bar u+d\bar d-2s\bar s)/\sqrt{6}, \qquad M_1=(u\bar u+d\bar d+s\bar s)/\sqrt{3}.
    \end{equation}
    Thus, the ideal mixing scenario for quarkonia would require  $\tan\theta=1/\sqrt{2}$, $\theta\simeq35.26^{\rm o}$, ensuring that only normal-quark pairs contribute to $M$ and only $s\bar s$ to $M'$. Namely:
    \begin{equation}
       \qquad M=(u\bar u+d\bar d)/\sqrt{2}, \quad  M'=-s\bar s.
    \end{equation}
    Beware that some authors define the mixing of the isoscalars with respect to the basis of ideally mixed states, instead of the singlet-octet basis, which can be done in several ways. The RPP in its review of the QM defines this rotation by another mixing angle $\alpha\simeq\theta+54.74^\circ$ such that 
    $M'=(\cos\alpha)(u\bar u+d \bar d)/\sqrt{2}-(\sin\alpha)s\bar s$. Replacing $\alpha$ by $\alpha-90^\circ$ gives the orthogonal combination. Ideal mixing occurs when $\alpha=90^\circ$, and $M'=-s\bar s$. However, other authors define the basis in reverse order and use $\beta=90^\circ-\alpha=35.26^\circ-\theta$, and ideal mixing corresponds to $\beta=0^\circ$. Here we will use the angle $\theta$.

    When ideal mixing occurs in quarkonia, one would expect the following hierarchy for the masses: $m\simeq m_3<m_K<m'$, where $m$ is the mass of the lightest isosinglet, $m_3$ is the isotriplet mass, $m_K$ is the mass of the strange isodoublets, and $m'$ is the mass of the heaviest isosinglet. Each ``$<$" symbol represent a difference of $O(100\,{\rm MeV})$. The $1^{--},1^{+\pm}, 2^{++}$ and $3^{--}$ are close to this ideal case, whereas the lightest scalar $f's$ and pseudoscalar mesons $\eta, \eta'$ are far from this scenario, for several reasons that we will discuss below in detail.

    For tetraquarks or molecules,  we have already discussed that the hierarchy $m_K<m_3$ is inverted. In addition, the ideal mixing scenario would also be different from the ordinary meson case, since now $M=u\bar d u\bar d$ and $M'=(u\bar u+d \bar d)s\bar s/\sqrt{2}$. The ideal-mixing tetraquark/molecule hierarchy would be $m<m_K<m'\simeq m_3$, where, once again, the ``$<$" symbol represents a difference of $O(100\,{\rm MeV})$.

   Assuming specific quark contents and that the isosinglets are sufficiently stable, the mixing angle can be calculated from the ratios of combinations of masses of the elements of the nonet.    
   Since these are just approximations, two angles are usually quoted: $\theta_{\rm lin}$ and $\theta_{\rm quad}$, depending on whether the angle is determined from a relation for $\tan\theta$ or $\tan^2\theta$. 
   Mixing angles can also be determined using relative decay rates for the physical states.
   These results are strongly model-dependent, and the sufficiently stable meson condition is not always met, particularly in the light-scalar sector.  
    \item {\bf Mixing between multiplets}.
    So far, we have discussed mixing within the same nonet. But there could be other mixings outside a nonet. The most relevant case is when the isoscalar members of a nonet could mix with an additional singlet. The latter could be coming from a glueball. This is believed to be the case of the scalars between 1.2 and 1.7 GeV, which can be grouped into a nonet and a singlet. Such a singlet is sometimes referred to as supernumerary and is believed to arise from a glueball state. Then, the $f_0(1370)$, $f_0(1500)$, and $f_0(1700)$  would be a mixture of the isosinglet and singlet from the nonet and the supernumerary singlet. The specific mixing pattern remains a matter of intense debate, which most likely involves the members of an additional nonet above 1.7 GeV.
    
    Moreover, it is even possible that not just isosinglets, but other members of one nonet could mix with the members of a nearby nonet with the same quantum numbers. As already commented, mesons are not stable states and have a decay width. If two mesons with the same quantum numbers have masses close enough to overlap significantly within their widths, they will definitely mix. As we will see below, a substantial mixing is believed to happen between the strange members $K_A$ and $K_B$ of the two different $J^P=1^+$ nonets that have been identified so far. Presumably, it also occurs in other cases for which the full nonets have not yet been observed.

    \item {\bf Mixing within different valence configurations}. Note that in this case, the mixing refers to the constituents, which, in principle, can combine to form a meson. In an appropriate reference frame one could even define a Fock space with a state basis of generic $\ket{q\bar q}, \ket{q\bar q q\bar q}, \ket{MM}, \ket{q\bar q g}, \ket{gg},...$ and each meson would be a linear combination of elements of such basis with the appropriate quantum numbers. After all, they would all mix through their decay products. For example, several models suggest that the lightest scalars ($0^{++}$) might contain an ordinary quark-model core \cite{Close:2002zu} surrounded by a meson cloud or simply that interactions at the meson-meson level dress a bare or ``preexisting"  quark state around 1 GeV \cite{Oller:1998zr}, modifying its mass and width.  We will also see below that the identification of glueballs is hindered by their mixing with other nearby states with different constituents. Indeed, it appears that none of the mesons we have observed is a pure glueball, and the most we can hope for is to determine how the components of a glueball are distributed among nearby mesons, or whether one of them is predominantly composed of a glueball component. 

    \item{\bf Mixing due to isospin or flavor breaking}. Finally, we should recall that neither isospin nor flavour symmetries are exact. Actually, mixing between mesons with different quantum numbers of isospin or flavor has been observed. 

    \begin{itemize}
        \item[-] \underline{ $K^0$ and $\bar K^0$ mixing:}  This mixing occurs because strangeness is not conserved by the electroweak force and the flavor and mass eigenvalues of the $d$ and $s$ quarks do not coincide. As a consequence, the lightest strange mesons, $K^0$ and $\bar K^0$, can mix too, which they do predominantly through their $\pi\pi$ decays: $K^0\leftrightarrow\pi\pi\leftrightarrow\bar K^0$. But then it is more convenient to define the CP eigenstates $\ket{K^0_\pm}=(\ket{K^0}\mp\ket{\bar K^0})/\sqrt{2}$, which satisfy $CP\ket{K_\pm}=\pm\ket{K_\pm}$. These states have very different lifetimes because the $K_+$ can decay to $\pi\pi$ in an $S$-wave, whereas $CP$-symmetry forbids the same decay for $K_-$, which mostly decays to $\pi\pi\pi$. Since $m_{K}\sim495\,$MeV and $m_{\pi}\sim140\,$MeV, the second decay is strongly suppressed by phase space. Thus, the $K^0_-$ lifetime should be about 570 times longer than that of $K_+^0$.
        In reality, $CP$ is also violated slightly by the electroweak interaction and thus neutral kaon states are not observed as $K^0_+$ and $K^0_-$, but in the slightly different ``long" and ``short" combinations $\ket{K^0_{\genfrac{}{}{0pt}{}{L}{S} }}=(\ket{K^0_\mp}+\bar\epsilon\ket{ K^0_\pm})/\sqrt{1+\vert \bar\epsilon\vert^2}$, where $\bar \epsilon$ is a $CP$-violating complex parameter of $O(10^{-3})$. The importance of this mixing is that this was the first system where $CP$ violation was observed. For our purposes, it is enough to note that this mixing exists and that $K_L$ and $K_S$ are the neutral strange mesons frequently observed in experiments. As we already commented $K_L$ is the meson with the longest lifetime $\tau_L\simeq 51.16\pm0.21\,$ns, whereas $\tau_S\simeq0.08954\pm0.00004\,$ns.
        
        \item[-] \underline{Unflavored neutral-meson mixing with different isospin:} This occurs between  $I=0$ and neutral $I=1$ mesons. The most relevant examples are $\eta-\eta'-\pi^0$, whose description would require two mixing angles, $\rho^0-\omega$ and $f_0(980)-a_0(980)$. For spectroscopy, this is a small effect, but it is significant for decay processes.
    \end{itemize}
    
    \end{itemize}

    All these mixings distort the naive expectations of the constituent QM, particularly the mass hierarchies and decay relations within nonets. Therefore, these hierarchies are merely semi-quantitative expectations. In addition, mixing also makes it harder to identify non-ordinary mesons when they do not have exotic numbers and can mix with ordinary mesons.

\section{Light-meson list and classification}

We will now review the light-meson list in the RPP \cite{ParticleDataGroup:2024cfk}, organized following their $J^{PC}$ numbers.
We will provide a, sometimes tentative, classification into SU(3)$_F$ multiplets and discuss their nature. 
For each meson, a detailed list of the measurements that have led to the estimates of its parameters can be found in the long ``Particle Listings" of the RPP. Only the resonances that are considered well-established are then collected in the RPP ``summary tables". Of course, the existence of many of these states, particularly those less established, has often been challenged in the literature, and other states have been proposed as well. A complete account of these discussions is out of the scope of this brief pedagogical review.

Thus, for concreteness, we will follow the RPP criteria and take for granted all light resonances that appear in the summary tables. In addition, we will also refer to the states omitted from the summary tables but present in the 
``Light Unflavored Mesons $(\mathsf{S=C=B=0})$" or the ``Strange Mesons $(\mathsf{S=\pm1, C=B=}0)$" sections of the RPP particle listings, because, very often, they are plausible candidates to complete multiplets, Regge trajectories, or illustrate some particular effect.
Due to our limitations in space and scope, we will ignore other states that the RPP lists in the “Further States” subsection of the  “Other Mesons” section because they are ‘‘observed by a single group, or states poorly established that thus need confirmation”.

Longer and more specialized reviews than this pedagogical introduction, including accounts on the evolution of some interpretations, can also be found in the RPP reviews on ``Spectroscopy of light mesons", on `` Scalar mesons below 1 GeV", and on the ``Quark Model". We also suggest the four chapters on the meson mass spectrum, light scalars, glueballs, and hybrids, in the review \cite{Gross:2022hyw}, as well as the sections on glueballs and hybrids in the review about new hadron states \cite{Chen:2022asf}.
A detailed account of ``Key Historical Experiments in Hadron Physics" and ``Hadron Production processes" can be found in this encyclopedia \cite{Amsler:2025wqz,Lenske:2025idu}.

In addition, we will provide an intuitive and pictorial view of the meson masses, widths (and their uncertainties), as well as how they may be classified into multiplets for each quantum number $J^P$.  As discussed above, the identification of the members of a multiplet is not solely based on their similar masses, but also on the numerical relations and ratios between their decays that are expected from SU(3)$_F$ symmetry when they belong to the same multiplet.
Discussing these in detail is beyond our limited scope. Thus, by default, when we group some
resonances in the same multiplet, it is also because their decays are in reasonable agreement with the SU(3)$_F$ expectations. Nevertheless, we will comment on specific cases of particular interest.

We will also discuss their nature as ordinary, non-ordinary, or exotic mesons, as well as their possible mixing schemes. We will consider the isospin limit, and thus the same mass and width for all members of the same isospin multiplet. The only exceptions are the lightest pseudoscalar mesons because, being stable under strong interactions, their parameters have been determined precisely enough to differentiate the states with different charges within the multiplet.

Let us then list and describe the light mesons for increasing $J$ and $P$.

\subsection{$J^{P}=0^{-}$: Pseudoscalar mesons}

\begin{figure} 
	\centering
    \includegraphics[width=0.495\textwidth]{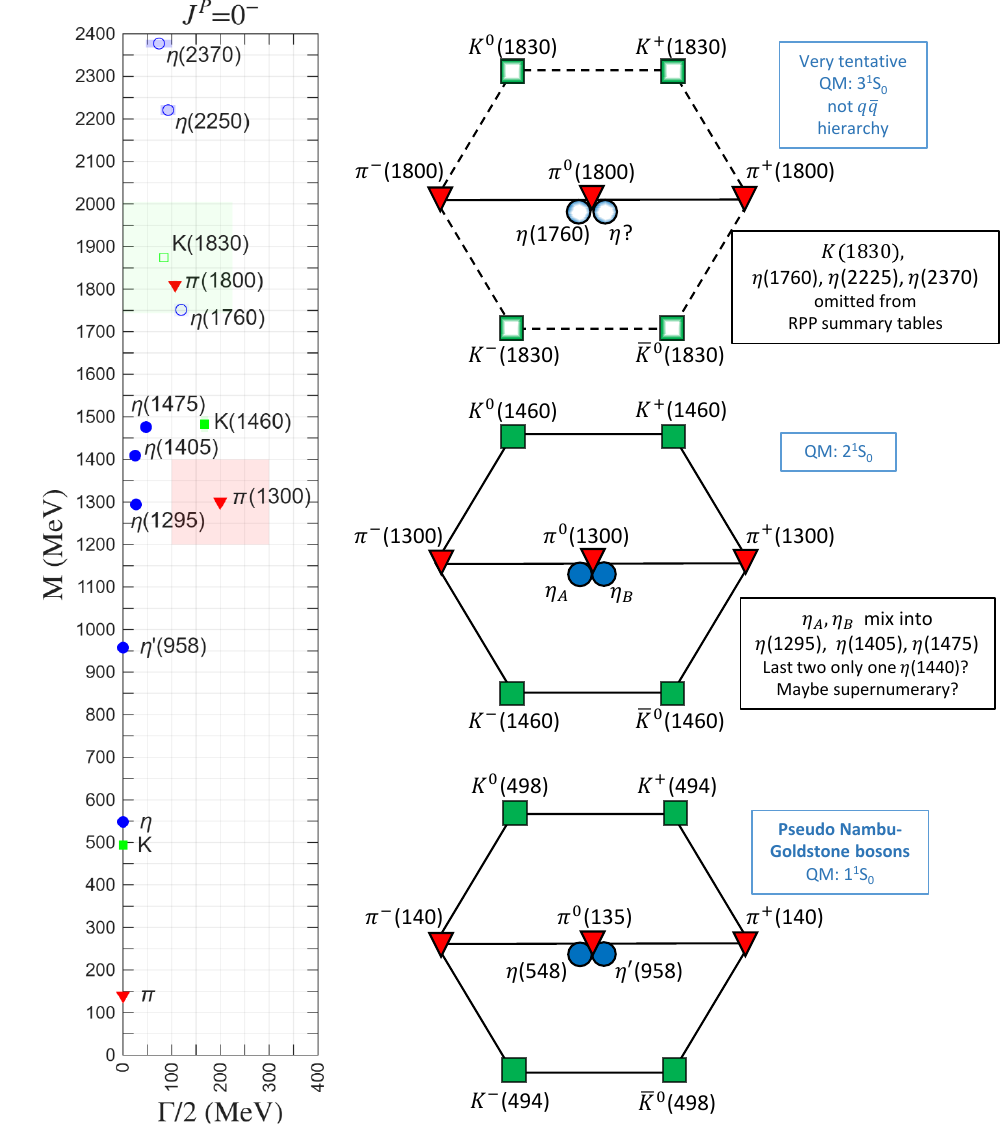}
    \includegraphics[width=0.495\textwidth]{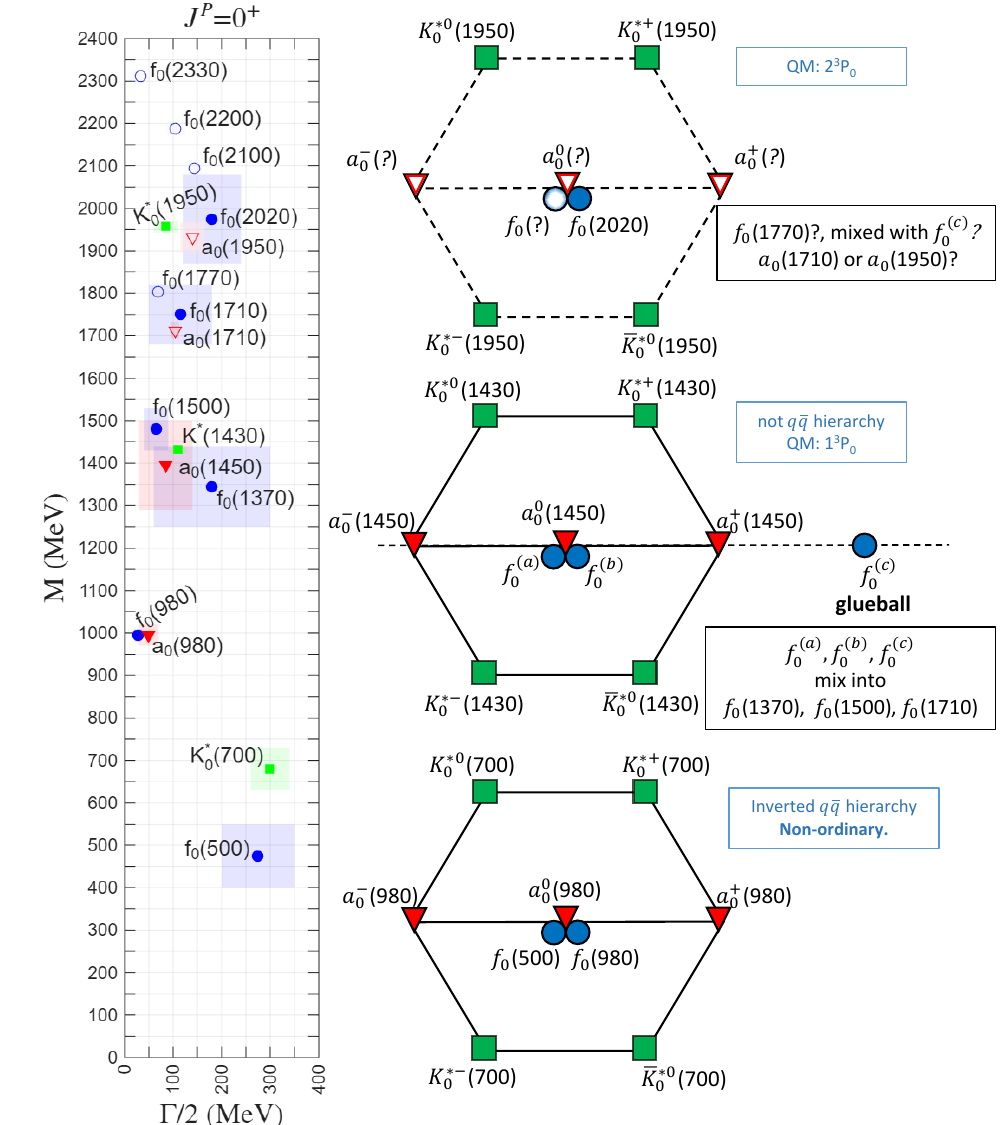}
	\caption{ 
    $J=0$ mesons in the $M,\Gamma/2$ plane and their possible multiplet assignments. Light-colored areas represent the uncertainties in the RPP parameters of each meson.  Hollow symbols represent states that are not well-established or confirmed, and are omitted from the RPP summary tables.    \textbf{Left:} Pseudoscalar $J^{P}=0^-$ mesons.
    The two light nonets are well identified except for the isoscalars of the second one. The third pseudoscalar nonet assignment is very tentative. There might be one supernumerary $\eta$ state above 2 GeV, a glueball candidate.
    \textbf{Right:} Scalar $J^{P}=0^+$ mesons. The two lightest nonets are well identified, as well as the existence of an extra isoscalar, widely accepted to be due to the presence of a glueball state that mixes with the other $f_0$ states nearby.
    There are more than enough states to form a third nonet, but it is not clear which ones to choose.
    We are not displaying the $f_0(2479)$ state above 2.4 GeV, as it is omitted from the RPP summary tables.
	\label{fig:J0}
    }
\end{figure}

The list of pseudoscalar mesons below 2 GeV and their mass and width (or lifetime) can be found in Table~\ref{tab:0-mesons}.  In the left panel of Fig.~\ref {fig:J0}, we show the position of these resonances, actually of their associated poles, in the $(M,\Gamma/2)$ plane, which allows us also to represent their uncertainties by a light-colored area. Also in that figure, we provide their tentative classification into two meson nonets, and the strong hint of a more massive third one.  Let us discuss the features of each three nonets separately.

\begin{table} %[b]
\centering
\caption{ Isospin $I$, mass $M$ and decay width $\Gamma$, or mean-life $\tau$, of the light pseudoscalar mesons, $J^{P}=0^-$, together with their plausible nonet assignments. For the well-established lightest nonet, listed in the left table,  we provide
different masses for the charged and neutral versions of each meson, as well as the lifetimes of $K_S$ and $K_L$, which are mixtures of $K^0$ and $\bar K^0$, without definite isospin. 
The tentative second and third nonets are listed in the table on the right.
As presented in the RPP, one $\eta$ resonance between 1250 and 1500 MeV would be supernumerary; other models suggest that only two $\eta$ states exist in that region. States marked with an asterisk are omitted from the RPP summary tables and only appear in the Particle Listings. $(^\dagger)$ For the values of the $K(1460)$ and $K(1830)$, the RPP does not provide an estimate or average; we have rounded to the MeV the most precise determination they list. The  $\eta(2370)$ was listed as the $X(2370)$ until its recent  $J^{PC}=0^{-+}$ determination by the BESIII collaboration \cite{BESIII:2023wfi}. 
\label{tab:0-mesons}}

\medskip

\begin{tabular}{lllll}
%\hline%
\toprule
$I$ & Name & M [MeV] & $\Gamma$ [MeV] & $\tau$ [ns] \\
%\hline
\midrule
1 & $\pi^0$ & 134.9768(5) & &$8.43(13) \times 10^{-8}$ \\
1 & $\pi^\pm$ & 139.57039(18) & &26.033(5)\\
0 & $\eta$ & 547.862(17) & 0.00131(5)  &\\
0 & $\eta'$ & 957.78(6) & 0.188(6) &\\
1/2 & $K^\pm$ & 493.677(15) & &12.38(2)\\
1/2 & $K^0/\bar{K}^0$ & 497.611(13) & &\\
\midrule
 & $K_L$ & & &51.16(21)\\
 & $K_S$ & &  &0.08954(4)\\
\midrule
 &  & &  &\\[9pt]
%\hline
%\bottomrule
\end{tabular} 
\qquad\quad
\begin{tabular}{llll}
%\hline%
\toprule
$I$ & Name & M [MeV] & $\Gamma$ [MeV] \\
%\hline%
\midrule
1 & $\pi(1300)$ & 1300$\pm$100 & 200 to 600 \\
0 & $\eta(1295)$ & 1294$\pm$4 & 55$\pm$5 \\ 
0 & $\eta(1405)$ & 1408.7$^{+2.0}_{-1.2}$ & 50.3$\pm$2.5 \\
0 & $\eta(1475)$ & 1475$\pm$4 & 96$\pm$9 \\
1/2 & $K(1460)$ & 1482$\pm$19 $(^\dagger)$ & 335$\pm$15 $(^\dagger)$ \\
%\hline%
\midrule
1 & $\pi(1800)$ & 1810$^{+9}_{-11}$ & 215$^{+7}_{-8}$ \\
0 & $\eta(1760)$ $(^*)$ & 1751$\pm$15 & 240$\pm$30 \\
0 & $\eta(2225)$ $(^*)$ & 2221$^{+13}_{-10}$ & 185$^{+40}_{-20}$ \\
0 & $\eta(2370)$ $(^*)$ & 2377$\pm$9 & 148$^{+80}_{-28}$ \\
1/2 & $K(1830)$ $(^*)$ & 1874$\pm43^{+59}_{-115}$ $(^\dagger)$  & 168$\pm90^{+280}_{-104}$ $(^\dagger)$  \\
%\hline
\bottomrule
\end{tabular}

\end{table}

\subsubsection{The lightest pseudoscalar nonet. Nambu-Goldstone bosons}

\medskip

\paragraph{Discovery, names and basic features} 
The left sub-table in Table~\ref{tab:0-mesons} provides the parameters of the lightest pseudoscalar nonet, namely, the pions, kaons, and etas.
These were the first mesons to be discovered because they are the longest-lived. For most purposes in hadron physics, they can be considered stable.
In 1935, Yukawa \cite{Yukawa:1935xg} proposed the existence of a meson as a mediator for what we now would call the residual strong force, which binds protons and nucleons within nuclei.  After some confusion with the muon, whose mass is relatively close but does not interact strongly, the existence of the charged pion was established in 1947 by the collaboration led by Powell \cite{Lattes:1947mw,Lattes:1947mx,Lattes:1947my} at U. Bristol. By studying cosmic ray tracks in photographic emulsions, they found that muons could appear as secondary particles in the decay of a primary track. In \cite{Lattes:1947my}, they decided to ``{\it represent the primary meson by the symbol $\pi$, and the secondary by $\mu$}" \footnote{It has been told \cite{Zee:2004} that the symbol $\pi$ was adopted by Yukawa too, due to its 
similarity with the Japanese symbol
\begin{CJK}{UTF8}{min} 介 \end{CJK} ``[Kai]", which may be translated by ``intermediate" or ``mediate" \cite{Zee:2004}. I have not been able to find written evidence of such claim. To my view, as the astronomer Giordano Bruno said: ``Se non \'e vero, \'e molto ben trovato" (De Gli Eroici Furori, 1585).}.
The pion was then identified as the mediator in Yukawa's theory.
Similarly, kaons---initially found in 1947 in cosmic ray observations by Rochester and Butler \cite{Rochester:1947mi}--- were called ``k-particles'' because they were identified as distinct mesons in the ``k-track'' event found in 1949 by the Bristol group \cite{Brown:1949mj}. Soon, pions and kaons were produced in accelerators, and the neutral pion was identified in 1949 by observing its two-photon decay. Another neutral meson decaying to three pions was discovered in 1961 by the team of Pevsner \cite{Pevsner:1961pa} at Johns Hopkins University, which received the name $\eta$ because it is the Greek symbol equivalent to the letter ``H" in their institution. 

Note that, since they are quasi-stable, the mass and width or lifetime of pions, kaons, and etas are very well determined, with precisions well below the MeV scale (or nanosecond scale).
With this accuracy, it is possible to clearly distinguish the parameters of the mesons within the same isospin multiplet but with different charges, which is out of reach for almost all other light mesons.

Following Table~\ref{tab:ordmesonnames}, in the QM $q\bar q$ interpretation, these mesons would carry the quantum numbers of the ground state $n=1$, $L=S=0$, namely $1^1 S_0$. Thus, they are expected to be the lightest. However, the first surprise is that the best-known mesons do not follow the naive $q\bar q$ mass hierarchy expected within the QM, but instead they are much lighter than expected, except the $\eta'(958)$. For instance, the pion mass is $\sim$140 MeV, which is a factor of two lighter than the naive estimation $\sim 700\,$MeV for a constituent $q\bar q$ with $L=0$.
That is indeed the mass of the $\rho(770)$ isotriplet vector meson, which is well described by $L=0$ quarkonia. 
Moreover, the mass difference between the pions and kaons is $\sim$350 MeV, also higher than expected.
Finally, the $\eta'(958)$ mass is significantly larger than the masses of the rest of the multiplet. These unexpected features are explained because these mesons are the so-called pseudo-Nambu-Goldstone bosons (NGB) \cite{Nambu:1960tm,Goldstone:1961eq,Goldstone:1962es} of the spontaneous chiral symmetry breaking of QCD, which we briefly introduce next.

\paragraph{Spontaneous chiral symmetry breaking and Nambu-Goldstone bosons.} 
Let us recall that quarks are  Dirac fermions, described in terms of 4-dimensional spinors $q$. They can be decomposed into
their right and left chiral components as $q=q_L+q_R$, where $q_{L,R}=P_{L,R}\,q$. Here, $P_{L,R}=(\mathbb{1}\mp\gamma^5)/2$ are the left and right chirality projectors, $\gamma^5=i\gamma^0\gamma^1\gamma^2\gamma^3$ and $\gamma^i$ with $i=0...3$ are the Dirac matrices.
Note that $P_L+P_R=\mathbb{1}$ and that under parity $P_L\leftrightarrow P_R$.
Let us now consider QCD only with the lightest $n_f=2$ or three flavors. In the limit when these lightest quarks were indeed massless, QCD would be invariant under independent SU($n_f$) global transformations, mixing either the left or the right quark components. This is called the  SU($n_f$)$_L\times$SU($n_f$)$_R$ chiral symmetry of the QCD lagrangian, and, given the smallness of the lightest quark current masses $m_q$, one could naively expect it to be a good approximation. In particular, since parity transforms left and right components into each other, if chiral invariance were a good symmetry, in the meson spectrum we would expect to find pairs of SU($n_f$) multiplets with opposite parity, almost degenerate up to small corrections of $O(m_q)$. For instance, thinking of $n_f=2$, we would expect pairs of isospin triplets with opposite parity, whose masses should not differ by more than a few MeV.

However, as we discuss in this review, in Nature we only observe light hadrons grouped in approximate isospin or SU(3)$_F$ flavor multiplets. Actually, $J^P$ and $J^{-P}$ multiplets do not appear in approximately degenerate pairs, but are separated by several hundred MeV instead. Therefore, SU($n_f$)$_L\times$SU($n_f$)$_R$ must be broken down to SU(3)$_F$, which is the subgroup where the left- and right-chirality components simultaneously undergo the same transformation.
When looking at SU(3)$_F$ as a subgroup of the chiral group, it is often called SU($n_f$)$_{L+R}=$SU($n_f$)$_V=$SU($n_f$)$_F$.
For the $n_f=2$ case, this subgroup is nothing but the SU(2) isospin group. 

The explicit $O(m_q)$ chiral-symmetry breaking due to the small current masses in the Lagrangian cannot account for the magnitude of the chiral symmetry breaking in the hadron spectrum.  
However, as explained in the Interlude, within quantum field theory, symmetries can also be broken spontaneously.

\begin{BoxA}[sec5:box2]{Interlude: Spontaneous symmetry breaking, Goldstone's Theorem and Nambu-Goldstone bosons}

In quantum field theory, particle states are understood as field excitations above the vacuum, which is defined as a state of minimum energy and also serves to define the zero-energy level. Typically, this vacuum corresponds to the unique configuration in which all the expectation values of all fields vanish.

However, it may also happen that the vacuum is not unique; instead, there may be multiple vacuum states with zero energy. In such cases, a symmetry can be broken even if it leaves the Lagrangian invariant. 
When formulating the whole theory, particles are then understood as excitations over a vacuum ``spontaneously" chosen among the many degenerate vacua. This spontaneous choice is not the familiar vacuum state where all expectation values vanish. Therefore, it must sustain at least a non-vanishing expectation value for some combination of fields, which is said to form a non-vanishing vacuum condensate. If this vacuum condensate is not invariant under a given symmetry of the Lagrangian, it is said that this symmetry is ``spontaneously broken" in the whole theory.

The generators of the symmetry group that do not commute with the condensate are called ``broken generators", and transform the new vacuum into one of the other degenerate states with zero energy.
However, these new states are now understood as particle excitations over the new vacuum. Intuitively, for each independent spontaneously broken generator, there must be a particle excitation, which, at rest, does not increase the energy of the state over the vacuum, which means that they are massless particles. Under very general conditions, which include Lorentz invariance, this observation can be formulated as the celebrated {\bf Goldstone's Theorem:} \cite{Goldstone:1961eq,Goldstone:1962es}.
Omitting technical subtleties, and in a very condensed form suitable for our introductory scope, it states that: 

  {\it ``For each spontaneously broken generator, there is a massless boson, with $J=0$, carrying the quantum numbers of the broken generator".}  

Before this general result, such massless modes were found by Nambu \cite{Nambu:1960tm}  in the hadron physics context and are therefore called Nambu-Goldstone boson (NGB). Since they are massless, they are separated from the other states in the spectrum by an energy gap, related to the scale of the condensate that triggers the spontaneous symmetry breaking.

\end{BoxA}

In fact, chiral symmetry is spontaneously broken in hadron physics. Let us provide a brief description of the relevant features for our discussion below, and refer the reader to another chapter of this Encyclopedia for a more detailed introduction \cite{Nefediev:2025zkv}.  
In the case of QCD, the spontaneous chiral symmetry breaking is due to the non-vanishing vacuum expectation value of the quark-antiquark condensate\footnote{Note that, as it happened with current quark masses, the quark condensate is renormalization-scale dependent, although in such a way to cancel the corresponding $m_q$ dependence to yield renormalization-scale independent meson masses.} $\langle q\bar q\rangle \simeq -(250\, {\rm MeV})^3$. This vacuum condensate is only invariant under the SU(3)$_F$ subgroup, which is still a good symmetry of the system, but ``spontaneously" breaks the remaining eight generators of SU(3)$_L\times$SU(3)$_R$.
Hence $n_f=3$ massless QCD remains symmetric under SU(3)$_F$ but, following Goldstone's Theorem, it has eight massless Nambu-Goldstone bosons (NGB) with zero spin, $J=0$.
Recall that SU(3)$_F=$SU(3)$_{L+R}$ and its eight generators transform the left and right components the same way and are thus parity invariant. Instead, the other eight broken generators change sign under parity. Therefore, the eight massless QCD NGBs should be pseudoscalars, $J^P=0^-$, and form an SU(3)$_F$ octet. Its eight members are nothing but the three pions, the four kaons, and the eta, which are not massless but much lighter than all other hadrons.

Of course, QCD is not massless, but current masses are so small that they only perturb this scenario by giving the NGBs a generic small mass $M^2\simeq B_0 m_q$ to leading order (LO) in $O(m_q)$. 
For not being exactly massless, the pions, kaons, and the eta are often called ``pseudo-NGB".
In the isospin limit, $B_0=-\langle q\bar q\rangle/f_\pi^2$, where $f_\pi\simeq93\,$MeV is the pion decay constant. 
The precise LO relation, taking into account the isospin structure in the isospin limit, leads to:
\begin{equation}
     \frac{M_\pi^2}{2 \hat m}=\frac{M_K^2}{\hat m+m_s}=\frac{3M_\eta^2}{2 (\hat m+2m_s)}=-\frac{\langle q\bar q\rangle}{f_\pi^2} \quad \Longrightarrow \quad 4 M_K^2-M_\pi^2-3 M_\eta^2=0. 
\end{equation}
The left-hand side is known as the Gell-Mann-Oakes-Renner relation \cite{Gell-Mann:1968hlm}. The right-hand side 
is the isospin-limit quadratic-mass version (see \cite{Glashow:1967rx} and chapter 19.7 of \cite{Weinberg:1996kr}) of the linear Gell-Mann-Okubo relations \cite{Gell-Mann:1961omu,Okubo:1961jc}, which reflects the NGB nature of pseudoscalar mesons, and predicts $M_\eta$ from the other two masses within a 3\% accuracy. In contrast, the linear-mass version prediction is $\sim 12\%$ off.

Apart from the existence of pseudo-NGB and these LO mass relations, there are further consequences of spontaneous chiral symmetry breaking. These can still be described in simple and intuitive models. However, it is important to note that introducing a scale, such as a constituent mass or a cut-off, typically leads to unwanted explicit chiral symmetry breaking. For this reason, the naive QM is usually modified into the chiral-QM, 
which is more complicated but still simple enough to provide useful insights. We refer to the reviews \cite{Klempt:2007cp,Entem:2025bqt,Nefediev:2025vmo}.
Nevertheless, all the effects of the spontaneous chiral symmetry breaking
 can be dealt with in a systematic and model-independent way employing the low-energy Effective Field Theory of QCD, called Chiral Perturbation Theory (ChPT), to which we dedicate a brief interlude, because it will be of interest later on. 

\begin{BoxA}[sec5:box3]{Interlude: Chiral Perturbation Theory (ChPT)}

Another consequence of spontaneous chiral symmetry breaking is that NGBs couple derivatively: crudely speaking, proportionally to their mass or momenta.
This derivative coupling and the existence of an energy gap separating the NGB from the other hadrons allow for the formulation of a low-energy Effective Field Theory of QCD \cite{Weinberg:1978kz} using pions, kaons, and the eta as degrees of freedom, which is known as Chiral Perturbation Theory (ChPT) \cite{Gasser:1983yg,Gasser:1984gg} (see the dedicated chapter in this Encyclopedia \cite{Meissner:2024ona}).

This technique provides a rigorous connection with QCD through a perturbative framework for calculating low-energy observables of NGBs, which are systematically obtained in a power series of NGB masses and momenta, generically denoted $(p/4\pi f_\pi)^n$, where $f_\pi$ is the pion decay constant.

Of particular interest for us is that the NGBs have derivative interactions among themselves and the
requirement of an Adler-zero below threshold in the NGB scattering amplitudes [20]. Weinberg obtained the leading-order calculation of those amplitudes at low energies in [22]. The results are called Low-Energy Theorems, because they only depend on the NGB masses and the symmetry breaking scale, which is parameterized by the pion decay constant $f_\pi$. Masses at LO in ChPT satisfy the Gell-Mann-Oakes-Renner and Gell-Mann-Okubo relations.

In contrast, higher-order calculations involve a set of low-energy constants (LECs) that depend on the specific QCD dynamics and can be determined phenomenologically, from experiment, or, for some of them, from lattice-QCD calculations. From the point of view of meson physics, these LECs represent, as contact terms, the low-energy interaction of more massive mesons with the NGB \cite{Donoghue:1988ed,Ecker:1988te}. Resonance poles cannot be reproduced with the perturbative expansion; only their low-energy tails.

Two remarks will be relevant below for light-meson spectroscopy. First, naively, the largest contribution to the LECs should come from the lightest resonances, which are not NGBs. However, it is found that their values are dominated by the lightest vector mesons  \cite{Donoghue:1988ed,Ecker:1988te}, despite being heavier than the lightest scalar mesons. Second, the LO dependence of the ChPT parameters is known and model independent, which provides a connection with a relevant QCD parameter.

Finally, being a low-energy expansion, the applicability of ChPT is limited to momenta well below the chiral scale that governs the loop expansion $4\pi f_\pi\simeq 1\,$GeV and the position of the first resonance pole appearing in the process under study.
The use of ChPT beyond its applicability region would lead to numerically large unitarity violations. However, ChPT can still be used to calculate the low-energy part (technically, the subtraction terms) of dispersion relations, restoring unitarity in single or coupled-channel formalisms (see \cite{Oller:2025leg}). This approach is generically referred to as Unitarized ChPT (UChPT) or chiral unitary approaches (UChPT) (see \cite{Dobado:1989qm,Dobado:1996ps,Oller:1997ti,Oller:1998hw} for seminal works). It can also be understood as a resummation within the Bethe-Salpeter formalism \cite{Nieves:1998hp}. The fact that they generate or reconstruct the resonance poles from ChPT has made this approach very relevant for light-meson spectroscopy (see \cite{Pelaez:2015qba,Oller:2020guq,Yao:2020bxx,Pelaez:2021dak} for reviews). 

\end{BoxA}

The singular behavior of the $\eta'(958)$ compared to its nonet partners is related to the fact that massless QCD is invariant under a larger group that includes an additional U(1)$_{L+R}$ symmetry. The additional generator outside this subgroup changes sign under parity. It generates the so-called ``axial" U(1)$_A$, which is not broken spontaneously, but due to a quantum field theory anomaly (In brief, it leaves the classical action invariant, but not the quantum field theory measure). 
As a consequence, the lightness of the $\eta'(958)$ mass is no longer protected by any NGB nature, and that is why it is much heavier than the rest of the lightest pseudoscalar octet.

Given the large mass difference between the $\eta$ and $\eta'$ and the fact that its nature is determined by its membership in the SU(3)$_F$ octet of NGB mesons, one would expect the mixing angle between the $\eta$ and $\eta'$ isosinglets to be rather small and far from ideal mixing. Indeed, the $\eta$ is very close to the octet state and the $\eta'$ to the singlet; the RPP estimates $\theta_{\rm quad}\simeq11.3^{\circ}$ \cite{ParticleDataGroup:2024cfk} in a $q\bar q$ assignment. 
Different phenomenological chiral models find angles, which, taking into account their uncertainties when they are provided, 
fall in a range between $\sim-9.5^\circ$ to $-15.5^\circ$ \cite{Feldmann:1999uf, Kroll:2005sd,Escribano:2005qq,Parganlija:2012fy}.  This range is not far from the findings of lattice-QCD calculations like
$\simeq -11^\circ$ \cite{Dudek:2013yja} and $\simeq-14.1\pm2.8^\circ$ \cite{Christ:2010dd}.
 Note that, outside the isospin limit, there is an additional small mixing with the $\pi^0$, giving rise to a more complicated mixing scheme, for which we refer the reader to \cite{Feldmann:1999uf,Kroll:2005sd,Escribano:2020jdy}.
 Moreover, when we introduced the isosinglet-singlet mixing, we only discussed quarkonia or tetraquark examples; however, the isosinglet could also have a glueball component.
The KLOE collaboration \cite{Ambrosino:2009sc} found in 2009
$\theta_{quad}\sim-14.3\pm0.6$ from a global fit, although they also introduce an additional angle for the mixing of the $s\bar s$ and glueball components, whose fraction in the $\eta'$ meson composition is estimated to be $0.12\pm0.04$.

\paragraph{The relevance of the lightest pseudoscalar mesons}
Finally, we should recall the importance of the members of this nonet. First, for Nuclear Physics, the pion plays a dominant role in the phenomenological description of nucleon-nucleon interactions, and so does the kaon when dealing also with hyperons (nucleons with strangeness).
Second, for QCD, they are essential in our understanding of spontaneous chiral symmetry breaking.
Third, being the lightest hadrons, they are the final products of almost all hadronic strong interactions, and being quasi-stable, they are the mesons directly observed in detectors.
Almost all other mesons that we describe below are seen as resonances in the production or scattering of pions and kaons as final products of other hadronic processes.

\subsubsection{The second and third pseudoscalar nonets}

In the right sub-table of Table~\ref{tab:0-mesons}, we provide the list of pseudoscalar mesons between 1 and 2.5 GeV that appear in the RPP long particle listings. Their positions in the $(M, \Gamma/2)$ plane, together with their uncertainties, are shown on the left of Fig.\ref{fig:J0} together with a pictorial illustration of their tentative nonet assignment.

Concerning the second nonet, which in the QM would correspond to the first radially excited state, $2^3S_0$, its isotriplet $\pi(1300)$ and two strange isodoublets $K(1430)$ are well established. Note that the uncertainty in the mass and width of the $\pi(1300)$ is very large, due to being so broad and decaying predominantly into 3$\pi$.
The situation with the isoscalars is quite confusing. 
The $\eta(1295)$ has been produced and observed in several processes, but its existence has been questioned \cite{Klempt:2007cp}.
In the RPP, there are also two other well-established states, the $\eta(1405)$ and $\eta(1475)$,
although only one state, called $\eta(1440)$, was listed until the 2004 RPP edition.
The analysis is further complicated due to the possible presence of a triangle singularity \cite{Wu:2011yx}, which could mimic the presence of one of these poles.
In such a case, it could not be interpreted as a resonance.
The mixing scheme between these two or three states is not clear to the point that, despite being absent from their Particle Listings for almost 20 years,
the RPP in its ``Light meson resonance review" still uses the $\eta(1440)$ as a member of the first radially excited nonet, ``standing" for the $\eta(1405)$ and the $\eta(1475)$. In contrast, their ``Quark Model" review does not show the $\eta(1405)$, claiming that together with the $\eta(1475)$ they could be the manifestation of a single state (see \cite{Du:2019idk}), but without mentioning the $\eta(1440)$. 
Clarifying this situation is relevant because if three states exist, one of them would be supernumerary, which could suggest the existence of a pseudoscalar glueball or even a hybrid state. However, we have already discussed that these are generically expected to have larger masses. To reflect this situation in the left panel of Fig.\ref{fig:J0} we have indicated that there are enough states to form the second pseudoscalar nonet and that they would be two states $\eta_A$ ad $\eta_B$ resulting from the mixture of the $\eta(1295)$ with either just one additional state, generically dubbed $\eta(1440)$, or two other isosinglets, $\eta(1405)$ and $\eta(1475)$, in which case one would be supernumerary.
Note that none of these states is considered part of the third nonet, because in that case their masses should be about 300 MeV heavier. It is worth noticing that the COMPASS@CERN Collaboration claims \cite{COMPASS:2025wkw} to have found a supernumerary $K(1690)$ state together with the $K(1460)$ and $K(1830)$, suggesting that it could be a crypto-exotic candidate, since the QM predicts just two ordinary $q\bar q$ states in that region.
All in all, there are enough candidates to build this second nonet and its mass hierarchy, particularly between the isotriplets and isodoublets, which fit reasonably well within the QM expectations for $q\bar q$ resonances, corresponding to the first radially excited states $2^1S_0$. 

Concerning the third nonet, illustrated in the left panel of Fig.\ref{fig:J0}, the isotriplet $\pi(1800)$ is well identified.
However, both the $\eta(1760)$ and $K(1830)$, as well as other possible candidates for the other isosinglet, the $\eta(2250)$ and $\eta(2370)$, are not so well-established states and, indeed, they are omitted from the RPP summary tables. Hence, we represent these states with empty symbols, rather than solid ones.
If these resonances were indeed to form a  $q\bar q$ nonet, corresponding to $3^1S_0$ states, they would not follow the expected QM mass hierarchy. Hence, we do not dare to identify the second isosinglet state. 

Let us remark that it was just in 2023 that the BESIII collaboration  \cite{BESIII:2023wfi}
was able to determine the $J^{PC}=0^{-+}$ quantum numbers of the $\eta(2370)$. Until the last RPP edition, it was still referred to as $X(2370)$. It is very suggestive that the $\eta(2370)$ mass is close to the one obtained for the lightest pseudoscalar glueball in lattice calculations \cite{Bali:1993fb,Morningstar:1999rf,Chen:2005mg,Gui:2019dtm, Gregory:2012hu}.

Therefore, although there are enough candidates for this third nonet, its identification should be considered very tentative, which is why the lines connecting the nonet are not continuous but dashed..

\subsection{$J^{P}=0^{+}$: Scalar mesons}

We list in Table~\ref{tab:0+mesons} the light scalar mesons below 2.5 GeV, and we show in the left panel of Fig.\ref{fig:J0} their position and uncertainties in the $(M,\Gamma/2)$ plane as well as their accepted assignments in two multiplets. 

\subsubsection{The lightest scalar nonet: Non-ordinary mesons}

This nonet has been the most controversial among light-meson resonances for over half a century. 
Even today, these are the only light mesons that have a dedicated review in the RPP entitled ``Scalar mesons below 1 GeV", which we highly recommend \cite{ParticleDataGroup:2024cfk}.
The subject of discussion has been both the existence of its two lightest members and the nature of all its members. This debate is amplified due to their prominent role in the early models used to understand the QCD spontaneous chiral symmetry breaking, as well as their role in nucleon-nucleon attraction. In addition, as seen in Table~\ref{tab:0+mesons} and Fig.\ref{fig:J0}, this nonet does not follow the expected $q\bar q$ mass hierarchy, but an inverted one. For this and other reasons to be discussed below, its members are believed to be predominantly non-ordinary mesons.

\begin{table} %[b]
\caption{ Isospin $I$, mass $M$ and decay width $\Gamma$ of the scalar mesons $0^+$, together with their plausible assignment into a nonet. In the left sub-table, we find a well-established nonet below 1 GeV and a second nonet and a singlet between 1 and 1.75 GeV, possibly mixed. Remarkably, there is one isospin $I=0$ state too many for the second nonet between the second and third nonets. It is believed to correspond to the presence of a glueball state that would mix with the isosinglets of the two nonets. 
Here we list, by default, the mass and width from the $T$-matrix pole provided by the RPP, rather than the Breit-Wigner parameters, since this parameterization is not applicable for almost all these mesons. Unfortunately, it is still being widely used in this context. 
The number of states is enough to form a third nonet, but the assignment of its members is very tentative.
We also list several isoscalar states above 2 GeV, which are omitted from the summary tables. 
$(^\dagger)$ The $K_0^*(1950)$ values are not listed as $T$-matrix poles.   $(^*)$ Resonances with an asterisk are omitted from the summary tables, and their values are not from $T$-matrix poles.  $(^\ddagger)$ The RPP does not provide an average or estimate for these values, and we take the latest and most precise BESIII results \cite{BESIII:2022iwi}.
\label{tab:0+mesons}}
\centering
\medskip
\begin{tabular}{llll} 
\toprule
$I$ & Name & $M$ [MeV] & $\Gamma$ [MeV] \\
\midrule
1 & $a_0(980)$ & 970-1020 &  60-140 \\
0 & $\sigma/f_0(500)$ & 400-550 &  400-700 \\
0 & $f_0(980)$ & 98-1010 & 40-70 \\
1/2 & $\kappa/K_0^*(700)$ & 630-730 & 520-680\\
\midrule
1 & $a_0(1450)$ & 1290-1500 & 60-280 \\
0 & $f_0(1370)$ & 1250-1440 & 120-600 \\
0 & $f_0(1500)$ & 1430-1530 & 80-180 \\
0 & $f_0(1710)$ & 1680-1820 & 100-360 \\
1/2 & $K_0^*(1430)$ & 1431$\pm$6 & 220$\pm$38 \\
\bottomrule
\end{tabular}
\qquad\qquad\qquad
\begin{tabular}{llll}
\toprule
$I$ & Name & $M$ [MeV] & $\Gamma$ [MeV] \\
\midrule
1 & $a_0(1710)$ $(^*)$ & 1713$\pm$19  & 107$\pm$15 \\
1 & $a_0(1950)$ $(^*)$ & 1931$\pm$36  & 271$\pm$51 \\
0 & $f_0(1770)$ $(^*)$ & 1804$\pm$16  & 138$^{+22}_{-25}$ \\
0 & $f_0(2020)$ & 870–2080 & 220-480 \\
1/2 & $K^*_0(1950)$  $(^\dagger)$ & 1950$\pm$14 & 170$\pm$50  \\
\midrule
0 & $f_0(2100)$ $(^*)$  & 2095$^{+17}_{-19}$ & 287$^{+32}_{-24}$ \\
0 & $f_0(2200)$ $(^*)$  & 2187$\pm$14 & 210$\pm$40 \\
0 & $f_0(2330)$ $(^{*\ddagger})$  & 2312$\pm7^{+7}_{-3}$ & 65$\pm10^{+3}_{-12}$ \\
0 & $f_0(2470)$ $(^*)$  & 2470$\pm4^{+4}_{-6}$ & 75$\pm9^{+11}_{-8}$ \\
\bottomrule
\end{tabular}

\end{table}

\paragraph{The relevance of the lightest scalar mesons}
A light ‘‘neutral scalar meson’’ was introduced by Johnson and Teller in 1955 \cite{Johnson:1955zz}
to explain part of the nucleon–nucleon (NN) attraction. Soon, it was incorporated by Schwinger \cite{Schwinger:1957em} into a unified
isospin framework,  a singlet that he called $\sigma$, warning that it would be highly unstable and not easily
observable, which is the reason for the long debate about its existence.
From the 1960s until present, phenomenological ‘‘one boson-exchange models’’ explain the main part of the NN strong attraction in the 1-2 fm range with an isoscalar–scalar meson with a mass around 500-600 MeV (see \cite{Nagasaki:1967bq,Machleidt:2000ge,Wiringa:1994wb}), or a ``correlated two-pion exchange" with the same quantum numbers.
Another scalar strange state, relatively similar to the $\sigma$ and called $\kappa$, was proposed by Dalitz in 1965 \cite{Dalitz:1966fd}, ``simply on the basis of SU(3)
symmetry”.

Also in the early 1960s, the role of the $\sigma$ in hadron physics became more relevant when Gell-Mann and Levy developed the simple and popular “Linear sigma model” (L$\sigma$M) of two-flavor chiral symmetry breaking \cite{Gell-Mann:1960mvl}. They gathered the $\sigma$ and the three pions inside a quadruple and built a Lagrangian invariant under the group of SO(4) rotations, isomorphic to the chiral SU(2)$\times$SU(2) group, which is spontaneously broken because the $\sigma$ field acquires a non-vanishing vacuum expectation value.
Subtracting this value from the $\sigma$ field, one is left with three massless pions and a massive $\sigma$ particle. Although it is beyond our scope, it is worth noting that L$\sigma$M is also very popular because it is renormalizable and also because the linear realization makes it relatively simpler to incorporate the interactions with nucleons.
Moreover, the L$\sigma$M can be formulated in terms of SU(2) matrices instead of SO(4) vectors, and then it can be easily generalized to three flavors and the SU(3)$\times$SU(3) case, which includes a $\kappa$ meson too. 
Similar bosonic states can be generated dynamically in the popular Nambu and Jona-Lasinio (NJL) model \cite{Nambu:1961fr,Nambu:1961tp}, which obtains the spontaneous symmetry breaking and the energy gap using four-fermion non-renormalizable interactions (originally, it was formulated with baryons; nowadays, with quarks).

However, although the $\sigma$ and $\kappa$ names are still being used in the literature, one should differentiate between the present $f_0(500)$ and $K_0^*(700)$ well-established scalar resonances and the $\sigma$ and $\kappa$ fields that appear in the simple NJL and linear $\sigma$ models. Neither the L$\sigma$M nor the NJL models are QCD.
As we saw in the previous section, the low-energy Effective Field Theory of QCD is Chiral Perturbation Theory, where, among other differences, the action of the chiral group is realized non-linearly, only in terms of NGBs. Nevertheless, the L$\sigma$M and NJL are interesting toy models, which capture some relevant features of the QCD spontaneous chiral symmetry breaking and, in the heavy-$\sigma$ limit, are generically consistent with ChPT to LO. 
However, the L$\sigma$M is known to yield higher-order low-energy ChPT constants incompatible with observations \cite{Gasser:1983yg,Pelaez:2015qba}. Of course, there are many modifications of these models, whose description lies beyond our limited scope, that supply some additional interesting features at the cost of losing their original simplicity and other caveats.

Fortunately, it is indeed possible to obtain a simultaneous description of the NGB scattering data, the higher orders of ChPT at low energies, and the poles associated with the scalar and even vector resonances, employing the so-called unitarized Chiral Perturbation Theory \cite{Dobado:1989qm,Dobado:1996ps} (UChPT). When applied to LO in ChPT,
it is often called the chiral unitary approach \cite{Oller:1997ti,Oller:1998hw}. This technique, developed around the 1990s, imposes unitarity by combining the ChPT expansion with dispersion theory, inheriting its powerful analytic properties. It played a relevant role in establishing the existence of the $T$-matrix poles associated with the lightest and widest two scalar mesons and their membership in the lightest nonet.   As discussed below, this allows us to study the nature of resonances through their dependence on QCD parameters, such as quark masses and the number of colors. 
Nevertheless, UChPT has several approximations, at high energies and in crossed channels, which make the approach relatively simple and, for the most part, algebraic. However, even if uncertainties from these approximations are numerically small, they make UChPT unsuitable for precision studies.

\paragraph{Struggling for recognition. Dispersion relations and $T$-matrix poles.}
The existence of the $f_0(980)$ and $a_0(980)$ was swiftly accepted, and their masses and widths have been relatively stable in the RPP since the 1970s, although under different names. This stability was undoubtedly due to the fact that they are relatively narrow, and peaks or interference dips are clearly identifiable in many processes.
On the contrary,  the $\sigma$ and $\kappa$ are very wide and very close in relative terms to the $\pi\pi$ and $K\pi$ thresholds, making their resonant shape unrecognizable in NGB scattering, 
and completely at odds with a BW description. Moreover, the scattering data collected in the early 1970s in that region, despite being plagued with systematic uncertainties and incompatibilities between different sets, were good enough to discard the presence of any familiar resonance shape.
Thus, the very wide $\sigma/f_0(500)$, then under a different name, disappeared in 1976 from the RPP. It was reinstated back in the RPP 1996 edition
as the $\sigma(400-1200)$---with an uncertainty of 800 MeV in its mass. Over the next years, additional experimental support came from decays of heavier particles into final states with at least two pions in the scalar–isoscalar channel, where, maybe not a peak, but some sort of bump could be observed. Further support also came from  UChPT. 

It was in the 2012 edition that the $\sigma$ became the $f_0(500)$ and was reported with reasonable uncertainties, somewhat bigger than the present ones. The decisive studies for this change employed crossing-symmetric coupled dispersion relations, generically called Roy equations \cite{Roy:1971tc}. 
For the pros and cons of this approach versus UChPT, see \cite{Pelaez:2021dak}.
When Roy equations were solved in combination with ChPT constraints, together with reliable data above the $\sigma$ resonance region and other $\pi\pi$ scattering partial waves, a $T$-matrix pole was found very close to the value listed in Table~\ref{tab:0+mesons} \cite{Ananthanarayan:2000ht,Colangelo:2001df,Caprini:2005zr}. A similar pole was confirmed \cite{Garcia-Martin:2011nna}, using a variation of Roy equations to constrain the scattering data in the $\sigma$ region and avoiding the use of ChPT \cite{Garcia-Martin:2011iqs}. All this led to its present name in the RPP.

Following a similarly tortuous path, the also very wide $K^*_0(700)$ obtained its current name in the 2018 RPP, but was still labeled under ``Needs Confirmation" until the 2020 edition ( see \cite{Pelaez:2015qba, Pelaez:2020gnd} for detailed accounts). Once again, it was the appearance of a pole when using crossing symmetric Roy-Steiner equations \cite{Steiner:1971ms,Buettiker:2003pp}, confirmed in different ways \cite{Descotes-Genon:2006sdr,Pelaez:2020uiw} with or without ChPT and/or scattering data input, which sealed the formal RPP recognition of the existence and parameters of this resonance (see \cite{Pelaez:2020gnd} for a review).
Both the $f_0(500)$ and $K^*_0(700)$ have been confirmed later with additional dispersive studies \cite{Danilkin:2020pak}.

In addition, for more than a decade, several lattice studies have confirmed the existence of $f_0(500)$ \cite{Briceno:2016mjc,Briceno:2017qmb,Guo:2018zss,Rodas:2023gma} and $K_0^*(700)$ \cite{Wilson:2014cna,Dudek:2014qha} $T$-matrix poles at unphysical pion masses, but gradually approaching the physical value. 
The dependence of resonance poles on the quark masses is also relevant for investigating their nature. Very recently, and once the existence of these poles was confirmed in the lattice, the precise determination of their parameters at unphysical masses has also been tackled with Roy-like dispersive methods \cite{Cao:2023ntr,Rodas:2023nec,Cao:2024zuy,Cao:2025hqm}.

Currently, the RPP provides a dedicated review on ``Scalar mesons below 1 GeV" with a detailed account of the $T$-matrix pole determinations for each member of the lightest scalar multiplet, giving particular weight to dispersive determinations. 
For these reasons and to minimize the model dependence, in our Table~\ref{tab:0+mesons}, we provide the scalar meson parameters in terms of $T$-matrix poles whenever they are reported in the RPP listings. 
Let us remark that, after the use of $ T$-matrix poles for these mesons, not only the lightest scalars, but many more resonance parameters are now reported in the RPP in terms of $T$-matrix poles.

\paragraph{Non-ordinary mesons. Inverted hierarchy, non-ordinary $N_c$ dependence and Regge trajectories.}

Once the members of the lightest scalar nonet have been established, it is clear that they do not follow the mass hierarchy of $q\bar q$ mesons in Fig.~\ref{fig:multiplets} (Left),
but are closer to the ``inverted" hierarchy of the tetraquark/molecule interpretation in 
Fig.~\ref{fig:multiplets} (Right).  Moreover, looking back at Table~\ref{tab:ordmesonnames}, in the QM $q\bar q$ assignment, they would be $1^3P_0$ states, with one more angular momentum unit than the lightest vector mesons and thus about 300 MeV heavier. Namely, they should have appeared above $\sim 1.1\,$GeV, not below. 
Even with \textit{ad hoc} flavor-dependent corrections to the NJL model, non-quarkonia states are needed to describe their masses \cite{Braghin:2020yri,Braghin:2022uih}.
 Naive tetraquarks also do not perform well, since four constituent-quark masses would add up to $\sim$1.2 GeV or more. However, because NGB masses are much lower than expected for quarkonia, the masses of the lightest scalars are very close to the mass of two NGBs.
This proximity is most evident in the $a_0(980)$ and $f_0(980)$ proximity to the $K\bar K$ threshold, and is very suggestive of a molecular nature. The $f_0(500)$ and $K_0^*(700)$ pole masses are, respectively closer to $2M_\pi$ and $M_K+M_\pi$ than to the 1 GeV scale. However, since they lie above these thresholds, they cannot be naively interpreted as meson molecules, but it seems natural that they might have a large meson-meson component.

Nevertheless, in 1976 Jaffe \cite{Jaffe:1976ig} showed that, within a ``bag model" based on the dominance of the magnetic contribution of gluon interactions, tetraquarks could appear well below 1 GeV while still displaying an inverted hierarchy. Hence, light scalars became the first non-ordinary-meson candidates. They would be cryptoexotics, since their quantum numbers can also be obtained from ordinary $q\bar q$ configurations, with which they will necessarily mix, thus complicating this simple picture.

Moreover, in the 1980s, it was shown that, although QM naive $q\bar q$ scalar states would naturally appear above 1 GeV, supplementing the model with unitarized meson-meson interactions could generate additional poles below 1 GeV \cite{vanBeveren:1986ea,vanBeveren:2006ua}, leaving the initial QM poles above. 
Indeed, it is now known that a simple unitarization of the universal ChPT LO calculation of $\pi\pi$ and $K\pi$ scattering generates both the $f_0(500)$ and $K_0^*(700)$ poles to a surprisingly good approximation \cite{Oller:1997ti,Pelaez:2015qba,Gross:2022hyw}. 
In general, all approaches incorporating the LO constraints and unitarity find these two states, as well as the rest of the lightest scalar nonet \cite{Black:1998zc,Black:1998wt,Black:2000qq}. 
These states are often dubbed ``dynamically generated", meaning that the formation and properties of these resonances are mostly due to meson-meson dynamics (loops), acting below the 1 GeV scale rather than quark-level dynamics that could provide a formation seed, also called ``preexistent" state, or an inner core \cite{Close:2002zu}. 

Interestingly, when using a unitarized LO chiral approach, it is possible to show the degeneracy of the octet members in the SU(3) limit. By studying the couplings defined from the poles of the physical states \cite{Oller:2003vf}, it can be concluded that the $f_0(500)$ is mostly singlet, whereas the $f_0(980)$ is mostly octet, with a mixing angle of $(19\pm5)^\circ$, far from ideal mixing. Given the huge width of the $f_0(500)$, it is very important to avoid inadvertently using the narrow-width approximation.

In contrast, other states, such as vector mesons, cannot be generated from the unitarization of NLO ChPT; they require explicit NLO contributions containing information on energy scales above meson dynamics. 
This explains one of the light-scalar meson puzzles, namely, why the values of the ChPT NLO low-energy constants are dominated by the vector-meson contributions, rather than the light-scalar contributions, despite the latter being lighter. The answer is that the dynamics governing the formation of the lightest scalars is dominated by meson-scale physics, which is contained in the ChPT LO. In contrast, as we will see soon, vector mesons are archetypal ordinary $q\bar q$ states. Hence, their formation dynamics are dominated by quark interactions at distances below the meson scale, encoded in ChPT through the NLO low-energy constants.

Actually, when considering UChPT to NLO \cite{Dobado:1996ps,Oller:1998hw,Nieves:1998hp,
GomezNicola:2001as,Pelaez:2004xp,Pelaez:2015qba}, NGB scattering data can be well described up to 1.2 GeV, and the resulting $T$-matrix poles are very consistent with the present values. Since this calculation includes the NLO low-energy constants, it is also capable of reconstructing the light vector meson poles. It is then possible to vary the ChPT parameters following the QCD LO $N_c$ dependence \cite{Pelaez:2003dy,Pelaez:2006nj}. As expected, vector-meson poles behave as ordinary mesons \cite{tHooft:1973alw,Witten:1980sp,Jaffe:2007id}. However, for $N_c$ not too far from its physical value of 3, the light-scalar pole displays a non-ordinary $N_c$ dependence, at odds \cite{Weinberg:2013cfa,Knecht:2013yqa} with the $q\bar q$ or pure tetraquark expected behavior. It is indeed dominated by the ChPT dynamics encoded in the scale $f_\pi$, which strongly supports a predominant non-ordinary, possibly meson-meson, nature for the lightest scalars. This behavior close to $N_c=3$ is very robust, as found even without ChPT in \cite{Harada:2003em}. However, for $N_c\gg3$, the behavior is not so well established, but combined with other techniques \cite{RuizdeElvira:2010cs} or at NNLO in UChPT \cite{Pelaez:2006nj}, it has been found that the scalar poles may behave again as $q\bar q$ states but with a much bigger mass. This suggests again the presence of a small admixture with some $q\bar q$ or tetraquark states with masses around 1 GeV, be it in the form of a QM seed, a preexisting state, or an inner core.
The mixing (via QCD non-perturbative instanton interactions) between a heavier quarkonia with $L=1$, and a light tetraquark or diquark-antidiquark nonet, has also been suggested as an explanation for the dynamics governing the formation of the lightest scalar mesons \cite{Maiani:2004uc,tHooft:2008rus}. 
The $f_0(500)$ is also obtained around 400 MeV within the Schwinger-Dyson functional approach \cite{Heupel:2012ua, Eichmann:2015cra} and its wave function seems to be dominated by $\pi\pi$ constituents.

Another piece of evidence for the non-ordinary nature of the $f_0(500)$ and $K^*_0(700)$ resonances is that they cannot be accommodated \cite{Anisovich:2000kxa} into linear $(M^2,J)$ Regge trajectories, which are indicative of a confining force at the GeV scale. Dispersive calculations \cite{Londergan:2013dza,Pelaez:2017sit,Carrasco:2015fva} using as input the pole positions ---without ignoring their widths--- yield a linear behavior for other ordinary low-lying mesons, but result in non-linear trajectories for light scalars, pointing once again to dynamics at the meson-meson scale. Regge trajectories will be discussed in more detail in Sec.\ref{sec:Regge}.

\subsubsection{Scalar mesons above 1 GeV: a well-established second nonet, hints of a third one, and a glueball}

\medskip

\paragraph{The second nonet and the extra singlet}
 We have already discussed that the lightest glueball candidate is expected to have the quantum numbers of an $f_0$ meson, and a mass around 1.7 to 1.9 GeV. 
 This expectation makes scalars in this region particularly interesting.

As seen in Table \ref{tab:0+mesons} and the right panel of Fig.\ref{fig:J0}, there are enough well-established states between 1.3 and 1.7 GeV to form another nonet and an additional singlet. 
The isotriplet $a_0(1450)$, and the two isodoublets $f_0(1370)$ and $f_0(1500)$ were first discovered at the Crystal Barrel proton-antiproton experiment running in the 1980s and 1990s at LEAR@CERN (see \cite{Amsler:1997up} for a review). 
Following this discovery, it was soon proposed \cite{Amsler:1995td,Amsler:1995tu} that the two of them could not be $q\bar q$ states and one was compatible with the expected glueball properties at the time.  The $f_0(1710)$, under a different name, was discovered at the Brookhaven National Laboratory (BNL) \cite{Etkin:1982se} in 1982, in $\pi p\to K_SK_Sn$. These three $f_0$ resonances have been observed in many other experiments. Until recently, the $f_0(1370)$ was still questioned, in part because it was not needed in the original 
$\pi\pi$ scattering experimental analysis \cite{Hyams:1973zf,Hyams:1975mc}. However, a recent dispersive reanalysis with analytic continuation techniques has shown \cite{Pelaez:2022qby} that it is indeed present in that data. Still, this resonance has the largest uncertainties in its parameters.

In general, these states have relatively large widths, disintegrating mostly into pairs of pseudoscalar mesons, or four mesons as in the case of the $f_0(1370)$ and $f_0(1500)$, which decay predominantly into $4\pi$.  
Moreover, their shapes in experiments can be deformed further by the presence of two-meson production thresholds like $\bar KK$, $\eta\eta$, $\eta\eta'$, $\sigma\sigma$, $\omega\omega$, $\rho\rho$, $N\bar N$, etc. 

The proximity of several wide $f_0$ resonances does not allow for a straightforward identification of the isoscalar in the octet, the singlet in the nonet, nor the singlet from the glueball. They are a mixture of all of them, and thus, in Fig.\ref{fig:J0}, we have not assigned the physical isoscalars to a specific multiplet, but rather pointed out their possible mixture.

\paragraph{Their composition and glueball content.}

 If this nonet were of the ordinary $q \bar q$ nature, it would correspond to a $1^3P_0$ configuration; but the measured masses do not follow very well the expected hierarchy of QM quarkonia. In particular, the $K^*_0(1430)$ and $a_0(1450)$ almost have the same mass, which suggests that these are not pure states of a given configuration or composition. We have already seen that NGB rescattering plays a dominant role in the formation of the lowest scalar nonet below 1 GeV, and it also distorts the multiplets above 1 GeV. We had already commented that the interplay between scalars above and below 1 GeV could be understood as a mixing between bare or preexisting states with a definite composition \cite{vanBeveren:1986ea,Black:1999yz,Oller:1998zr}.
A relatively recent unquenched lattice study \cite{Brett:2019tzr,Morningstar:2024vjk} suggests that no scalar state below 2 GeV is predominantly a $q\bar q$ nor a glueball state, and a large, if not dominant, molecular component is present. 

Implementing meson-meson interactions correctly is essential for describing the data in this region, preferably using dispersive methods or unitarization.
 Within unitarized NJL-L$\sigma$M quark-models, two nonets and a singlet can provide a consistent description of data below $\sim$1.7 GeV \cite{Close:2002zu}. 
 As we know, one of them could be ``preexistent" \cite{vanBeveren:1986ea,vanBeveren:2006ua,Oller:1998zr}.
A detailed description may require introducing four-meson decays as quasi-two-body decays ($\sigma\sigma$, $\rho\rho$, etc.), i.e., treating the $\sigma$ and $\rho$ as quasi-stable states. Indeed, some of these scalars also appear in unitarized vector-vector scattering chiral amplitudes \cite{Molina:2008jw}, suggesting a large meson-meson component, or even a largely molecular nature \cite{Molina:2008jw,Geng:2008gx,Brett:2019tzr}.

Coupled-channel analyses are therefore essential, and the more data that can be fitted, the better they become. One can also compare different decays and, from all the information, infer mixing patterns. Over the years, phenomenological analyses have become more elaborate. 
Since there is no space here to explain them all, next we will briefly review a few of the most complete and refer the reader to specialized reviews \cite{Klempt:2007cp,ParticleDataGroup:2024cfk,Chen:2022asf,Gross:2022hyw}
for further details on this topic and even historical accounts of its evolution.

In particular, using LO NGB unitarized interactions, the consistent description of NGB scattering and Crystal Barrel and WA102 $p\bar p$ annihilation data up to 2 GeV, requires two preexisting or bare chiral octets and a singlet \cite{Albaladejo:2008qa}. These multiplets are dressed and mixed by the pseudoscalar-pseudoscalar interactions to yield the measured masses and widths. Quasi-two-body chiral interactions with $\sigma\sigma$ and $\rho\rho$ are also included. Chiral symmetry is a relevant ingredient of this analysis, due to an expected chiral suppression in the production from the glueball of $u\bar u$ or $d\bar d$, relative to $s \bar s$, \cite{Chanowitz:2005du,Chanowitz:2007ma}. Conclusions are drawn from poles, without introducing mixing angles. It turns out that the $f_0(1710)$ and an important contribution to the $f_0(1500)$ are of glueball nature, whereas the $f_0(1370)$ predominantly stems from the octet. 

Concerning the radiative $J/\psi$ decays, recently \cite{Sarantsev:2021ein}, the BESIII data have been analyzed within a $\pi\pi, K \bar K, \eta\eta, \omega\phi$ coupled-channel formalism constrained by the CERN-Munich data
on $pN$ scattering (meson-meson scattering indeed), from the GAMS collaboration at CERN, from BNL, and $p\bar p$ annihilation at LEAR, finding as many as ten isoscalar-scalar mesons. Based on their approximate proximity to $q\bar q$ radial Regge trajectories (see section about Regge trajectories below) the $f_0(500)$, $f_0(1370)$, $f_0(1710)$, $f_0(2020)$ and $f_0(2200)$ come out as mostly singlet, whereas the $f_0(980)$, $f_0(1500)$, $f_0(1770)$, $f_0(2100)$ and $f_0(2330)$ as mostly octet. The octets should barely appear in radiative $J/\psi$ decays, but, in practice, they do, which is interpreted as a signal of mixing with a glueball. The mass and width of the unmixed or preexisting glueball are determined by a Breit-Wigner fit to reproduce all the $f_0$ yields, and it appears around 1865 MeV with a width of $\simeq 370$ MeV \cite{Klempt:2021wpg}. Its total production rate, or yield from radiative $J/\psi$ decays would be $(5.8\pm1.0)\times10^{-3}$, remarkably close to the lattice-QCD prediction \cite{Gui:2012gx} $(3.8\pm0.9)\times10^{-3}$.
In real life, this glueball state becomes part of the surrounding $f_0$ states, with a maximum of $\sim25\%$ glueball content for the $f_0(1770)$ and $\sim15\%$ for the $f_0(1710)$, $f_0(2020)$, and $f_0(2100)$.
Gluons also couple strongly to $\eta$ and $\eta'$ through the strong anomaly \cite{Novikov:1979uy,Gershtein:1983kc,Ball:1995zv}
and the analysis \cite{Frere:2015xxa} of $J\psi$ decays including $\eta\eta'$ states, performed before BESIII,
also suggested that the $f_0(1710)$ has the largest glueball component.

The Joint Physics Analysis Collaboration (JPAC) has also performed a study
\cite{Rodas:2021tyb} of the radiative $J/\psi\to \gamma\pi^0\pi^0$ and $\gamma K_S K_S$ partial waves measured by BESIII \cite{BESIII:2015rug,BESIII:2018ubj}, using a large set of parameterizations satisfying unitarity and analyticity to reduce model dependence. Apart from the two $\pi\pi$ and $\bar KK$ coupled channels,
they require an additional $\rho\rho$ channel, which provides a substantial contribution. They determined precisely the properties of four scalar and three tensor resonances from 1.25 to 2.5 GeV. In the scalar sector, they identify poles for the $f_0(1500)$, $f_0(1710)$, $f_0(2020)$
and $f_0(2330)$ and obtain their couplings. 
The  $f_0(1710)$, although with a mass $\sim 1769\,$MeV, larger than in the RPP, is the one with the largest glueball component (one may wonder if this could be the $f_0(1770)$).
The very BESIII collaboration obtains a similar conclusion \cite{Jin:2021vct}, by just noting that the production rate they estimate for the $J/\psi\to\gamma f_0(1500)$ is ten times smaller than that for $J/\psi\to\gamma f_0(1710)$, which is $\sim2.2\times10^{-3}$, once again relatively close to the lattice-QCD result we quoted above. 

Unfortunately, the scalar-meson spectrum above 1700 MeV is not well established. 
The RPP in their ``Spectroscopy of light-meson resonances" assigns the $a_0(1950)$ and $f_0(2020)$ to a third multiplet, but we have taken one tentative step further and incorporated also the $K^*_0(1850)$.  Note that the $f_0(1770)$ has a reasonable mass to complete the assignment, but we have seen that it may well have a large glueball component. Moreover, much as the $\rho\rho$ may be important around 1.5 GeV, the $K^*\bar K^*$ molecular component may be relevant around 1.7 GeV. In the QM, this nonet would correspond to $2^3P_0$, $ q\bar q$ configuration, but it is evident that the actual masses do not follow the $q\bar q$ hierarchy. Moreover, the recently discovered $a_0(1710)$ state has been added to the listings in the latest RPP update, but is still omitted from the summary tables. It has a mass $\simeq$1704 MeV according to BaBar \cite{BaBar:2021fkz}, or $\simeq$1817 MeV according to BESIII \cite{BESIII:2022npc}. Assigning this state to the tentative third nonet would make it appear much more $q\bar q$-like in the mass hierarchy, but a priori, there is no reason for it to be pure $q\bar q$. Other interpretations have been put forward, including an additional nonet \cite{Gross:2022hyw} (but a $K^*_0$ around 1700 MeV would be missing), or molecular states \cite{Geng:2008gx,Zhu:2022wzk}. For the possible repercussions of the recent $a_0(1710)$ discovery, and future research prospects, see \cite{Oset:2023hyt}.

In summary,  the existence of a second scalar nonet above 1 GeV is well established, but does not follow the naive $q \bar q$ mass hierarchy. Therefore, it may have a sizable meson-meson component, not only of pseudoscalar pairs, but also vector pairs and possibly lighter scalar pairs. The extent to which these contributions are important or even dominant remains a matter of debate.

Concerning the extra singlet due to the presence of a glueball, although the discovery of the $f_0(1500)$ made it the most promising glueball candidate at the time, nowadays it is widely agreed that the constituent glueball state is mixed between several $f_0$ resonances observed in the 1.5 to 2.2 GeV range. 
Over the years, and with the recent discovery of additional scalar states, the tendency is to consider that the $f_0(1710)$, or possibly the $f_0(1770)$, is the state with the largest glueball contribution.

There are hints of a third nonet, but the assignment of its members is very speculative. Additional states above 2 GeV have been reported, but further confirmation would be welcome. 

 From the theory side, analyses are becoming more complex and complete, incorporating more data and theoretical constraints. 
There is also considerable progress in lattice QCD towards unquenched calculations with realistic quark masses. Fortunately, from the experimental side, more results are expected from BESIII soon, and possibly from PANDA as well. Given the precision and quantity of the data, simple models should be set aside and theoretically sound analyses should be used to interpret existing and new results.

\subsection{$J^P=1^{-}$: Vector mesons}

In this case, we will discuss separately the observed resonances with non-exotic quantum numbers $J^{PC}=1^{--}$ from those with exotic quantum numbers $1^{-+}$, which are hybrid candidates.

\begin{figure}
	\centering
    \includegraphics[width=0.495\textwidth]{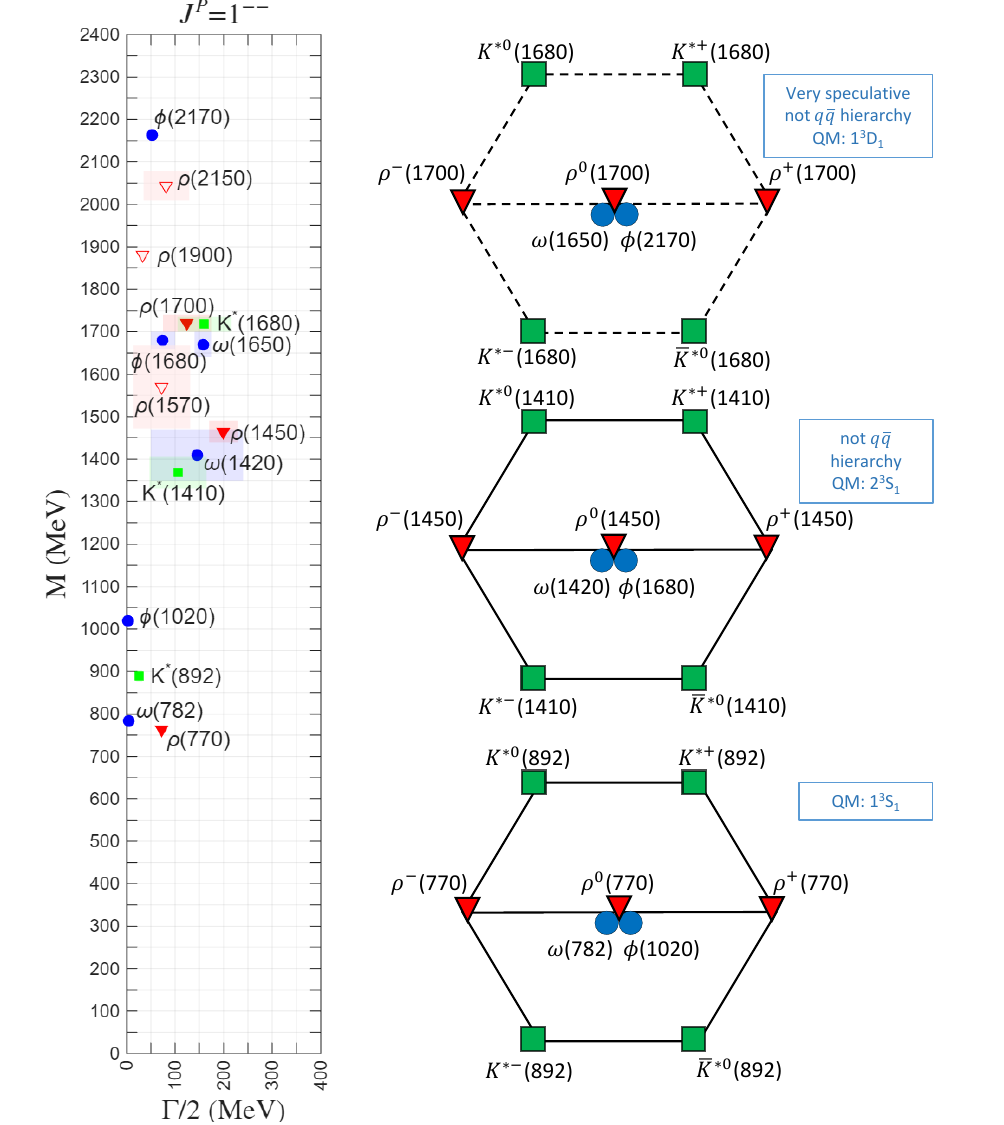}
    \includegraphics[width=0.495\textwidth]{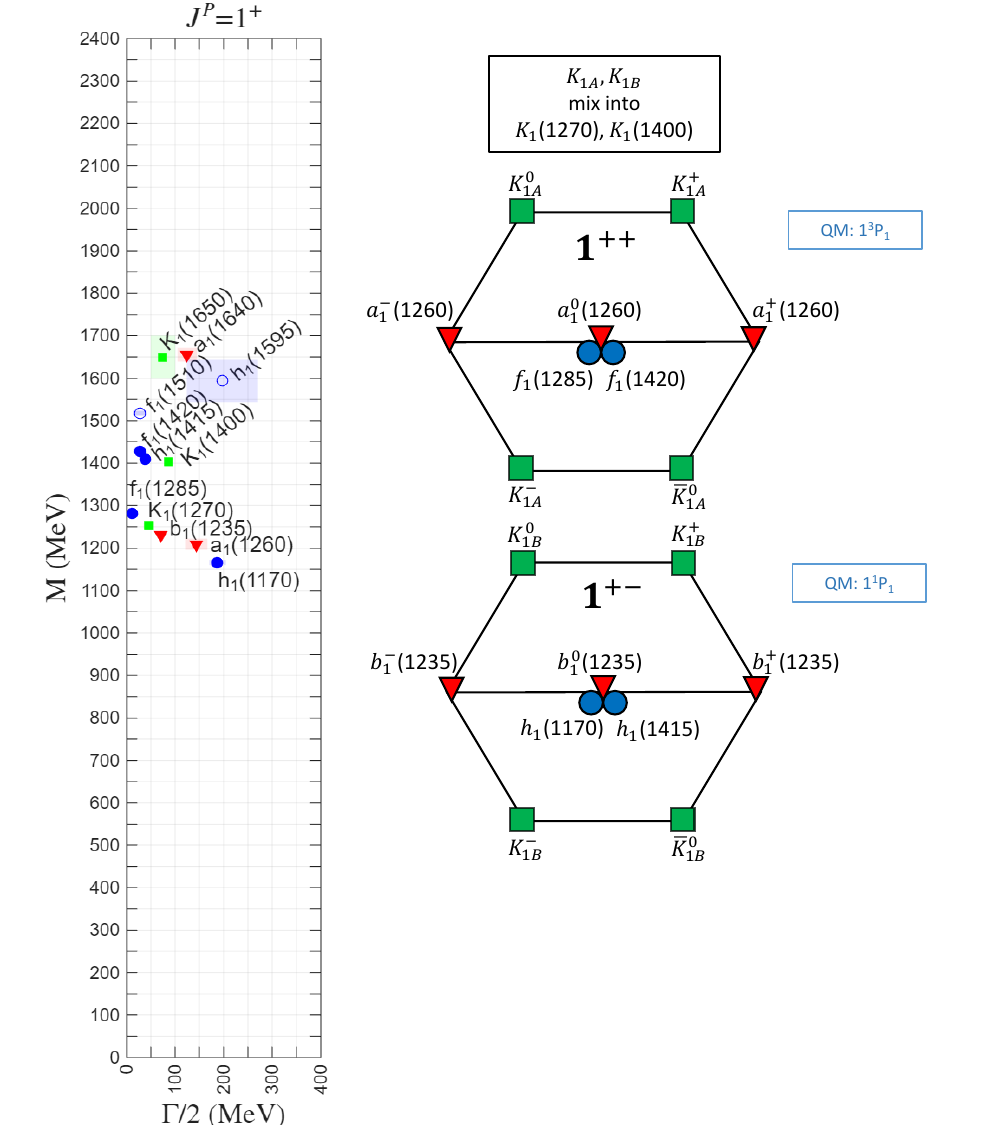}
	\caption{Non-exotic $J=1$ resonances in the RPP summary tables and their tentative nonet assignment. We also show their position and uncertainties (light-colored areas) in the $(M,\Gamma/2)$ plane, where hollow symbols stand for resonances omitted from the RPP summary tables.
    \textbf{Left:} Three $J^{PC}=1^{--}$ vector nonets could be formed. The lowest one is very well determined and is the archetypal QM nonet of ordinary quarkonia, following the expected mass hierarchy and an almost ideal mixing scheme for their isosinglets. In the QM, it corresponds to the $1^3S_1$ configuration. In contrast, the other two nonets do not follow the expected $q\bar q$ hierarchy, and the identification of their members, as well as the QM configuration, is more tentative.   
    In particular, the third nonet is very speculative. 
   \textbf{Right:} axial-vector mesons $J^{P}=1^{+}$. We display two multiplets: the slightly lighter one with $J^{PC}=1^{+-}$ and a slightly heavier one with $J^{PC}=1^{++}$, as determined by the quantum number $C$ of their non-strange components. The strange members of these two nonets, i.e. $K_{1A}$ and $K_{1B}$ mix into the physical $K_1(1270)$ and $K_1(1400)$ resonances.
	\label{fig:J1}}
\end{figure}

We show in the left panel of  Fig.\ref{fig:J1} the position of the $J^{P}=1^{--}$ vector resonances in the $(M,\Gamma/2)$ plane, together with their uncertainties as light-colored areas.
The figure also shows the tentative classification into three nonets of the non-exotic $1^{--}$ vectors, whose parameters are given in Table~\ref{tab:1-mesons}. 
We do not attempt any multiplet assignment for the few $1^{-+}$ states reported in the RPP, whose parameters we provide in Table~\ref{tab:1-+mesons}.

\subsubsection{The lightest $1^{--}$ vector nonet: The archetypal quark-model quarkonium}

These are the meson resonances that were discovered first, and that is why they are often referred to without their mass attached, namely $\rho$, $\omega$, $\phi$, etc.
In particular, already in 1957, Nambu \cite{Nambu:1957wzj} proposed the existence of a neutral vector meson to explain certain features of the proton and neutron form factors (like the repulsive core). He called it $\rho_0$, but nowadays it is called the $\omega(782)$ meson. It was discovered in 1961 at Berkeley \cite{Maglich:1961rtx} as a ``three-pion" resonance in $p\bar p$ annihilation.
Soon after Nambu's prediction, several works \cite{Frazer:1959gy,Breit1960,Bowcock1960,Sakurai:1960ju,Bergia:1961zz,Derado1960} suggested the existence of a ``two-pion" resonance, nowadays called the $\rho(770)$, which was 
discovered very soon after \cite{Derado1960,Stonehill:1961zz,Erwin:1961ny}.

\begin{table} \caption{ Isospin $I$, mass $M$, and decay width $\Gamma$ of the vector mesons $1^-$, together with their plausible nonet assignments (separated by a horizontal line).
$(T)$ stands for the $T$-matrix pole parameters. The RPP also provides different
Breit-Wigner parameters for the charged and neutral $\rho$ and $K^*(892)$ mesons, and even in different production mechanisms, typically differ by $O(1\,{\rm MeV})$. 
Resonances with an asterisk $(^*)$ are omitted from the RPP summary table. In addition, the $\rho(1900)$ and $\rho(2150)$ appear in the long Particle Listings without estimated or averaged values for the mass and width.  $(^\dagger)$ For them, we provide the BESIII results from
 \cite{BESIII:2022wxz} (only statistical error) and \cite{BESIII:2023sbq}, respectively.
\label{tab:1-mesons}}
\centering
\medskip
\begin{tabular}{llll}
\toprule
$I$ & Name & $M$ [MeV] & $\Gamma$ [MeV] \\
\midrule
1 & $\rho(770)$ & 761-765 $(T)$ & 142-148 $(T)$ \\
0 & $\omega(782)$ & 782.66$\pm$0.13 & 8.68$\pm$0.13 \\
0 & $\phi(1020)$ & 1019.460$\pm$0.016 & 4.249$\pm$0.013 \\
1/2 & $K^{*}(892)$ & 890$\pm$14 $ (T)$ & 52$\pm$12 $(T)$ \\
\midrule
1 & $\rho(1450)$ & 1465$\pm$25 & 400$\pm$60 \\
0 & $\omega(1420)$ & 1410$\pm$60 & 290$\pm$190 \\
0 & $\phi(1680)$ & 1680$\pm$20 & 150$\pm$50 \\
1/2 & $K^*(1410)$ & 1368$\pm$38 $(T)$ & 212$^{+96}_{118}$ $(T)$ \\
\midrule
 &  &  &  \\
 &  &  &  \\
\end{tabular}
\qquad\qquad\qquad
\begin{tabular}{llll}
\toprule
$I$ & Name & $M$ [MeV] & $\Gamma$ [MeV] \\
\midrule
1 & $\rho(1570)$ $(^*)$& 1570$\pm$98 & 144$\pm$118 \\
1 & $\rho(1700)$ & 1720$\pm$20 & 250$\pm$100 \\
0 & $\omega(1650)$ & 1670$\pm$30 & 315$\pm$35 \\
1/2 & $K^*(1680)$ & 1718$\pm$18 & 320$\pm$110 \\
\midrule
1 & $\rho(1900)$ $(^*)$ & 1880$\pm$10 $(^\dagger)$ & 65$\pm$15 $(^\dagger)$\\
1 & $\rho(2150)$ $(^*)$ & 2044$\pm$35 & 163$\pm$93\\
%1 & $\rho(1900)$ $(^*)$ & 1840-1950& 5-180\\
0 & $\omega(2220)$ $(^*)$ & 2188$\pm$21& 105$\pm$34\\
0 & $\phi(2170)$ & 2164$\pm$5 & 88$^{+26}_{-21}$ \\
\midrule
& & & \\
& & & \\
\end{tabular}
\end{table}

Both the $\rho(770)$ and $K^*(892)$ give rise to beautiful BW peaks in meson-meson interactions, which have been thoroughly studied with dispersive approaches, and thus we provide $T$-matrix pole parameters. Their model-dependent BW parameters are known with a precision of a few MeV (see the RPP). However, the precision required for certain calculations, such as hadronic contributions to the anomalous magnetic moments of leptons, necessitates more elaborate parameterizations.

The relevance of these states, apart from providing the repulsive core of the nucleon-nucleon potential, is that the neutral states share the quantum numbers of the photon. Very naively, this means that the exchange of these resonances dominates the electromagnetic interaction of hadrons. Indeed, in 1961, Sakurai \cite{Sakurai:1960ju} proposed a simple but very popular model for the interactions of photons and hadrons called ``Vector Meson Dominance" \cite{Sakurai:1960ju,Sakurai:1972wk} (see  \cite{Schildknecht:2005xr} for a review), where the photon becomes one of these neutral mesons before interacting with other hadrons. Intuitively, these mesons would be the hadronic components of the physical photon when it is dressed with strong interactions. It is a surprisingly good first approximation when describing electromagnetic form factors at the meson scales, and due to its simplicity, it is still used to obtain estimates or when no other methods are available. Of course, it is not QCD, and it fails at high energies, where quark degrees of freedom become important, as in deep inelastic scattering. In an effective Lagrangian formulation of VMD, vector mesons are identified with the dynamical gauge bosons of a ``Hidden local symmetry"  of a nonlinear sigma model \cite{Bando:1987br}.

These light vector states follow the QM $q\bar q$ expectations and mass hierarchy very nicely for a $1^3S_1$ configuration. Having $L=0$ and $S=1$, the $\rho(770)$ mass is roughly twice the constituent quark mass. 

The members of this nonet are therefore the first example we meet of genuine quarkonia within light mesons. There are, of course, meson-meson contributions, etc, but they are predominantly $q \bar q$. Therefore, their formation is dominated by physics at the quark scale. Being the lowest nonet encoding information from such scales, its low-energy contributions saturate the numerical values of most ChPT NLO low-energy constants \cite{Donoghue:1988ed,Ecker:1988te} (those that survive in the chiral limit or are related to the electromagnetic form factors). Consequently, apart from dominating electromagnetic hadron interactions, light vectors also dominate most low-energy NGB interactions.

This nonet is also a perfect example of ideal mixing, with $\theta_{\rm lin}=36.5^\circ$. Consequently, the $\phi$ is almost an $s \bar s$ meson, and since it is just 30 MeV above the $K\bar K$ threshold, it becomes very narrow. The $\omega$ is lighter and decays predominantly into three pions, which have a weaker coupling and have less phase space than two pions, in which the $\omega$ cannot decay in the isospin limit due to Bose statistics. The fact that these two states are almost stable makes the mixing formalism very consistent. Out of the isospin limit, the $\rho$ can mix with the $\omega$ (and also with the $\phi$, but this is a smaller effect), which is usually parameterized in terms of the mixing amplitude (see \cite{Gardner:1997ie}), which lies far beyond our scope.

\subsubsection{Heavier $1^{--}$ vector nonets}

The next $1^{--}$ vector nonet is also well identified. In the QM $q \bar q$ assignment, it would correspond to the radial excitations $2^3S_1$ of the previous nonet. However, it does not follow the $q \bar q$ mass hierarchy, nor the naive $q\bar q$ expectations for their decays.
Since a hybrid can be constructed with the same quantum numbers, it has been proposed that this nonet could be a quarkonia-hybrid mixture \cite{Barnes:1996ff,Close:1997dj}.  

There are enough resonances to build a third nonet. Here we follow the RPP proposal.
In the QM, it would not be the next radial excitation of the previous nonets, but a $1^3D_1$ configuration $1^3D_1$. The well-established $K^*(1680)$ is somewhat light for a $1^3D_1$ state, but clearly needs some partners. The $\rho(1700)$ would be surprisingly close to the $K^*(1680)$ for a $q \bar q$ octet. In contrast, the $\rho(1570)$, which 
is also light for a $1^3D_1$ state, is omitted from the summary tables, and could be just a misidentified decay mode of the $\rho(1700)$. The $\phi(2170)$ would be too heavy, but maybe not if it belongs in a multiplet with the $\rho(1900)$, also omitted from the summary tables, which could be due to $N \bar N$ threshold effects.
The $\phi(2170)$ has also been interpreted as a tetraquark ($s\bar s s \bar s$ \cite{Ke:2018evd} or $su \bar s \bar u$ \cite{Agaev:2019coa}).

In summary, the identification of the members of the second nonet is widely accepted, but their nature is unclear. The members and nature of the third nonet are still a matter of discussion.
Apart from $q\bar q $ states, alternative interpretations are possible, in terms of crypto-exotic hybrids and tetraquarks, and of course their mixing with quarkonia, which would deform considerably the naive $q \bar q $ expectations..

\subsubsection{At least one exotic $1^{-+}$ meson}

In its latest edition, the RPP summary tables include one candidate with such numbers: The $\pi_1(1600)$, whose  $T$-matrix pole position is given in Table~\ref{tab:1-+mesons}. 
Once these quantum numbers are identified, it cannot be a quarkonia state, and of course, it cannot be a glueball if it has isospin 1.

As we had already seen, the lightest exotic hybrid is expected to have $J^{PC}=1^{-+}$ and the $\pi_1(1600)$ is therefore a strong candidate for this non-ordinary exotic configuration. However, such non-ordinary numbers can also be obtained with molecular states \cite{Yan:2023vbh}.

\begin{table} [h]
\caption{ Isospin $I$, mass $M$ and decay width $\Gamma$ of the exotic vector mesons $J^P=1^{-+}$. (T) stands for the $T$-matrix pole parameters. 
 $(^*)$ The $\eta_1(1885)$, so far only observed at BESIII\cite{BESIII:2022riz}, is omitted from the RPP summary tables but appears in the listings.
  \label{tab:1-+mesons}
 }
\centering
\medskip
\begin{tabular}{lllll}
\toprule
$I$ & $C$ & Name & $M$ [MeV] & $\Gamma$ [MeV] \\
\midrule
1 & + & $\pi_1(1600)$ &  1480-1680 (T) & 300-600 (T)\\
0 & + & $\eta_1(1885)$ $(^*)$ & 1885$\pm$9$^{+6}_{-1}$ & 188$\pm$18$^{+3}_{-8}$ \\
\bottomrule
\end{tabular}
\end{table}

This resonance has been observed in various processes, including pion diffraction, antiproton-nucleon annihilation, and the decays of heavy mesons.
For a few years, two states $\pi_1(1400)$ and $\pi_1(1600)$ were reported, and their existence was the subject of intense debate. Even in the same experiment, COMPASS@CERN, two different masses were found when analyzing $\eta\pi$ or $\eta'\pi$ decays. This situation is still reflected in the different masses for these modes provided in the RPP when no $T$-matrix poles are used.
These puzzles were first clarified by the very COMPASS collaboration \cite{COMPASS:2018uzl,COMPASS:2021ogp}.
The JPAC Collaboration \cite{JPAC:2018zyd} carried out a decisive analysis using a
dispersively constrained unitary coupled channel analysis of $ \eta\pi$ and $\eta'\pi$ in the $J=1$ and $J=2$ amplitudes provided by COMPASS \cite{COMPASS:2014vkj}.
They found two poles for the $a_2(1320)$ and $a_2(1700)$ resonances in the spin-2 sector,
but just a single resonance pole with isospin 1, located at $(1564\pm89)-i(246\pm56)\,$MeV, quite consistent with the $\pi_1(1600)$.
These results were confirmed in a later work \cite{Kopf:2020yoa}, which combined COMPASS and Crystal Barrel data; however, the two-pole scenario could not be completely ruled out. The large decay ratio $\Gamma_{\eta'\pi/\Gamma{\eta\pi}}\sim5.5$ found in \cite{Kopf:2020yoa} matches the expectations of a strong anomaly-mediated process 
coupling the $\eta,\eta'$ state to the glue content  \cite{Frere:1988ac,Frere:2025mpf}.

Of course, once an exotic isotriplet meson has been identified, its partners are expected to lie nearby, which may help to determine their nature. 
In this respect, BESIII has reported an enhancement around 1885 MeV in the $J/\psi$ decay into $\eta\eta'$ with $J=1$ \cite{BESIII:2022iwi}, soon identified as a resonance with exotic $J^{PC}=1^{-+}$
quantum numbers. The RPP lists this as $\eta_1(1885)$ resonance, whose parameters are also listed in Table~\ref{tab:1-+mesons}, although it is omitted from the summary tables. If confirmed, it would be a plausible candidate for one of the two isoscalar partners of the $\pi_1(1600)$. The relatively large 
$\eta\eta'$ decay of the $\eta_1(1885)$ also matches the expectations of the expected anomaly enhancement due to a glue content \cite{Chen:2022qpd}.

Note that since these states have exotic quantum numbers, their names do not appear in the naming scheme illustrated in Table~\ref{tab:ordmesonnames}.

\subsection{$J^P=1^+$: Axial-vector mesons}

We list the axial-vector mesons and their parameters in Table~\ref{tab:1+mesons}.
Their tentative multiplet assignments are shown in the right panel of Fig.\ref{fig:J1},
together with their position in the $(M,\Gamma/2)$ plane. The uncertainties in these positions are shown as light-colored areas.

\begin{table} %[b]
\caption{ Isospin $I$, mass $M$ and decay width $\Gamma$ of the axial-vector mesons $J^P=1^+$, together with plausible nonet assignments.\label{tab:1+mesons}
(T) stands for the $T$-matrix pole parameters. 
\textbf{Left:} Below 1600 MeV, there are two nonets with opposite charge conjugation $C$, which, however, is a number that is only well defined for neutral particles with $I=0$. The absence of this number for the strange and anti-strange isodoublets allows for the mixing of the $K_1$ members of both multiplets. The $h_1(1415)$ was $h_1(1380)$ until recently.
\textbf{Right:} Some additional $1^-$ states above 1500 MeV. Those marked with $(^*)$ are omitted from the RPP summary tables.}
\centering
\medskip
\begin{tabular}{lllll}
\toprule
$I$ & $C$ & Name & $M$ [MeV] & $\Gamma$ [MeV] \\
\midrule
1 & + & $a_1(1260)$ & 1209$^{+13}_{-10}$ (T) & 576$^{+90}_{-24}$ (T) \\
0 & +&$f_1(1285)$ & 1281.8$\pm$0.5 & 23.0$\pm$1.1 \\
0 &+ &$f_1(1420)$ & 1428.4$^{+1.5}_{-1.3}$ & 56.7$\pm$3.3 \\
1/2 & &$K_{1A}$ & \multicolumn{2}{l}{ Mixture of $K_1(1270)$ and $K_1(1400)$} \\
\midrule
1 &- &$b_1(1235)$ & 1229.5$\pm$3.2 & 142$\pm$9 \\
0 &- &$h_1(1170)$ & 1166$\pm$6 & 375$\pm$ 35\\
0 & -&$h_1(1415)$ & 1409$^{+9}_{-8}$ & 78$\pm$11 \\
1/2 & &$K_{1B}$ &  \multicolumn{2}{l}{ Mixture of $K_1(1270)$ and $K_1(1400)$} \\
\midrule
 & &$K_1(1270)$& 1253$\pm$7& 90$\pm$20 \\ 
 & &$K_1(1400)$& 1403$\pm$7& 174$\pm$13\\%[3pt]
\bottomrule
\end{tabular}
\qquad
\begin{tabular}{lllll}
\toprule
$I$ & $C$ & Name & $M$ [MeV] & $\Gamma$ [MeV] \\
\midrule
1 & + & $a_1(1640)$ & 1655$\pm$16 & 250$\pm$40 \\
1 & + & $f_1(1510)$ $(^*)$ & 1518$\pm$5 & 73$\pm$25 \\
\midrule
1/2 & &$K_1(1650)$ & 1650$\pm$50 & 150$\pm$50\\
\midrule
0 & - & $h_1(1595)$ $(^*)$ & 1594$\pm$15$^{+10}_{-60}$ & 394$\pm$60$^{+70}_{-100}$\\
\bottomrule
&&&&\\
&&&&\\
&&&&\\
&&&&\\
&&&&\\
&&&&\\
\end{tabular}
\end{table}

An axial vector, or pseudovector, behaves as a vector under rotations but does not change sign under parity.  Axial-vector neutral states are $C$ eigenstates whose $C$-parity eigenvalues can be either $+1$ or $-1$. Particles with opposite $C$ parity must belong to different multiplets and cannot mix with each other. This is why, contrary to other tables, we also provide $C$ in the table and separate the multiplets into those with $C=1$ and those with $C=-1$. As usual, the notation is abused, and the same $C$-parity is assigned to the whole isotriplet, even if the charged members of the isotriplet change into one another under $C$.   However, $C$-parity is not defined for the $1^+$ strange isodoublets. Therefore, 
the observed $K_1$ states are a mixture of the $K_{1A}$ and $K_{1B}$ mathematical combinations that would belong to the $C=+1$ and $C=-1$ nonets, respectively. They are customarily defined as follows:
\begin{equation}
    \ket{K_1(1400)}=\cos\theta_{K_1}\ket{K_{1A}}-\sin\theta_{K_1}\ket{K_{1B}},\quad
    \ket{K_1(1270)}=\sin\theta_{K_1}\ket{K_{1A}}+\cos\theta_{K_1}\ket{K_{1B}}.
\end{equation}
Assuming these nonets are quarkonia, the mixing angle is  $\theta_{K_1}=(33.6\pm 4.3)^\circ$ \cite{Cheng:2011pb,Divotgey:2013jba}.

 If they were $q \bar q$, in the QM the $1^{+-}$ and $1^{++}$ nonets would correspond to $1^1P_1$ and $1^3 P_1$ configurations, respectively. Thus, they only differ in one unit of $S$, which is consistent with their relatively similar mass. Nominally, the $J^{+-}$ nonet is only slightly lighter than the $J^{++}$ one. In contrast, both have one more unit of angular momentum than the lightest vector $1^{--}$, and thus they are expected to be heavier than the $\rho(770)$ nonet by $\sim$400-600 MeV, which is indeed the case. It is difficult to determine whether the members of the nonets follow the expected mass hierarchy due to the mixing of the $K_1$ states. In the $q\bar q$ assignment the isoscalar mixing angles are $\theta_{+-}\sim28^\circ$ and $\theta_{++}\sim23^\circ$
 \cite{Cheng:2011pb}.

 However, it should be remarked that, within a chiral unitary approach 
 \cite{Geng:2006yb} with vector degrees of freedom besides the NGBs, the description of the data requires the presence of two nearby $T$-matrix poles for the $K_1(1270)$, besides the one for the $K_1(1400)$.  In principle, the existence of two different resonances, instead of just the $K_1(1270)$, could be tested with more precise charmonium decays.
 In this respect, it should be noted that there is another very close isosinglet, the $f_1(1510)$, although it is omitted from the summary tables. Moreover, it has been argued that the $f_1(1410)$  is not
a resonance, but the manifestation
of the $\pi a_0(980)$ and $K^* \bar K$ decay modes of the $f_1(1285)$ at energies higher than the nominal mass, with the first mode magnified by a triangle singularity \cite{Debastiani:2016xgg}. If any of these caveats is confirmed, the interpretation of this nonet should be revisited.

 In any case, it is quite clear that the masses of the lightest axial mesons lie between 1150 and 1500 MeV, which is very relevant for our understanding of chiral symmetry breaking. As we have already discussed, if the chiral SU(3)$_L\times$SU(3)$_R$ symmetry were exact, the $1^{-}$ vector nonet should have a degenerate $1^+$ axial-vector chiral partner. However, they differ by 400 MeV or more, and such a big breaking scale cannot be explained by the explicit breaking due to the small $u$ and $d$ masses. As we know, chiral symmetry is spontaneously broken.

It is worth noticing that there are four other nearby states above 1.5 GeV, listed in the right sub-table of Table~\ref{tab:1+mesons}, although two of them are omitted from the summary tables. The RPP review on ``Spectroscopy of light scalar mesons"
suggests that the well-established $a_1(1640)$ is a radial excitation of the $a_1(1260)$, i.e. $2^3P_1$. The $f_1(1510)$ decays seem consistent with a predominant $s\bar s$ nature, but it would be too light to be a radial excitation of the first nonet.

 In summary, enough axial resonances have been observed to form two nonets with opposite $C$-parity, whose interpretation as quarkonia seems plausible. Nevertheless, some caveats have been raised, such as the existence of additional poles for the $K_1$, and alternative interpretations have been proposed in the literature.

\subsection{$J^{P}=2^{+}$: Tensor mesons. Hints of a glueball}

\begin{figure}
	\centering
    \includegraphics[width=0.495\textwidth]{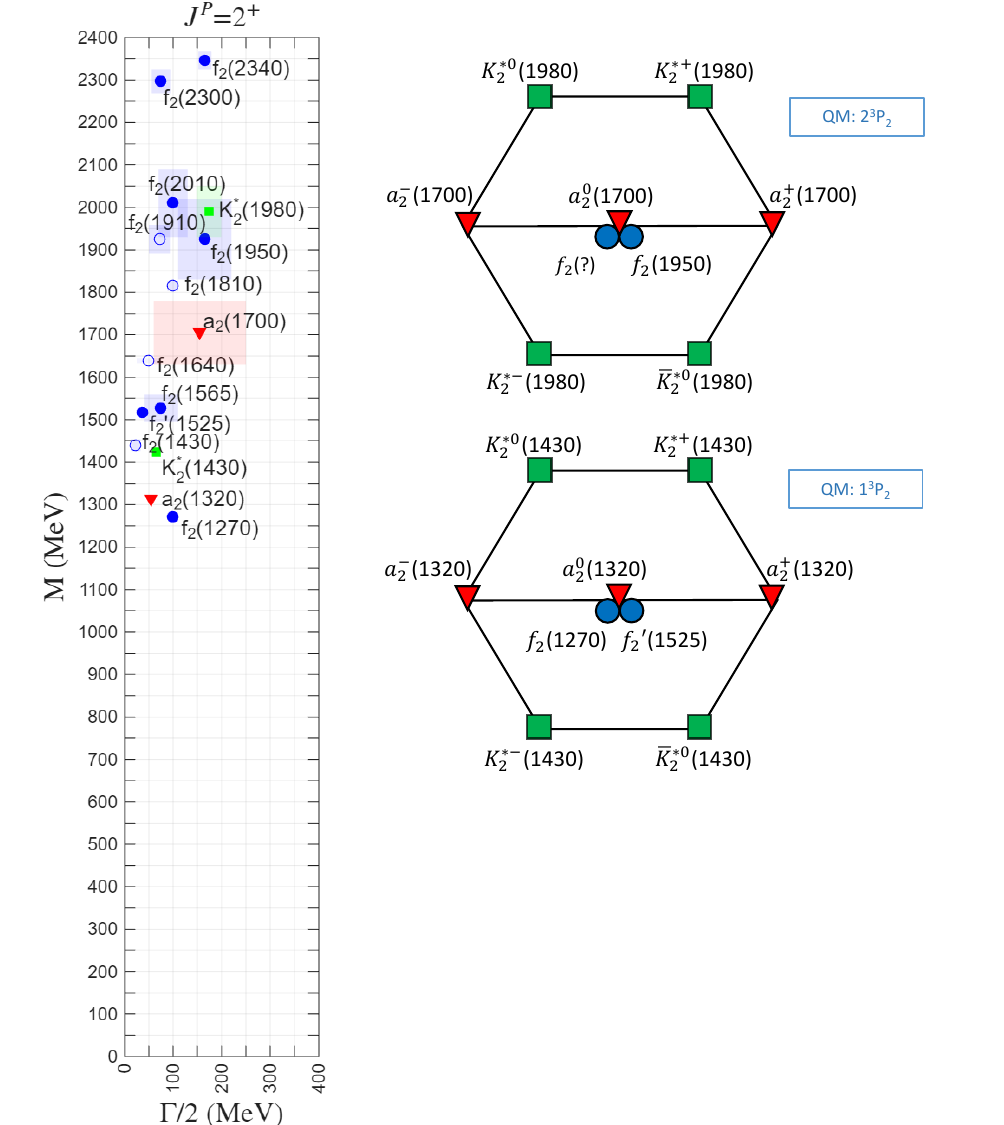}
    \includegraphics[width=0.495\textwidth]{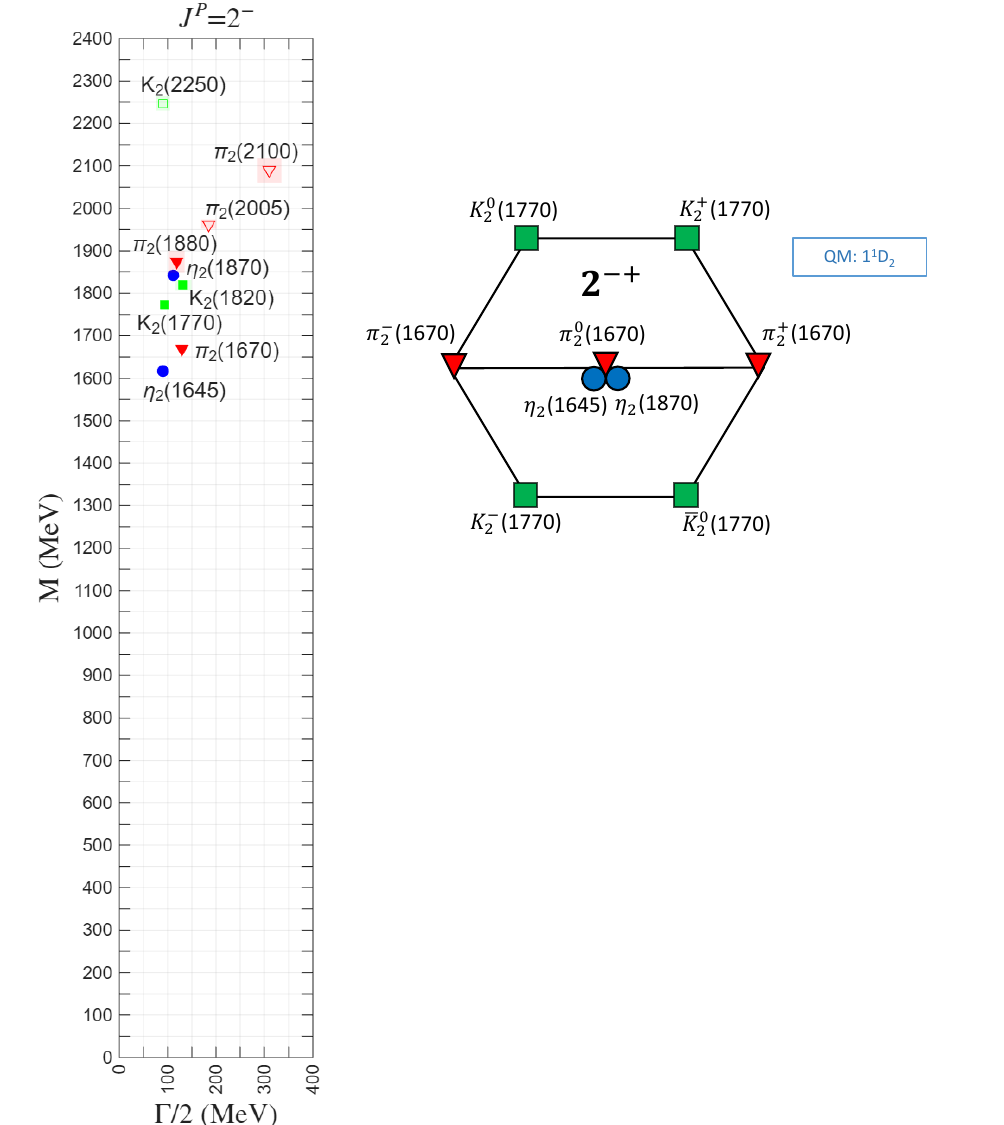}
	\caption{Tensor $J=2$ mesons in the RPP and their tentative nonet assignments, together with their position in the $(M,\Gamma/2)$ plane, where we also show their estimated uncertainties as light-colored areas.
    \textbf{Left:} $J^{PC}=2^{++}$ tensor mesons. Enough states have been observed to form two nonets, which follow the quarkonia mass hierarchy. The second would be a radial excitation of the first. Note the overpopulation of $f_2$ states, four of them omitted from the summary tables  (hollow symbols), some of which might be the same state. Those near or above 2 GeV are glueball candidates or are expected to have sizable glueball components. We are not plotting another resonance, listed as $f_J(2200)$, but omitted from the summary tables, whose $J^{PC}$ quantum numbers are unknown yet, but could be $2^{++}$ or$4^{++}$.
    \textbf{Right:} Pseudotensor $J^{P}=2^{-}$ resonances. Only $2^{-+}$ isotriplets and isosinglets are reported in the RPP. The nonet $2^{-+}$ follows the expected $q\bar q$ mass hierarchy and would correspond to $1^1D_2$ states. The RPP also reports two other well-established nearby states, the $K_2(1820)$ and $\pi_2(1880)$, as well as three other resonances omitted from the summary tables.  }
	\label{fig:J2}
\end{figure}

The list and parameters of the observed mesons with $J^{P}=2^+$ are given in the left sub-table of Table~\ref{tab:2+mesons}. In the left panel of Fig.\ref{fig:J2}, we show their tentative classification into nonets as well as their positions in the $(M,\Gamma/2)$ plane, whose uncertainties are represented by light-colored areas.

\begin{table} %[b]
\caption{ Isospin $I$, mass $M$ and decay width $\Gamma$ of tensor mesons $J=2$, together with plausible nonet assignments. Here (T) stands for $T$-matrix pole parameters.
Resonances with an $(^*)$  are omitted from the RPP summary table.
Left, $J^{PC}=2^{++}$ resonances: The lowest tensor nonet below 1.55 GeV is rather well established.  Enough candidates exist between 1.5 and 2 GeV to form another nonet.  
Very likely, out of these twelve reported $f_2$ resonances, several of them are the same state, due to low statistics and/or unaccounted systematic uncertainties in different production mechanisms. $(^\dagger)$ The RPP does not provide an estimate of the $f_2(1430)$ parameters and we provide the latest value they list \cite{BaBar:2021fkz}.
$(^\ddagger)$ The RPP lists different parameter estimates for different decay modes of the $f_2(1910)$, and we provide the range that covers them.
\textbf{Right:} The pseudo-tensor nonet with $J^P=2^-$ and two additional states above 1800 MeV that may be part of yet another nonet. Only states with $C=+$ appear in the RPP Listings.
\label{tab:2+mesons}
}
\centering
\medskip
\begin{tabular}{lllll}
\toprule
$J^P$& $I$ & Name & $M$ [MeV] & $\Gamma$ [MeV] \\
\midrule
$2^+$& 1 & $a_2(1320)$ & 1305–1321 (T) & 104-116 (T) \\
$2^+$& 0 & $f_2(1270)$ & 1260-1283 (T) & 180-220 (T) \\
$2^+$& 0 & $f_2'(1525)$ & 1517.3$\pm$2.4 & 72$^{+7}_{-6}$ \\
$2^+$& 1/2 & $K^*_2(1430)$ & 1424$\pm$4 (T) & 132$\pm$4 (T) \\
\midrule
$2^+$& 1 & $a_2(1700)$ & 1630–1780 (T) & 120-500 (T) \\
$2^+$& 0 & $f_2(1565)$ & 1495-1560 (T) & 80-220 (T) \\
$2^+$& 0 & $f_2(1430)$ $(^*)$ & 1440$\pm$14  $(^\dagger)$ & 46$\pm20$ $(^\dagger)$ \\
$2^+$& 0 & $f_2(1640)$ $(^*)$ & 1639$\pm$6 & 100$^{+60}_{-40}$ \\
$2^+$& 0 & $f_2(1810)$ $(^*)$ & 1815$\pm$12 &197$\pm$22 \\
$2^+$& 0 & $f_2(1910)$ $(^*)$ & 1925$\pm$34 $(^\ddagger)$ & 144$\pm$44 $(^\ddagger)$ \\
$2^+$& 0 & $f_2(1950)$ & 1830–2020 (T) & 220-440 (T) \\
$2^+$& 1/2 & $K^*_2(1980)$ & 1990$^{+60}_{-50}$ & 348$^{+50}_{-30}$ \\
\midrule
$2^+$& 0 & $f_2(2010)$ & 2010$^{+60}_{-80}$ & 200$\pm$60 \\
$2^+$& 0 & $f_2(2150)$ & 2157$\pm$12 & 152$\pm$30 \\
$2^+$& 0 & $f_2(2300)$ & 2297$\pm$28 & 150$\pm$40\\
$2^+$& 0 & $f_2(2340)$ & 2346$^{+21}_{-10}$ & 331$^{+27}_{-18}$\\
\bottomrule
\end{tabular}
%--------------------------------------------------
\qquad
\begin{tabular}{llllll}
\toprule
$J^P$& $I$ & $C$ &Name & $M$ [MeV] & $\Gamma$ [MeV] \\
\midrule
$2^-$& 1 &+ & $\pi_2(1670)$ &1670.6$^{+2.9}_{-1.2}$ & 258$^{+8}_{-9}$ \\
$2^-$&0 &+ & $\eta_2(1645)$ & 1617$\pm$5 & 181$\pm$11 \\
$2^-$&0 & +& $\eta_2(1870)$ & 1842$\pm$8 & 225$\pm$14 \\
$2^-$&1/2 & & $K_2(1770)$ & 1773$\pm$8 & 186$\pm$14 \\
\midrule
$2^-$& 1&+ & $\pi_2(1880)$ & 1874.6$^{+26}_{-5}$ & 237$^{+33}_{-30}$ \\
$2^-$& 1/2& & $K_2(1820)$ & 1819$\pm$12 & 264$\pm$34 \\
\midrule
$2^-$&1 &+& $\pi_2(2005)$ $(^*)$ & 1963$^{+17}_{-27}$ & 370$^{+16}_{-90}$ \\
$2^-$&1 &+& $\pi_2(2100)$ $(^*) $ & 2090$\pm$29 & 620$\pm$50 \\
$2^-$&1/2 & & $K_2(2250)$ $(^*)$   & 2247$\pm$17 & 180$\pm$30 \\
\midrule
& & & & \\
& & & & \\
& & & & \\
& & & & \\
& & & & \\
& & & & \\[9pt]
\end{tabular}
\end{table}

\paragraph{Two quarkonia-like nonets}
There are enough observed states to form two nonets. 
In the QM, the lightest would correspond to $1^3P_2$. As quarkonia, the masses are expected to be similar to the other first radial excitations of the
$L=1$ multiplets. By comparing with the second scalar nonet and the two lowest axial nonets (either with $C=1$ or $C=-1$), we see that this is indeed the case.
The masses within the nonet follow the expected $q \bar q$ hierarchy, with a close to ideal-mixing scenario. The masses within the nonet also follow the expected $q\bar q$
 hierarchy. In the quarkonia scenario, using a chiral Lagrangian model for spin-2 mesons to explain the pattern of decays, the mixing angle is found to be $\theta=30.10^\circ$ \cite{Jafarzade:2022uqo}, once again very close to ideal mixing.  The $f_2'(1525)$ is therefore close to $s\bar s$, explaining why it decays $\sim 90\%$ of the time to $K\bar K$. However, 
an interpretation in terms of vector-meson molecules
has also been proposed \cite{Molina:2008jw,Geng:2008gx,MartinezTorres:2020hus}.

The $f_2(1270)$ state, whose width is $\sim$85\% due to $\pi\pi$ decay and gives rise to a beautiful BW-like shape in $\pi\pi$ scattering, plays a prominent role in Regge Theory. As we will see in the section dedicated to Regge trajectories, its trajectory has the largest intercept. Therefore, it is the subleading trajectory in hadron-hadron scattering only after the Pomeron, which is why it is often called the $P'$ trajectory.

Concerning the second nonet, it can be naturally interpreted as the next radial excitation; indeed, it appears around 400-500 MeV higher. It also follows the expected $q \bar q$ mass hierarchy. The RPP, in the note on ``Light scalar resonances", assigns the $f_2(1950)$ to the second nonet, the other isoscalar being the $f_2(1640)$,
despite it is omitted from the summary tables, However, since there are three other $f_2$ nearby states, also omitted from the summary tables, we refrain from identifying the other isosinglet of this nonet. Presumably, the isoscalars are mixtures of singlet and octet states as well as a glueball singlet, as we discuss next.

\paragraph{Overpopulation of $f_2$ resonances and hints of a tensor glueball.}
As we have already discussed, the lightest tensor glueball is expected to have the $f_2$ quantum numbers at around 2.3 GeV. The situation in this respect is far from clear.

First of all, apart from the well-identified ground states $f_2(1270)$ and $f'_2(1525)$, there are ten additional $f_2$ states below 2.4 GeV reported by the RPP, although four of them are omitted from the summary tables. There is an additional $f_J(2200)$, whose parameters we list in Sec.\ref{sec:other}, that could have $J=2$ or $4$.
Presumably, some of them are the same resonance.
For instance, it has been suggested \cite{Klempt:2007cp} that the $f_2(1640)$ and $f_2(1565)$ may be the same state, and the $f_2(1810)$ lies suspiciously close to the well-established $f_2(1950)$ (although it is claimed to differ in its decays). It has even been suggested that the $f_2(1565)$ mostly observed in $p\bar p$ annihilation could be baryonium (see the RPP note on ``Spectroscopy of light meson resonances"). 
Anyhow, two isosinglets belong to the second nonet, even if it is unclear which ones they are. In addition, the quarkonium $1^3F_2$ nonet and the next radial excitation $3^2P_2$ are roughly expected in that region, which could
account for four other states. There are still many states left to explain.

Second, as was the case with the scalar glueball, the tensor glueball is most likely distributed among the observed $f_2$ states, mixed with other internal configurations.

In this respect, using chiral lagrangians with vector and spin-2 degrees of freedom, it has been recently argued \cite{Vereijken:2023jor} that the $f_2(1950)$ is a good candidate to be predominantly glueball. However, there is marginal evidence for the $f_2(1640)$, $f_2(1910)$ and $f_2(1950)$ in radiative $J/\psi$ decays. In contrast, the $f_2(2010)$, $f_2(2300)$ and $f_2(2340)$ have been observed by BESIII \cite{BESIII:2016qzq} in $J/\psi\to\gamma\phi\phi$ and $J/\psi\to\gamma\eta\eta$. Assuming a two-step decay  $J/\psi\to\gamma f_2(2340)\to J/\psi\to\gamma MM$, BESIII claims that the first step $J/\psi\to\gamma f_2(2340)$ is compatible with the quenched lattice-QCD result for the tensor glueball \cite{Yang:2013xba}. Three states above 2 GeV had also been observed before in other gluon-rich processes like $\pi^- p\to \phi\phi n$
\cite{Etkin:1987rj},  $p\bar p$ annihilation \cite{Booth:1985kr} and $pp$ central collisions \cite{WA102:1998nzq}. All these pieces of evidence support the presence of at least one $2^{++}$ glueball state, although distributed among several $f_2$ states near 1.9-2.3 GeV. A recent analysis \cite{Klempt:2022qjf} of data on the $J/\psi\to\gamma f_2$ decay versus $f_2$ decays to $\pi\pi$ or $\bar KK$, reveals the existence of a 180 MeV-wide enhancement around 2210 MeV. This enhancement could be interpreted as the genuine tensor glueball, i.e., `` before mixing", which is distributed among the nearby physical $f_2$ resonances, mixed with other $q \bar q$ components, much as it happened in the scalar case.

In summary, there are enough well-established states to identify two $2^{++}$ nonets. The second might well be a radial excitation of the first. Besides the four $f_2$ resonances that should be assigned to those two nonets, there are another eight, possibly nine, reported $f_2$ isoscalars, four of them omitted from the RPP summary tables. It is very plausible that those in the 1.9-2.4 GeV region might have large glueball components, and indeed, a few of them appear in gluon-rich processes. Nevertheless, the situation is far from settled. The discovery of additional partners, additional high-statistics data, and more elaborate studies would be welcome.

\subsection{$J^P=2^{-}$: Pseudo-tensor mesons}

In the right sub-table of Table~\ref{tab:2+mesons}, we provide the list of the observed mesons with $J^{P}=2^-$ and their parameters. In the right panel of Fig.\ref{fig:J2}, we show their positions in the $(M,\Gamma/2)$ plane, whose uncertainties are represented by light-colored areas as well as their tentative classification into nonets. 

There are enough states to form a nonet. Note that the observed isotriplet and the isoscalars have a $C=+1$ value. 
In the QM, a $q\bar q$ nonet with $C=-1$ is also allowed (see Table~\ref{tab:ordmesonnames}), with expected masses comparable to those with $C=+1$. However, no isotriplets or isoscalars with $C=-1$ are reported in the RPP listings.

In the quarkonia scenario, the $2^{-+}$ nonet corresponds to the $1^1D_2$ configuration. The states that have been reported appear in the expected mass region and follow a fair QM mass hierarchy. Surprisingly, using a chiral Lagrangian model for spin-2 mesons, the mixing angle is found to be $\theta=-6.7^\circ$ \cite{Koenigstein:2016tjw}, very far from ideal mixing.
In contrast, the lattice-QCD value $33^\circ$ \cite{Dudek:2013yja} is very close to ideal mixing, although it has been calculated with a range of unphysical quark masses, yielding pion masses of $\sim$400 MeV or larger.

Note that the RPP lists two very close strange resonances, $K_2(1770)$ and $K_2(1820)$. This decision was justified in their 2004 edition, claiming that, at the time, two different experiments \cite{Aston:1993qc,ACCMOR:1981yww} agreed on the two masses for their two-state analysis but not in the masses for their one-state analysis. The two-resonance solution was later confirmed by LHCb \cite{LHCb:2016axx} and, more recently, by COMPASS \cite{COMPASS:2025wkw}. We have assigned 
the $K_2(1770)$ to the $2^{-+}$ nonet following the 
RPP note on ``Spectroscopy of light meson resonances", where they assign the $K_2(1820)$ to the $2^{--}$. However, if these two nonets exist, the most likely situation is that these two $K_2$ nearby states result from the mixing of two $K_{2A}$ and $K_{2B}$ states, each belonging to the $2^{-+}$ and a $2^{--}$ octet. This is similar to the mixing mechanism we described for the $K_2(1770)$ and $K_2(1820)$ in the $1^{--}$ and $1^{-+}$
nonets.

In addition, there is another well-established isotriplet $\pi_2(1880)$, a bit light for a radial excitation. Since neither the $\eta_2(1645)$ nor the $\eta(1870)$ decays measured at Crystal Barrel \cite{CrystalBarrel:1996bnu}
corresponded to $s\bar s$ states, the $\eta_2(1870)$ might be the partner of the $\pi_2(1880)$. In such a case, another $\eta_2$ would be missing for the lightest nonet. Moreover, the $\pi_2(1880)$ is somewhat narrow for a radial excitation, and it has been proposed as a non-exotic hybrid candidate \cite{Klempt:2007cp}, together with the $\eta_2(1870)$.

Above 2 GeV, there are three more resonances, which are omitted from the summary tables: two more $2^{-+}$ isotriplets $\pi_2(2005)$ and $\pi_2(2100)$, which look wide enough for radial excitations, and the strange $K_2(2250)$. 
Given the situation, and unless more partners are found and their decays analyzed in detail, it makes no sense to try to identify further $2^-$ multiplets.

\subsection{$J^{P}=3^{-}$}

\begin{figure}
	\centering
    \includegraphics[width=0.495\textwidth]{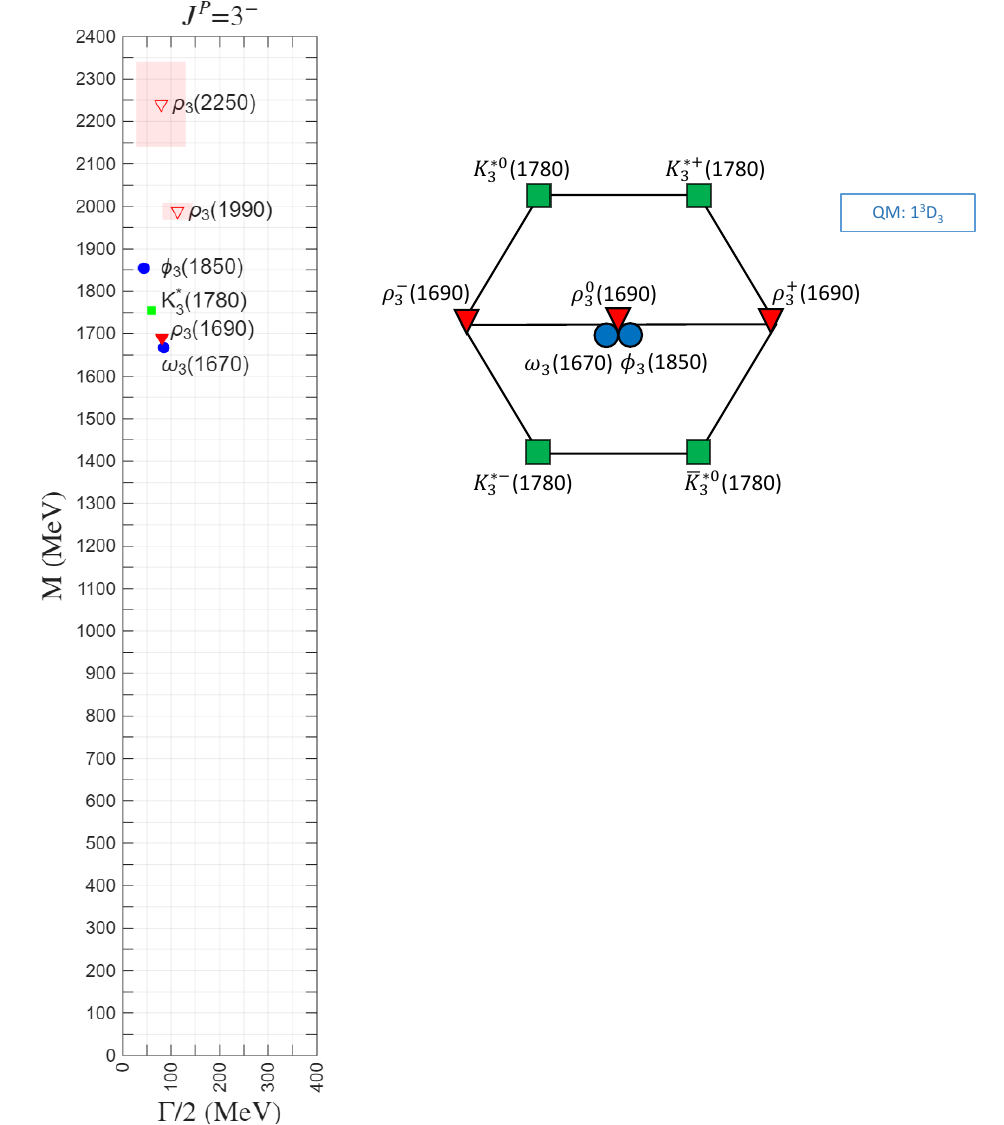}
    \includegraphics[width=0.495\textwidth]{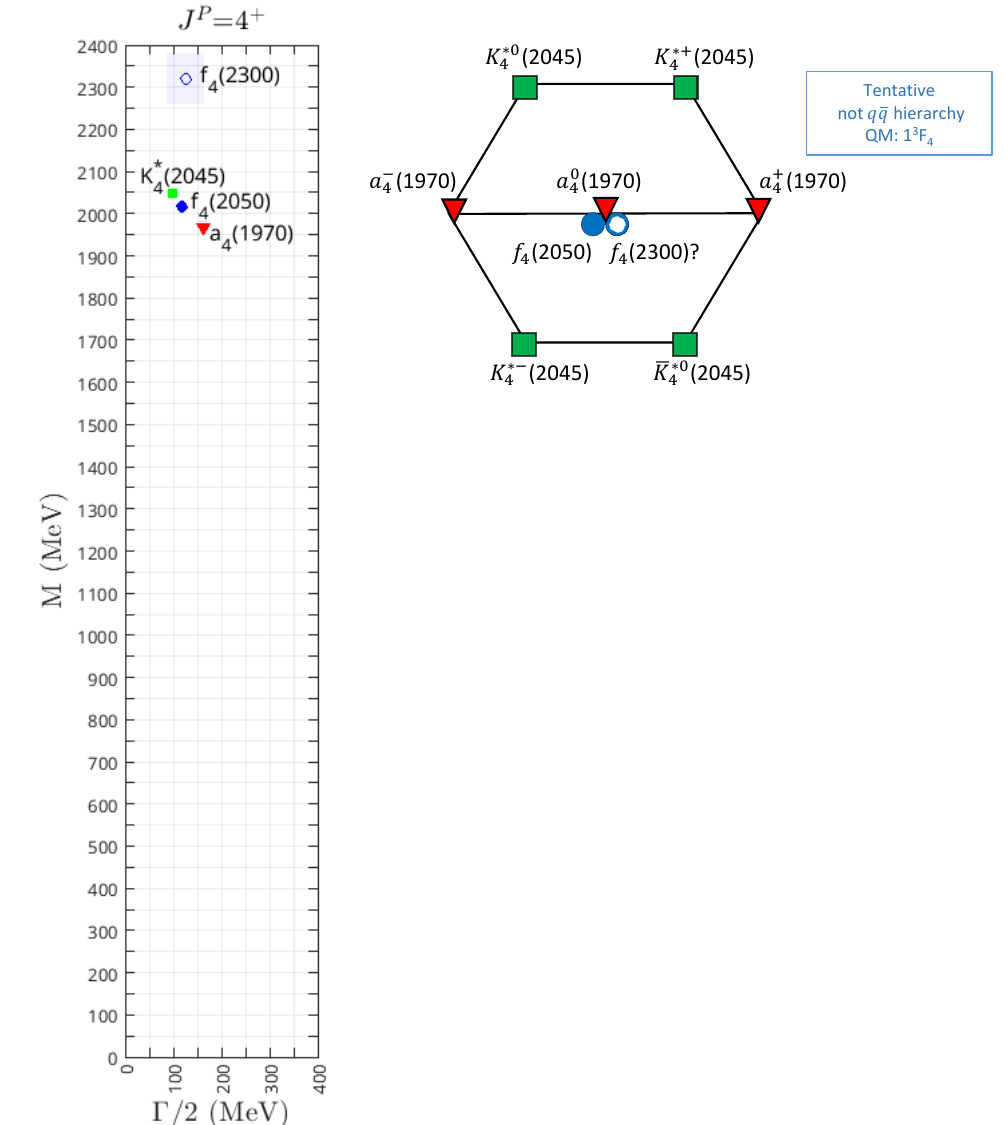}
	\caption{ Tensor $J^{PC}=3^{--}$ (Left) and $J^{PC}=4^{++}$ (Right) in the RPP particle listings and their tentative nonet assignments, together with their position in the $(M,\gamma/2)$ plane, where their estimated uncertainties are represented as light-colored areas. 
    Hollow symbols stand for resonances omitted from the RPP summary tables.
    We are not plotting another resonance, listed as $f_J(2200)$, but omitted from the RPP summary tables, whose $J^{PC}$ quantum numbers, $2^{++}$ or $4^{++}$, remain undetermined.}
	\label{fig:J34}
\end{figure}

The resonances reported in the RPP with $J^P=3^-$ are listed in the left sub-table of Table~\ref{tab:3-mesons} and their tentative nonet assignment is shown in the left panel of Fig.\ref{fig:J34}, together with their position in the $(M,\Gamma/2)$ plane, whose uncertainty is shown as light-colored areas.

\begin{table} %[b]
\caption{ Isospin $I$, mass $M$, and decay width of tensor mesons with $\Gamma$ of $J^P=3^-$ listed in the RPP (Left) and $\Gamma$ of $J^P=4^+$ (Right), and their possible plausible nonet assignment.  The RPP omits the resonances with an asterisk $(^*)$ from summary tables and does not provide an estimate or average for their parameters; the ranges we provide for the mass and width cover the values reported by the RPP in their Particle Listings. \label{tab:3-mesons}}
\centering
\medskip
\begin{tabular}{lllll}
\toprule
$J^{P}$& $I$ & Name & $M$ [MeV] & $\Gamma$ [MeV] \\
\midrule
$3^-$&1 & $\rho_3(1690)$ & 1688.8$\pm$2.1 & 161$\pm$10 \\
$3^-$&0 & $\omega_3(1670)$ & 1677$\pm$4 & 168$\pm$10 \\
$3^-$&0 & $\phi_3(1850)$ & 1854$\pm$7 & 87$^{+28}_{-23}$ \\
$3^-$&1/2 & $K^*_3(1780)$ & 1754$\pm$13 & 238$\pm$28 \\
\midrule
$3^-$&1 & $\rho_3(1990)$ $(^*)$& 1968-2007 & 164-287 \\
$3^-$&1 & $\rho_3(2250)$ $(^*)$& 2140-2340 & 56-260 \\
\bottomrule
\end{tabular}
%\end{table}
%\begin{table} \caption{ Isospin $I$, mass $M$ and decay width $\Gamma$ of $J^P=4^+$ tensor mesons, which form a plausible nonet assignment.\label{tab:4+mesons}}
%\centering
%\medskip
\qquad
\qquad
\begin{tabular}{lllll}
\toprule
$J^{P}$& $I$ & Name & $M$ [MeV] & $\Gamma$ [MeV] \\
\midrule
$4^+$&1 & $a_4(1970)$ & 1967$\pm$16 & 324$^{+15}_{-18}$ \\
$4^+$&0 & $f_4(2050)$ & 2018$\pm$11 & 237$\pm$18 \\
$4^+$&0 & $f_4(2300)$ $(^*)$ & 2266-2385 & 270-335 \\
$4^+$&1/2 & $K^*_4(2045)$ & 2048$^{+8}_{-9}$ & 199$^{+27}_{-19}$ \\
\midrule
&&&&\\
&&&&\\[3pt]
\end{tabular}
\end{table}

Only $3^-$ isotriplets and isoscalars with $C=-1$ are reported in the RPP. 
There are enough well-established resonances observed around 1.7-1.8 GeV to identify them as a $3^{--}$ nonet, which in the QM $q\bar q$ assignment corresponds to a $1^3D_3$ configuration. They appear roughly in the expected mass range, i.e., about 400 MeV higher than the $1^3P_2$ nonet.
In addition, they follow the quarkonia mass hierarchy. 
Their $q \bar q$ nature is confirmed by studying their strong and radiative decays employing a chiral lagrangian for spin-2 and spin-3 mesons, showing an overall agreement with data \cite{Jafarzade:2021vhh}. The mixing angle is found to be $\theta=31.8^\circ$ \cite{Jafarzade:2021vhh}, once again close to the ideal mixing value. Still, a vector-vector molecular interpretation has been proposed too \cite{Roca:2010tf,Yamagata-Sekihara:2010muv}.

The RPP also lists two more isotriplets, the $\rho_3(1990)$ and $\rho_3(2250)$, omitted from their summary tables. In the quarkonia interpretation, the lighter one could be a radial excitation $2^3D_2$ of the $\rho_3(1690)$,
whereas the $\rho(2250)$ may belong to a nonet with $L=4$, which we have not listed in Table~\ref{tab:ordmesonnames}. This corresponds to the first $L=4$ state we discuss, with $S=1$, i.e., $1^3G_3$. It is not far from the $\rho_5(2350)$, which is another candidate for $L=4$, to be discussed below.

\subsection{$J^{P}=4^{+}$}

We list, in the right sub-table of Table~\ref{tab:3-mesons}, the $4^+$ states reported in the RPP, and we show their very tentative nonet assignment in Fig.\ref{fig:J34} as well as their positions in the $(M,\Gamma/2)$ plane, whose uncertainties are represented by light-colored areas. 

The $4^+$ isotriplets and isosinglets listed in the RPP all have $C=+1$ and there are enough states to form a $J^{PC}=4^{++}$ nonet. 
If these were quarkonia, they would correspond to $1^3F_4$.
However, their masses do not follow the expected $q\bar q$ hierarchy very well. In particular, 
the isotriplet is lighter than the two isodoublets and only 70-80 MeV lighter 
than the strange members of the multiplet.
A vector-vector molecular interpretation has been proposed in \cite{Roca:2010tf,Yamagata-Sekihara:2010muv}.
In addition, following the RPP, we have assigned to the nonet the $f_4(2300)$, even though it is omitted from the RPP summary tables and is much heavier than the other tentative members of the nonet.
Finally, in the RPP listings, but omitted from the summary tables, there is another resonance with positive parity, the $f_J(2200)$, whose spin remains undetermined, but it could be 2 or 4.
If it is the latter, it could also be a candidate for the heavier isoscalar in the $4^{--}$ nonet. 
In summary, the identification of the isoscalar members of the $4^{--}$ nonet should be considered very tentative.

\subsection{Other unflavored or strange resonances above 2 GeV}
\label{sec:other}

For the sake of completeness, we provide in Table~\ref{tab:above2} a few more states above 2 GeV that appear in the RPP particle listings for light unflavored or strange mesons, but are omitted from the summary tables. All their quantum numbers are non-exotic. In the left sub-table, we list the isoscalars and isotriplets, also indicating their $C$-parity. In the right sub-table, we list the strange ones.
Note that in the QM quarkonia scenario, some of them require $L=4,5$.  Once again, a vector-vector molecular interpretation has been proposed for some of them \cite{Roca:2010tf,Yamagata-Sekihara:2010muv}.

It is tempting to pair the $\rho_5(2350)$ with the $K_5^*(2380)$ in the same $J^{PC}=5^{--}$; however, we refrain from doing so because their masses are too close.

In a very recent study of $K$-meson spectroscopy, the COMPASS Collaboration finds \cite{COMPASS:2025wkw}  the
$K_3$ and $K_4$ masses around 200 MeV lighter than in Table~\ref{tab:above2}. They suggest that the resonances they observe may correspond to the $J=3^+$ and $4^-$ QM ground states, whereas those listed in the RPP would correspond to the first excitations.

%\textcolor{red}{Combined analyses \cite{Anisovich:2002su} of Anisovich of Crystal Barrel, ver entradas de eho3(1990), aunque no veo la rho3(1990) en ningún sitio}

Finally, let us recall that in this introductory review we have limited the discussion to light meson resonances that appear in the RPP particle listings in the ``Light Unflavored Mesons $(\mathsf{S=C=B=0})$" or the ``Strange Mesons $(\mathsf{S=\pm1, C=B=}0)$" sections. We consider them well-established if they also appear in the RPP's "Summary tables." 
Of course, there have been many more proposals in the literature, which lie beyond our limited scope. We leave the interested reader to learn about such proposals in the RPP particle listings section on ``Other Mesons", which contains a subsection on ``Further States", and follow the references therein. According to the RPP authors, those further states are  {\emph ``observed by a single group,  or states poorly established that thus need confirmation"}. Eventually, some of them will be observed by another group and confirmed, or discarded. 

\begin{table}
\caption{ Other light-unflavored or strange mesons with quantum numbers identified and referenced in the particle listings of the RPP, but omitted from their summary tables. For some parameters of these resonances, the RPP does not provide averages or estimates.
\label{tab:above2}}
\centering
\medskip
\begin{tabular}{lllll}
\toprule
$J^{PC}$&$I$ & Name & $M$ [MeV] & $\Gamma$ [MeV] \\
\midrule 
$2^{++}$ or $4^{++}$ &0 & $f_J(2200)$  & 2231.1$\pm$3.5 & 23$^{+8}_{-7}$ \\
\midrule
$6^{++}$&0 & $f_6(2510)$ & 2470$\pm$50 & 260$\pm$40 \\
\midrule
$5^{--}$&1 & $\rho_5(2350)$ & - & - \\
\bottomrule
&  &    &  &   \\[5pt]
\end{tabular}
\qquad
\begin{tabular}{lllll}
\toprule
$J^P$&$I$ & Name & $M$ [MeV] & $\Gamma$ [MeV] \\
\midrule 
$3^+$&1/2 & $K_3(2320)$ & 2324$\pm$24 & 150$\pm$30 \\
\midrule
$5^-$&1/2 & $K^*_5(2380)$ & 2382$\pm$33 & 178$\pm$69 \\
\midrule
$4^-$&1/2 & $K_4(2500)$ & 2490$\pm$20 & - \\
\midrule
$?^?$&1/2 & $K(3100)$ & $\sim3100$ & - \\
\bottomrule
\end{tabular}
\end{table}

\subsection{Regge trajectories.}
\label{sec:Regge}

We have already seen the usefulness of promoting energy from a real to a complex variable when relating resonances to their associated poles in the complex plane, which we used to define the pole mass and width as $s_p=(M-i\Gamma/2)^2$. Analyticity in complex variables also provides strong constraints on amplitudes in the form of dispersion relations. 
In 1959, Regge \cite{Regge:1959mz} went one step forward and introduced the study of the analytic properties of scattering amplitudes or the $S$-matrix in terms of complex angular momenta. The trajectories followed by the position $\alpha_i(s_p)$ of resonance poles in the complex $J$ plane, or Regge poles, are called Regge trajectories. When $\alpha_i$ becomes an integer $J$ (half-integer), this Regge pole corresponds to a meson (baryon) resonance of spin $J$. For textbook introductions, see \cite{Collins:1977jy,Gribov:2003nw}.

Motivated by the properties of Regge poles, in 1962, Chew and Frautschi \cite{Chew:1962eu} suggested, with a Regge-potential scattering formula, the universality of the slope $\alpha'=d\alpha(s)/ds|_{s=0}\sim1/(50m_\pi^2)\sim 1\,$GeV$^{-2}$ in hadron trajectories.  Indeed, neglecting the width of resonances, i.e., $\alpha_i(s)\sim {\rm Re}\,\alpha_i(M^2)$, they looked for this universal behavior by plotting the positions of the known resonances in the $(M^2,J)$ plane. 
From that moment on, this kind of plot became a ``Chew-Frautschi" plot. 
Abusing the  ${\rm Re}\,\alpha_i(M^2)=J$ identification, which only holds for integer or half-integer $J$, it is then customary to plot $J$ versus $M^2$ and also call ``Regge trajectory" the resulting line.
Back in 1962, Chew and Frautschi found tentative evidence of this universal behavior. However, the limited information at the time did not allow them to infer `` a strict linear relation of the trajectories". 
Nevertheless, over time, it has been found that a linear behavior such as
\begin{equation}
    {\rm Re}\,\alpha_i(M^2)\simeq \alpha_i(0)+\alpha'\, M^2,
\end{equation}
with a universal slope $\alpha'\simeq 0.8\,$GeV$^{-2}$, is a fairly good approximation for many mesons and baryons, with only a few exceptions.
These linear Regge trajectories were later understood in terms of a confining mechanism, where the quark and antiquark were bound by a confining field contained within a flux tube, resulting in relativistic, string-like dynamics. The universal slope would correspond to a universal string tension, independent of the masses of the constituents. In the case of QCD, this string picture is just a toy model, although it lies at the origin of String Theory, whose discussion extends far beyond our scope.

 In practice, each tower of states in a linear trajectory is usually called a ``Reggeon", characterized by the common quantum numbers like charge, isospin, flavor, etc. The value of $\alpha_i(0)$ is called the intercept, and, due to analyticity and crossing symmetry, it is very relevant for {\it high energy} physics. In that limit, the crossed exchange of the tower of states in a given trajectory $i$ gives rise to a contribution that behaves as $s^{\alpha_i(0)}$ in the corresponding total cross section. 
For this reason, the trajectories with the largest intercept are called ``leading trajectories", whereas the other trajectories with similar quantum numbers but smaller intercept are called ``daughter trajectories". Note that unitarity imposes $\alpha_i(0)\leq1$. As a matter of fact, there is another Regge exchange called the Pomeron or $P$, which behaves as $\sim s\log^2(s)$, which dominates the high-energy cross section when it is present. However, its slope does not correspond to the same mechanism of quarkonia trajectories, and, together with the so-called odderon trajectory, lies beyond the scope of this mini-review. We refer the reader again to the textbooks \cite{Collins:1977jy,Gribov:2003nw}.

\begin{figure}
%\centering
   \includegraphics[width=\textwidth]{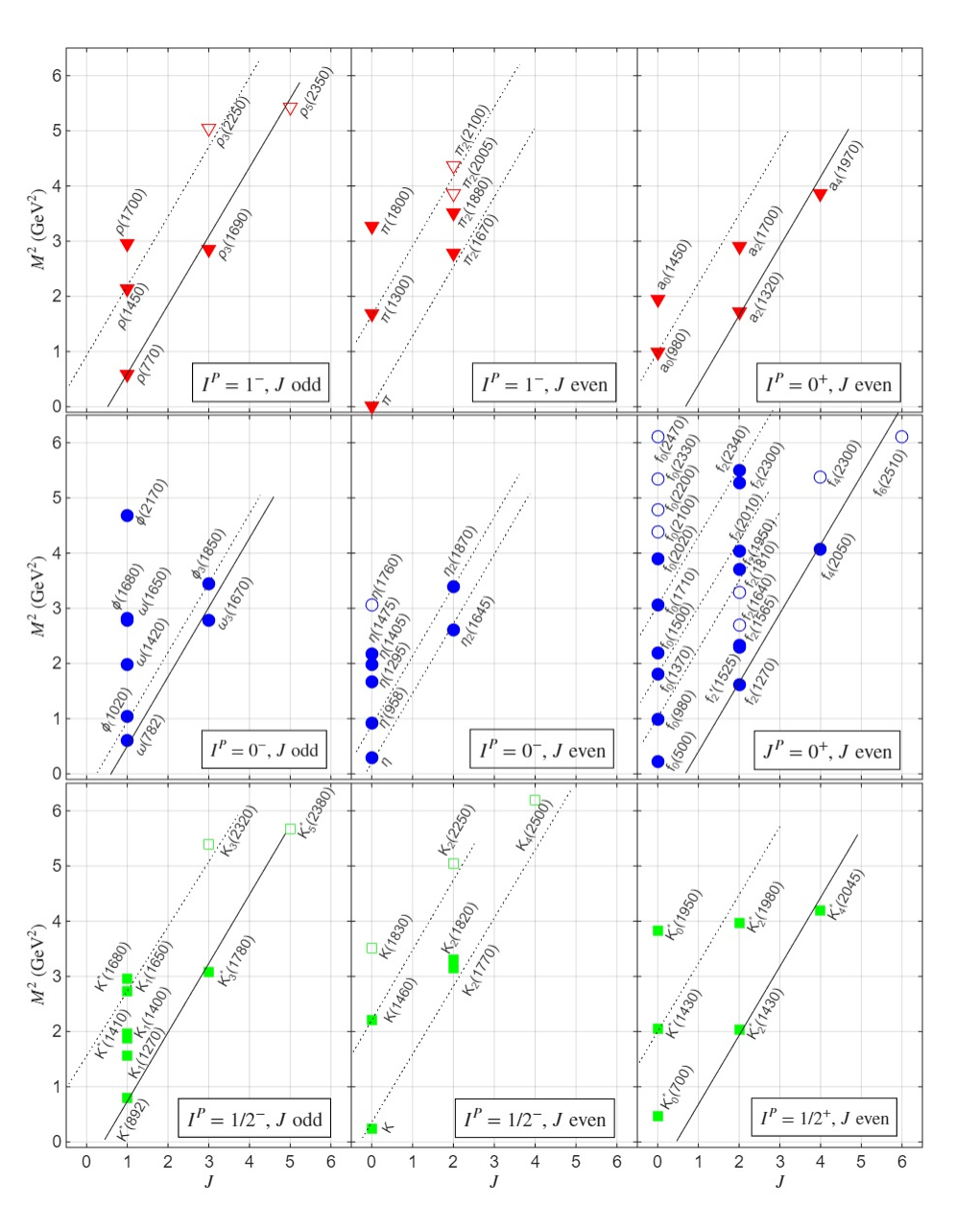}
	\caption{Examples of meson Regge trajectories: Solid symbols indicate the position in the $(J, M^2)$ plane of the light resonances that appear both in the RPP particle listings and summary tables, whereas hollow symbols stand for some resonances omitted from the summary tables. We also illustrate ideal linear Regge trajectories with a universal slope $\alpha'=0.8\,$GeV$^{-2}$. These are not fits, but still allow for the clear identification of the leading trajectories with intercept $\alpha_M(0)\gtrsim0.5$ (solid lines) as well as some daughter trajectories with smaller intercept close to zero (dotted lines). In the upper left panel, we only show the $C=-1$ states.
    }
	\label{fig:Regge}
\end{figure}

In Fig. \ref{fig:Regge}, we show the present Chew-Frautschi plots of the light-meson resonances appearing in the summary tables of the RPP, together with some additional states omitted from the summary tables but present in the particle listings. 
Note that, for technical reasons beyond the scope of this review, only mesons of either even or odd spin can belong to the same $(J, M^2)$ trajectory.
We have also plotted several guiding lines with a universal slope of $\sim 0.8\,$GeV$^2$, corresponding to the leading trajectories (continuous lines) and a few daughter trajectories. 
The leading trajectories with $\alpha_i(0)\gtrsim 0.5 $ correspond to the $f_2(1270)$, $\rho(770)$, $\omega(782)$ and $a_2(1320)$, ordered from largest to smaller intercept. The $f_2(1270)$ trajectory is the leading Reggeon, whose contribution is largest at high energies after the Pomeron. This is why it is often referred to as the $P'$ trajectory. Fig.~\ref{fig:Regge} illustrates that the confinement scenario where the color field is concentrated on a flux-tube or string-like configuration, with a $\sim1\,$GeV tension, i.e., a quark-level scale, is qualitatively correct for most light mesons. Most, but not all of them.

Indeed, a thorough analysis of the systematics of QM $q\bar q$ states, not only for $(J, M^2)$ Regge trajectories but also for radial $(n, M^2)$ trajectories, was carried out in 2000 in \cite{Anisovich:2000kxa}. The latter kind of trajectories do not connect direct observables but numbers interpreted through the QM, and we refer the reader to specialized reviews.
Linear trajectories fitted most of the resonances fairly well
with a universal slope $\alpha'\simeq0.8\pm0.1\,$GeV$^{-2}$. 
It is worth noticing that in \cite{Anisovich:2000kxa} the $\sigma/f_0(500)$-meson was not included
``supposing $\sigma$ is alien to this classification".
A relativistic QM calculation \cite{Ebert:2009ub} of Regge trajectories, including strange light resonances, also requires the lightest $q\bar q$ scalar mesons in linear trajectories to lie above 1 GeV. 
The difficulty in assigning partners to the $f_0(500)$ and $K^*_0(700)$ in a linear Regge trajectory is illustrated in Fig.\ref{fig:Regge}, particularly in the case of the latter.

Moreover, one should recall that Regge trajectories describe the dependence of the complex pole position $\alpha_i$ in the complex plane, not just its real part. The complex part of $\alpha_i$ is related to the width of the state (and should not be considered the uncertainty in the mass, as is very often incorrectly done), and should not be neglected for resonances as wide as the $f_0(500)$ or $K_0^*(700)$. Actually, using dispersion relations obtained from the complex angular momentum strong analyticity constraints, it is possible to calculate, not fit, the Regge trajectory of resonances that decay predominantly into one channel \cite{Londergan:2013dza,Pelaez:2017sit,Carrasco:2015fva}. The resulting trajectories are very approximately real and linear for well-established ordinary resonances like the $\rho(770)$, $f_2(1270)$, and $K^*(892)$ or even for the $K_0^*(1430)$. However, those of the $\sigma/f_0(500)$ and the $K^*_0(700)$ have a significant imaginary part, which is why we do not even attempt to draw their trajectories in Fig.\ref{fig:Regge}, and their real part is highly non-linear. At low masses, they both exhibit trajectories similar to those of a Yukawa potential, with slopes much smaller than the GeV, once again indicating that meson-meson interactions are responsible for the dynamics of their formation.  All in all, these non-linear trajectories constitute another piece of evidence for the non-ordinary nature of the lightest scalars and the relevance of meson-scale dynamics in their formation.

Note that, despite having $I^P=1^-$ and $J=1$, we have not included the hybrid candidate $\pi_1(1600)$ in the upper left Fig.\ref{fig:Regge}, which contains only $C=-1$ states.

%%%%%%%%%%%%%%%%%%%%%%%%%%%%%%%%%%%%%%%%%
%% Mandatory: A concluding paragraph summing up your main points in the chapter
%% Optional: Also include big questions in the field that are still to be answered. What topics/methods/questions are researchers like to focus on next?
\section{Conclusions}
\label{sec:conclusions}

In this introductory review of light meson resonances, we have first introduced the basic concepts in a pedagogical manner and then listed and classified the reasonably well-established states. 

We have seen that mesons are hadrons, i.e., states made of quarks, $q$, antiquarks, $\bar q$, and gluons, $g$, whose internal dynamics are governed by the fundamental strong force, described by Quantum Chromodynamics (QCD).  Quarks and gluons carry a color charge, but QCD confines it within colorless bound states, i.e., hadrons. Mesons are hadrons with integer spin, and are often called meson resonances because they are unstable under strong interactions. Most of them disintegrate with extremely small lifetimes into pions and kaons---which normally are the only mesons to reach the detectors---or other, more stable particles.
In practice, resonances are detected indirectly as peaks, or more subtle deformations, in the measured distributions of other particles. These peaks or distorted structures 
have a width that is larger for the more unstable resonances.
Being so unstable, meson peaks and deformations often overlap with one another, making the identification of mesons and determining their properties a very challenging task. For this reason, instead of just peaks and deformations, it is very convenient to employ a more rigorous and sophisticated mathematical formalism in terms of complex poles associated with resonances whenever they appear in an amplitude describing a process. The pole positions rigorously define the mass and width of the resonance, as well as their residues at different amplitudes that determine their partial decays.

We have thus seen that mesons are bosons, characterized by their integer spin, $J$, parity, $P$, and charge conjugation, or $C$-parity. Besides their gluon content, light mesons are those primarily made of quarks and antiquarks with the lightest flavors; up, down, and strange, abbreviated as $u,d$, and $s$, respectively, which provide the meson isospin, $I$, and strangeness, $\mathsf{S}$.

Strong interactions are almost invariant under isospin symmetry, i.e., complex rotations of $u$ and $d$ quarks. Mesons with the same $J^{PC}$ are then classified into multiplets of isospin $I$, with almost the same properties under the strong force. Similarly, complex ``flavor" rotations of $u,d$, and $s$ leave meson properties virtually the same. This is known as flavor symmetry, which is not as good as isospin symmetry, but also allows us to group isospin multiplets with different strangeness into larger flavor multiplets, again with the same $J^{PC}$, and still fairly similar properties under strong interactions. 

The usual perturbative expansion is not applicable in the hadronic realm for QCD. This is why the Quark Model (QM), which considers that hadrons are just made of the ``valence" constituents needed to provide the quantum numbers, became so popular. In its simplest version, the so-called constituent quarks, antiquarks, and gluons are just effective degrees of freedom that incorporate QCD non-perturbative effects into their $\sim$300~MeV masses, and are subject to simple potentials that bind them together into hadrons. We have discussed that the QM is not QCD, but it is a very useful classification scheme, as it incorporates the relevant QCD symmetries, particularly isospin and flavor symmetry. 

In the QM, ``ordinary" mesons---the most common--- are made of one constituent quark and one antiquark, and are called quarkonia or $q\bar q$. Quarkonia multiplets are nonets composed of an octet, whose members transform among themselves under flavor symmetry, and a singlet, invariant under flavor-symmetry transformations. Their quantum numbers, including $P$ and $C$, are fully determined from the isospin, strangeness, angular momentum $L$, spin $S$, and total angular momentum $J$ of the constituent $q \bar q$ pair, whose configuration is denoted by $n^{2S+1}L_J$, $n$ being the radial number. Let us recall that $n$, $L$, and $S$ are not directly observed, but rather inferred from the observable quantum numbers within the QM quarkonia interpretation. Note that  
not all combinations of $I$, $\mathsf{S}$ and $J^{PC}$ are possible for $q \bar q$ states.

Very unfortunately, in many popular-science descriptions of mesons, they are often identified solely with quarkonia. However, from the very moment the quark model was proposed in the 1960s, it was clearly stated that mesons with different compositions may also exist. Indeed, other compositions are expected with relatively light masses, such as tetraquarks ($q q\bar q\bar q$), meson molecules (two mesons bound by meson-scale interactions), hybrids ( $q\bar q g$), and glueballs (made of constituent gluons). These non-ordinary mesons can have exotic quantum numbers not allowed for quarkonia; otherwise, they are referred to as crypto-exotic.

Light-meson spectroscopy aims to determine the existence of light meson resonances, classify them into multiplets, and determine their nature or composition. These studies should help us understand the underlying QCD dynamics and the still-unknown mechanism of hadron confinement. Thus, in the second part of this review, we listed the known light-meson resonances and 
briefly reviewed their spectroscopic classification, which in some cases is merely tentative. For this purpose, we used the QM as a guideline, paying particular attention to exotic and non-ordinary states. Unfortunately, the identification and interpretation of meson states are hindered by the ubiquitous mixing that occurs between states with the same quantum numbers.

The first surprise is that the lightest nonet, made of perfectly identified pseudoscalar $J^{-+}$ mesons, does not follow the QM $1^1S_0$ quarkonia expectations. This is because the pions, kaons, and etas are understood as Nambu-Goldstone bosons (NGB), which are collective excitations that result from the spontaneous chiral symmetry breaking of QCD, which we also explained very briefly. The spontaneous chiral symmetry breaking is due to the existence of degenerate vacua in QCD, a very non-perturbative effect that, in the limit of massless quarks, would render the NGB massless and separated by a large mass gap $\sim$500 MeV from the rest of the mesons. Since, in real life, quarks have tiny masses, the NGBs also have masses, but are considerably smaller than for the rest of the hadrons, and well below quarkonia expectations. Hence, chiral symmetry breaking deforms and drags down to much lower masses the $1^1S_0$ $q \bar q$ multiplet, whose members behave as NGBs.

The second surprise is that the features of the next-to-lightest meson nonet, composed of $0^{++}$ scalars, are also significantly different from those of quarkonia. Most strikingly, they are much lighter than expected, and the internal mass hierarchy is inverted compared to $q\bar q$ expectations, instead being closer to that of tetraquarks or meson molecules. We have seen how spontaneous chiral symmetry breaking, this time through the effect of NGB interactions at the meson scale, is responsible for the formation of these predominantly non-ordinary mesons. However, it seems likely that they may have a small mixing with heavier quarkonia or tetraquark components, which would have appeared around 1 GeV, as naively expected, were it not for the effect of meson-meson interactions. The lightest states within this light-scalar nonet are extremely wide, decaying very strongly and rapidly into NGB pairs. For this reason, their identification has been controversial for decades, until powerful dispersive mathematical techniques were used to analyze NGB-scattering data. Their understanding has also required the combination of such dispersive techniques and Chiral Perturbation Theory, which is the low-energy effective theory of QCD, which we introduced briefly. One of them, traditionally known as the $\sigma$ meson, now referred to as the $f_0(500)$, plays a very significant role in nucleon-nucleon attraction and the spontaneous chiral symmetry breaking pattern of QCD, and has given rise to several popular toy or simplified models of this breaking.

Apart from these first two surprises, there are still many well-identified light-meson nonets whose 
behavior is fairly close to the quarkonia expectations, making the QM still useful and widely used.
These include the QM ground states $1^3S_1$, $1^1P_1$, $1^3P_2$, $1^1D_2$, $1^3D_1$ and $1^3D_3$, which correspond to $J^{PC}=1^{--},1^{+-},2^{++},2^{-+}, 1^{--}$ again, and $3^{--}$. Still, since QM expectations are semi-quantitative, it is not ruled out that some of these resonances may have sizable or even dominant components of a different nature, compatible with their $J^{PC}$ numbers.

Other nonets are more uncertain; their member identification is still controversial. Even more so, the interpretation of their nature.  The assignment of isoscalar mesons (those with zero charge and isospin) is often tentative. In particular, the several $\eta$ states around the 1.3 to 1.5 GeV region and their assignment to the second $0^{-+}$ ($2^1S_0$) nonet,  the $f_1$ assignment to the $1^{++}$ ($1^3P_1$) nonet, the overpopulation of $f_2$ resonances and their $2^{++}(2^3P_2)$ assignment, as well as the $f_4$ states of the $4^{++}(1^3F_4)$ nonet, are matters of ongoing debate.

Of particular interest is the scalar $0^{++}$ nonet around 1.2 to 1.7 GeV. As quarkonia, it would be the ground $1^3P_0$ scalar state. However, it comes somewhat heavier and does not quite follow the $q\bar q$ expectations, which could be due to its interplay with the lowest non-ordinary
scalar nonet below 1 GeV. However, the most interesting feature is the presence of an additional singlet beyond the nonet, which is interpreted in terms of a glueball state. Of course, it is mixed with the other nearby isoscalars, which are composed of constituent quarks. While immediately after its discovery, the $f_0(1500)$ was considered a candidate with the largest glueball component, the tendency over the last years has been to consider that it is the $f_0(1710)$ or even the recently reported $f_0(1770)$.
Actually, with the recently discovered $a_0(1710)$ and further isoscalar $f_0$ tensors, the identification of the members of a possible third nonet is very speculative. Nevertheless, we have briefly described the strong indications that glueball components are needed to explain the decay pattern of several of these $f_0$ resonances. The mass of this glueball state before mixing would be around 1.7 to 1.8 GeV, which is fairly consistent with non-perturbative lattice QCD calculations.
Similar analyses on the many isoscalar $f_2$ isoscalar states also suggest the presence of a tensor glueball, presumably mixed with other configurations. 

Most remarkably, a well established state, the $\pi_1(1600)$ has been reported with exotic quantum numbers $J^{PC}=1^{-+}$. This resonance cannot be, nor can it mix with, quarkonia. In contrast, it lies fairly close to the predictions for a ground-state hybrid meson, i.e., a state whose valence constituents are a quark-antiquark pair and a gluon. 
If it is indeed a hybrid, it should have partners to complete a multiplet.
We have also seen that a possible isoscalar partner, the $\eta_1(1885)$, has also been reported, but it is not as firmly established as the $\pi_1(1600)$ yet.

There are many other well-established and still-to-be-confirmed mesons, which, at present, cannot be assigned to multiplets because their properties are not well known or because their partners have not been found yet.

To conclude, considerable progress has been made in our understanding of light mesons, with the Quark Model and the quarkonia interpretation playing a relevant guiding role. However, apart from deepening our knowledge of ordinary quark-antiquark states, we have learned that there are other non-ordinary and even exotic states. Still, several big questions in the field remain to be answered:

On the experimental side:
\begin{itemize}

    \item To determine whether there are two or three pseudoscalar $\eta$ states around 1.3-1.5 GeV, to confirm recent claims for a $K(1640)$ and to clarify their possible glueball or crypto-exotic nature.

    \item The confirmation of states already observed, particularly the  $f_0(1770)$, $a_0(1710)$, $a_0(1950)$, $\eta_1(1885)$, as well as the several $f_2$ resonances around 2 GeV, which are directly involved in the identification of mesons with non-ordinary nature like glueballs and hybrids.

    \item The discovery of additional states to complete the tentative nonet identification, addressing the currently very confusing situation around and above 2 GeV.
    
    \item To identify almost from scratch other nonets with quantum numbers for which none or very few candidates have been observed but are allowed as quarkonia. It is also crucial to search for partners of exotic mesons and investigate their decays.

    \item To improve the search for additional decay modes and obtain precise determinations on decay rates and branching ratios to identify the nature of different resonances and the mixing patterns within multiplets.

\end{itemize}

There is an enormous worldwide experimental effort, with dedicated programs addressing these questions, particularly focused on the search and identification of glueballs and exotics. This effort includes present large facilities like the Gluex \cite{E12-06-102,GlueX:2014hxq,Adhikari:2020cvz} and CLAS12 \cite{E12-11-005} MesonEx photoproduction experiments at the Thomas Jefferson National Laboratory (JLab \cite{Arrington:2021alx}) in the US; the BESIII $e^+e^-$ experiment in Beijing \cite{BESIII:2020nme}, China, and their tens of billions sample of $J/\psi$ decays that give unprecedented large statistics to study the whole light-meson spectrum, and particularly glueballs in the $J/\psi$  radiative decays. As we have seen, other recent experiments, such as COMPASS at CERN, continue to present interesting results, for instance, with studies on light-strange resonances. In addition, there is also a planned $p\bar p$ annihilation PANDA experiment \cite{PANDA:2021ozp} at the GSI, in Germany, also aiming at a wide spectroscopic coverage of parity, spin, gluon, and quark flavors, where glueball searches could be favored by large production rates of $f_0$ states above 1.5 GeV. 

On the phenomenological side, the large amount of existing data and the many more that will be obtained soon will require the use of much more elaborate analysis tools (see, for instance \cite{Battaglieri:2014gca,JPAC:2021rxu}), possibly including deep learning techniques (see, for example \cite{Sombillo:2020ccg,Sombillo:2021rxv,Ng:2021ibr,Malekhosseini:2024eot}). It will become increasingly necessary to abandon oversimplified models, such as simple Breit-Wigner parameterizations, which are suitable for analyzing isolated resonances but are inappropriate for wide resonances, particularly when they overlap with each other or are produced near thresholds or other types of singularity. 

On the more theoretical side, multiple channels must be incorporated into the analysis, and model dependence must be reduced by using various parameterizations or dispersive constraints. Chiral Lagrangians and other tools are essential for systematically imposing QCD symmetries, as well as for relating multiple processes with a few parameters. Although impressive progress has been achieved in lattice-QCD, which is discussed in other chapters of this encyclopedia, the use of dynamical quarks and more realistic quark masses should definitely be explored in the near future to relate spectroscopic observations to the fundamental strong interaction.

\begin{ack}[Acknowledgments]%
I wish to thank F.K. Guo for the invitation to write this review, as well as for his many comments and suggestions. I also appreciate the assistance of F.J. Llanes-Estrada, E. Oset, and A. Rodas, who provided help with references, comments, suggestions, and clarifications of several concepts. Special thanks to I. Leyva for her invaluable help with the figures; without her support, this review would not have been possible. This work is supported by the Spanish Grant PID2022-136510NB-C31 funded by MCIN/AEI/ 10.13039/501100011033 and the European Union’s Horizon 2020 research and innovation program under grant agreement No.\ 824093 (STRONG2020).
\end{ack}

%%%%%%%%%%%%%%%%%%%%%%%%%%%%%%%%%%%%%%%%%%%%
%% Optional: A list of references to other relevant works/articles/websites which are not cited in the text but that would further enhance a reader's understanding of this topic
%\seealso{article title article title}

%%%%%%%%%%%%%%%%%%%%%%%%%%%%%%%%%%%%%%%%%
%% Mandatory: Bibliography using BibTeX 
\bibliographystyle{Numbered-Style} %% for Numbered Reference Style
%\bibliography{reference}

\begin{thebibliography*}{100}
\providecommand{\bibtype}[1]{}
\providecommand{\url}[1]{{\tt #1}}
\providecommand{\urlprefix}{URL }
\expandafter\ifx\csname urlstyle\endcsname\relax
  \providecommand{\doi}[1]{doi:\discretionary{}{}{}#1}\else
  \providecommand{\doi}{doi:\discretionary{}{}{}\begingroup \urlstyle{rm}\Url}\fi
\providecommand{\bibinfo}[2]{#2}
\providecommand{\eprint}[2][]{\url{#2}}
\makeatletter\def\@biblabel#1{\bibinfo{label}{[#1]}}\makeatother

\bibtype{Article}%
\bibitem{ParticleDataGroup:1986kuw}
\bibinfo{author}{M. Aguilar-Benitez}, et al. (\bibinfo{collaboration}{Particle Data Group}), \bibinfo{title}{{Review of Particle Properties. Particle Data Group}}, \bibinfo{journal}{Phys. Lett. B} \bibinfo{volume}{170} (\bibinfo{year}{1986}) \bibinfo{pages}{1--350}.

\bibtype{Article}%
\bibitem{ParticleDataGroup:2024cfk}
\bibinfo{author}{S. Navas}, et al. (\bibinfo{collaboration}{Particle Data Group}), \bibinfo{title}{{Review of particle physics}}, \bibinfo{journal}{Phys. Rev. D} \bibinfo{volume}{110} (\bibinfo{number}{3}) (\bibinfo{year}{2024}) \bibinfo{pages}{030001}, \bibinfo{doi}{\doi{10.1103/PhysRevD.110.030001}}.

\bibtype{Article}%
\bibitem{Guo:2019twa}
\bibinfo{author}{Feng-Kun Guo}, \bibinfo{author}{Xiao-Hai Liu}, \bibinfo{author}{Shuntaro Sakai}, \bibinfo{title}{{Threshold cusps and triangle singularities in hadronic reactions}}, \bibinfo{journal}{Prog. Part. Nucl. Phys.} \bibinfo{volume}{112} (\bibinfo{year}{2020}) \bibinfo{pages}{103757}, \bibinfo{doi}{\doi{10.1016/j.ppnp.2020.103757}}, \eprint{1912.07030}.

\bibtype{Article}%
\bibitem{Mai:2025wjb}
\bibinfo{author}{M. Mai}, \bibinfo{title}{{Theory of resonances}}  (\bibinfo{year}{2025}), \eprint{2502.02654}.

\bibtype{Article}%
\bibitem{Oller:2025leg}
\bibinfo{author}{J.~A. Oller}, \bibinfo{title}{{Coupled-channel formalism}}  (\bibinfo{year}{2025}), \eprint{2501.10000}.

\bibtype{Article}%
\bibitem{Pelaez:2015qba}
\bibinfo{author}{J.~R. Pelaez}, \bibinfo{title}{{From controversy to precision on the sigma meson: a review on the status of the non-ordinary $f_0(500)$ resonance}}, \bibinfo{journal}{Phys. Rept.} \bibinfo{volume}{658} (\bibinfo{year}{2016}) \bibinfo{pages}{1}, \bibinfo{doi}{\doi{10.1016/j.physrep.2016.09.001}}, \eprint{1510.00653}.

\bibtype{Article}%
\bibitem{Oller:2019opk}
\bibinfo{author}{J.~A. Oller}, \bibinfo{title}{{Coupled-channel approach in hadron\textendash{}hadron scattering}}, \bibinfo{journal}{Prog. Part. Nucl. Phys.} \bibinfo{volume}{110} (\bibinfo{year}{2020}) \bibinfo{pages}{103728}, \bibinfo{doi}{\doi{10.1016/j.ppnp.2019.103728}}, \eprint{1909.00370}.

\bibtype{Article}%
\bibitem{Yao:2020bxx}
\bibinfo{author}{D.-L. Yao}, \bibinfo{author}{L.-Y. Dai}, \bibinfo{author}{H.-Q. Zheng}, \bibinfo{author}{Z.-Y. Zhou}, \bibinfo{title}{{A review on partial-wave dynamics with chiral effective field theory and dispersion relation}}, \bibinfo{journal}{Rept. Prog. Phys.} \bibinfo{volume}{84} (\bibinfo{number}{7}) (\bibinfo{year}{2021}) \bibinfo{pages}{076201}, \bibinfo{doi}{\doi{10.1088/1361-6633/abfa6f}}, \eprint{2009.13495}.

\bibtype{Article}%
\bibitem{Pelaez:2021dak}
\bibinfo{author}{J.~R. Pel\'aez}, \bibinfo{author}{A. Rodas}, \bibinfo{author}{J. Ruiz~de Elvira}, \bibinfo{title}{{Precision dispersive approaches versus unitarized chiral perturbation theory for the lightest scalar resonances $\sigma /f_0(500) $ and $\kappa /K_0^*(700) $}}, \bibinfo{journal}{Eur. Phys. J. ST} \bibinfo{volume}{230} (\bibinfo{number}{6}) (\bibinfo{year}{2021}) \bibinfo{pages}{1539--1574}, \bibinfo{doi}{\doi{10.1140/epjs/s11734-021-00142-9}}, \eprint{2101.06506}.

\bibtype{Article}%
\bibitem{Heisenberg:1932dw}
\bibinfo{author}{W. Heisenberg}, \bibinfo{title}{{\"Uber den Bau der Atomkerne (On the structure of atomic nuclei)}}, \bibinfo{journal}{Z. Phys.} \bibinfo{volume}{77} (\bibinfo{year}{1932}) \bibinfo{pages}{1--11}, \bibinfo{doi}{\doi{10.1007/BF01342433}}.

\bibtype{Article}%
\bibitem{Nakano:1953zz}
\bibinfo{author}{T. Nakano}, \bibinfo{author}{K. Nishijima}, \bibinfo{title}{{Charge Independence for V-particles}}, \bibinfo{journal}{Prog. Theor. Phys.} \bibinfo{volume}{10} (\bibinfo{year}{1953}) \bibinfo{pages}{581--582}, \bibinfo{doi}{\doi{10.1143/PTP.10.581}}.

\bibtype{Article}%
\bibitem{Nishijima:1955gxk}
\bibinfo{author}{K. Nishijima}, \bibinfo{title}{{Charge Independence Theory of V Particles}}, \bibinfo{journal}{Prog. Theor. Phys.} \bibinfo{volume}{13} (\bibinfo{number}{3}) (\bibinfo{year}{1955}) \bibinfo{pages}{285--304}, \bibinfo{doi}{\doi{10.1143/PTP.13.285}}.

\bibtype{Article}%
\bibitem{Gell-Mann:1956iqa}
\bibinfo{author}{M. Gell-Mann}, \bibinfo{title}{{The interpretation of the new particles as displaced charge multiplets}}, \bibinfo{journal}{Nuovo Cim.} \bibinfo{volume}{4} (\bibinfo{number}{S2}) (\bibinfo{year}{1956}) \bibinfo{pages}{848--866}, \bibinfo{doi}{\doi{10.1007/BF02748000}}.

\bibtype{Unpublished}%
\bibitem{Gell-Mann:1961omu}
\bibinfo{author}{M. Gell-Mann}, \bibinfo{title}{{The Eightfold Way: A Theory of strong interaction symmetry}} \bibinfo{year}{1961}, \bibinfo{doi}{\doi{10.2172/4008239}}, \bibinfo{note}{report CTSL-20, TID-12608}.

\bibtype{Article}%
\bibitem{Neeman:1961jhl}
\bibinfo{author}{Y. Ne'eman}, \bibinfo{title}{{Derivation of strong interactions from a gauge invariance}}, \bibinfo{journal}{Nucl. Phys.} \bibinfo{volume}{26} (\bibinfo{year}{1961}) \bibinfo{pages}{222--229}, \bibinfo{doi}{\doi{10.1016/0029-5582(61)90134-1}}.

\bibtype{Book}%
\bibitem{Gell-Mann:1964ook}
\bibinfo{author}{M. Gell-Mann}, \bibinfo{author}{Y. Ne'eman}, \bibinfo{title}{{The Eightfold way: a review with a collection of reprints}}, \bibinfo{publisher}{Frontiers in Physics. CRC Press} \bibinfo{year}{1964}, ISBN \bibinfo{isbn}{0367091747,978-0367091743}.

\bibtype{Article}%
\bibitem{Gell-Mann:1964ewy}
\bibinfo{author}{M. Gell-Mann}, \bibinfo{title}{{A Schematic Model of Baryons and Mesons}}, \bibinfo{journal}{Phys. Lett.} \bibinfo{volume}{8} (\bibinfo{year}{1964}) \bibinfo{pages}{214--215}, \bibinfo{doi}{\doi{10.1016/S0031-9163(64)92001-3}}.

\bibtype{Unpublished}%
\bibitem{Zweig:1964ruk}
\bibinfo{author}{G. Zweig}, \bibinfo{title}{{An SU(3) model for strong interaction symmetry and its breaking. Version 1}} \bibinfo{year}{1964}, \bibinfo{doi}{\doi{10.17181/CERN-TH-401}}, \bibinfo{note}{see also Version 2 in Developments in the quark theory of hadrons. Vol. 1. 1964 - 1978, pages 22--101. Eds: Lichtenberg, D. B. and Rosen, S. P., doi:10.17181/CERN-TH-412}.

\bibtype{Article}%
\bibitem{CMS:2025kzt}
\bibinfo{author}{A. Hayrapetyan}, et al. (\bibinfo{collaboration}{CMS}), \bibinfo{title}{{Observation of a pseudoscalar excess at the top quark pair production threshold}}  (\bibinfo{year}{2025}), \eprint{2503.22382}.

\bibtype{Article}%
\bibitem{Gross:2022hyw}
\bibinfo{author}{F. Gross}, et al., \bibinfo{title}{{50 Years of Quantum Chromodynamics}}, \bibinfo{journal}{Eur. Phys. J. C} \bibinfo{volume}{83} (\bibinfo{year}{2023}) \bibinfo{pages}{1125}, \bibinfo{doi}{\doi{10.1140/epjc/s10052-023-11949-2}}, \eprint{2212.11107}.

\bibtype{Article}%
\bibitem{Fritzsch:1973pi}
\bibinfo{author}{H. Fritzsch}, \bibinfo{author}{M. Gell-Mann}, \bibinfo{author}{H. Leutwyler}, \bibinfo{title}{{Advantages of the Color Octet Gluon Picture}}, \bibinfo{journal}{Phys. Lett. B} \bibinfo{volume}{47} (\bibinfo{year}{1973}) \bibinfo{pages}{365--368}, \bibinfo{doi}{\doi{10.1016/0370-2693(73)90625-4}}.

\bibtype{Article}%
\bibitem{Politzer:1973fx}
\bibinfo{author}{H.~D. Politzer}, \bibinfo{title}{{Reliable Perturbative Results for Strong Interactions?}}, \bibinfo{journal}{Phys. Rev. Lett.} \bibinfo{volume}{30} (\bibinfo{year}{1973}) \bibinfo{pages}{1346--1349}, \bibinfo{doi}{\doi{10.1103/PhysRevLett.30.1346}}.

\bibtype{Article}%
\bibitem{Gross:1973id}
\bibinfo{author}{D.J. Gross}, \bibinfo{author}{F. Wilczek}, \bibinfo{title}{{Ultraviolet Behavior of Nonabelian Gauge Theories}}, \bibinfo{journal}{Phys. Rev. Lett.} \bibinfo{volume}{30} (\bibinfo{year}{1973}) \bibinfo{pages}{1343--1346}, \bibinfo{doi}{\doi{10.1103/PhysRevLett.30.1343}}.

\bibtype{Article}%
\bibitem{Prelovsek:2025gmd}
\bibinfo{author}{Sasa Prelovsek}, \bibinfo{title}{{Lattice QCD calculations of hadron spectroscopy}}  (\bibinfo{year}{2025}), \eprint{2505.10002}.

\bibtype{Article}%
\bibitem{Entem:2025bqt}
\bibinfo{author}{D.~R. Entem}, \bibinfo{author}{F. Fern\'andez}, \bibinfo{author}{P.~G. Ortega}, \bibinfo{author}{J. Segovia}, \bibinfo{title}{{The Constituent Quark Model}}  (\bibinfo{year}{2025}), \eprint{2504.07897}.

\bibtype{Article}%
\bibitem{Nefediev:2025vmo}
\bibinfo{author}{A. Nefediev}, \bibinfo{title}{{Quark models: What can they teach us?}}  (\bibinfo{year}{2025}), \eprint{2507.19256}.

\bibtype{Article}%
\bibitem{Hanhart:2025bun}
\bibinfo{author}{C. Hanhart}, \bibinfo{title}{{Hadronic molecules and multiquark states}}  (\bibinfo{year}{2025}), \eprint{2504.06043}.

\bibtype{Article}%
\bibitem{Guo:2017jvc}
\bibinfo{author}{Feng-Kun Guo}, \bibinfo{author}{Christoph Hanhart}, \bibinfo{author}{Ulf-G. Mei{\ss}ner}, \bibinfo{author}{Qian Wang}, \bibinfo{author}{Qiang Zhao}, \bibinfo{author}{Bing-Song Zou}, \bibinfo{title}{{Hadronic molecules}}, \bibinfo{journal}{Rev. Mod. Phys.} \bibinfo{volume}{90} (\bibinfo{number}{1}) (\bibinfo{year}{2018}) \bibinfo{pages}{015004}, \bibinfo{doi}{\doi{10.1103/RevModPhys.90.015004}}, \bibinfo{note}{[Erratum: Rev.Mod.Phys. 94, 029901 (2022)]}, \eprint{1705.00141}.

\bibtype{Article}%
\bibitem{Isgur:1984bm}
\bibinfo{author}{N. Isgur}, \bibinfo{author}{J.~E. Paton}, \bibinfo{title}{{A Flux Tube Model for Hadrons in QCD}}, \bibinfo{journal}{Phys. Rev. D} \bibinfo{volume}{31} (\bibinfo{year}{1985}) \bibinfo{pages}{2910}, \bibinfo{doi}{\doi{10.1103/PhysRevD.31.2910}}.

\bibtype{Article}%
\bibitem{Chanowitz:1982qj}
\bibinfo{author}{M.~S. Chanowitz}, \bibinfo{author}{S.~R. Sharpe}, \bibinfo{title}{{Hybrids: Mixed States of Quarks and Gluons}}, \bibinfo{journal}{Nucl. Phys. B} \bibinfo{volume}{222} (\bibinfo{year}{1983}) \bibinfo{pages}{211--244}, \bibinfo{doi}{\doi{10.1016/0550-3213(83)90635-1}}, \bibinfo{note}{[Erratum: Nucl.Phys.B 228, 588--588 (1983)]}.

\bibtype{Article}%
\bibitem{Barnes:1982tx}
\bibinfo{author}{T. Barnes}, \bibinfo{author}{F.~E. Close}, \bibinfo{author}{F. de Viron}, \bibinfo{title}{{Q anti-Q G Hermaphrodite Mesons in the MIT Bag Model}}, \bibinfo{journal}{Nucl. Phys. B} \bibinfo{volume}{224} (\bibinfo{year}{1983}) \bibinfo{pages}{241}, \bibinfo{doi}{\doi{10.1016/0550-3213(83)90004-4}}.

\bibtype{Article}%
\bibitem{Lacock:1996ny}
\bibinfo{author}{P. Lacock}, \bibinfo{author}{C. Michael}, \bibinfo{author}{P. Boyle}, \bibinfo{author}{P. Rowland} (\bibinfo{collaboration}{UKQCD}), \bibinfo{title}{{Hybrid mesons from quenched QCD}}, \bibinfo{journal}{Phys. Lett. B} \bibinfo{volume}{401} (\bibinfo{year}{1997}) \bibinfo{pages}{308--312}, \bibinfo{doi}{\doi{10.1016/S0370-2693(97)00384-5}}, \eprint{hep-lat/9611011}.

\bibtype{Article}%
\bibitem{Dudek:2011bn}
\bibinfo{author}{J.~J. Dudek}, \bibinfo{title}{{The lightest hybrid meson supermultiplet in QCD}}, \bibinfo{journal}{Phys. Rev. D} \bibinfo{volume}{84} (\bibinfo{year}{2011}) \bibinfo{pages}{074023}, \bibinfo{doi}{\doi{10.1103/PhysRevD.84.074023}}, \eprint{1106.5515}.

\bibtype{Article}%
\bibitem{Dudek:2013yja}
\bibinfo{author}{J.~J. Dudek}, \bibinfo{author}{R.~G. Edwards}, \bibinfo{author}{P. Guo}, \bibinfo{author}{C.~E. Thomas} (\bibinfo{collaboration}{Hadron Spectrum}), \bibinfo{title}{{Toward the excited isoscalar meson spectrum from lattice QCD}}, \bibinfo{journal}{Phys. Rev. D} \bibinfo{volume}{88} (\bibinfo{number}{9}) (\bibinfo{year}{2013}) \bibinfo{pages}{094505}, \bibinfo{doi}{\doi{10.1103/PhysRevD.88.094505}}, \eprint{1309.2608}.

\bibtype{Article}%
\bibitem{Woss:2020ayi}
\bibinfo{author}{Antoni~J. Woss}, \bibinfo{author}{Jozef~J. Dudek}, \bibinfo{author}{Robert~G. Edwards}, \bibinfo{author}{Christopher~E. Thomas}, \bibinfo{author}{David~J. Wilson} (\bibinfo{collaboration}{Hadron Spectrum}), \bibinfo{title}{{Decays of an exotic $1{-+}$ hybrid meson resonance in QCD}}, \bibinfo{journal}{Phys. Rev. D} \bibinfo{volume}{103} (\bibinfo{number}{5}) (\bibinfo{year}{2021}) \bibinfo{pages}{054502}, \bibinfo{doi}{\doi{10.1103/PhysRevD.103.054502}}, \eprint{2009.10034}.

\bibtype{Article}%
\bibitem{Meyer:2015eta}
\bibinfo{author}{C.~A. Meyer}, \bibinfo{author}{E.~S. Swanson}, \bibinfo{title}{{Hybrid Mesons}}, \bibinfo{journal}{Prog. Part. Nucl. Phys.} \bibinfo{volume}{82} (\bibinfo{year}{2015}) \bibinfo{pages}{21--58}, \bibinfo{doi}{\doi{10.1016/j.ppnp.2015.03.001}}, \eprint{1502.07276}.

\bibtype{Article}%
\bibitem{Chen:2022asf}
\bibinfo{author}{H.-X. Chen}, \bibinfo{author}{W. Chen}, \bibinfo{author}{X. Liu}, \bibinfo{author}{Y.-R. Liu}, \bibinfo{author}{S.-L. Zhu}, \bibinfo{title}{{An updated review of the new hadron states}}, \bibinfo{journal}{Rept. Prog. Phys.} \bibinfo{volume}{86} (\bibinfo{number}{2}) (\bibinfo{year}{2023}) \bibinfo{pages}{026201}, \bibinfo{doi}{\doi{10.1088/1361-6633/aca3b6}}, \eprint{2204.02649}.

\bibtype{Article}%
\bibitem{Bali:1993fb}
\bibinfo{author}{G.~S. Bali}, \bibinfo{author}{K. Schilling}, \bibinfo{author}{A. Hulsebos}, \bibinfo{author}{A.~C. Irving}, \bibinfo{author}{Christopher Michael}, \bibinfo{author}{P.~W. Stephenson} (\bibinfo{collaboration}{UKQCD}), \bibinfo{title}{{A Comprehensive lattice study of SU(3) glueballs}}, \bibinfo{journal}{Phys. Lett. B} \bibinfo{volume}{309} (\bibinfo{year}{1993}) \bibinfo{pages}{378--384}, \bibinfo{doi}{\doi{10.1016/0370-2693(93)90948-H}}, \eprint{hep-lat/9304012}.

\bibtype{Article}%
\bibitem{Morningstar:1999rf}
\bibinfo{author}{C.~J. Morningstar}, \bibinfo{author}{M.~J. Peardon}, \bibinfo{title}{{The Glueball spectrum from an anisotropic lattice study}}, \bibinfo{journal}{Phys. Rev. D} \bibinfo{volume}{60} (\bibinfo{year}{1999}) \bibinfo{pages}{034509}, \bibinfo{doi}{\doi{10.1103/PhysRevD.60.034509}}, \eprint{hep-lat/9901004}.

\bibtype{Article}%
\bibitem{Chen:2005mg}
\bibinfo{author}{Y. Chen}, et al., \bibinfo{title}{{Glueball spectrum and matrix elements on anisotropic lattices}}, \bibinfo{journal}{Phys. Rev. D} \bibinfo{volume}{73} (\bibinfo{year}{2006}) \bibinfo{pages}{014516}, \bibinfo{doi}{\doi{10.1103/PhysRevD.73.014516}}, \eprint{hep-lat/0510074}.

\bibtype{Article}%
\bibitem{Gui:2019dtm}
\bibinfo{author}{L.-C. Gui}, \bibinfo{author}{J.-M. Dong}, \bibinfo{author}{Y. Chen}, \bibinfo{author}{Y.-B. Yang}, \bibinfo{title}{{Study of the pseudoscalar glueball in $J/\psi$ radiative decays}}, \bibinfo{journal}{Phys. Rev. D} \bibinfo{volume}{100} (\bibinfo{number}{5}) (\bibinfo{year}{2019}) \bibinfo{pages}{054511}, \bibinfo{doi}{\doi{10.1103/PhysRevD.100.054511}}, \eprint{1906.03666}.

\bibtype{Article}%
\bibitem{Athenodorou:2020ani}
\bibinfo{author}{A. Athenodorou}, \bibinfo{author}{M. Teper}, \bibinfo{title}{{The glueball spectrum of SU(3) gauge theory in 3 + 1 dimensions}}, \bibinfo{journal}{JHEP} \bibinfo{volume}{11} (\bibinfo{year}{2020}) \bibinfo{pages}{172}, \bibinfo{doi}{\doi{10.1007/JHEP11(2020)172}}, \eprint{2007.06422}.

\bibtype{Article}%
\bibitem{Gregory:2012hu}
\bibinfo{author}{E. Gregory}, \bibinfo{author}{A. Irving}, \bibinfo{author}{B. Lucini}, \bibinfo{author}{C. McNeile}, \bibinfo{author}{A. Rago}, \bibinfo{author}{C. Richards}, \bibinfo{author}{E. Rinaldi}, \bibinfo{title}{{Towards the glueball spectrum from unquenched lattice QCD}}, \bibinfo{journal}{JHEP} \bibinfo{volume}{10} (\bibinfo{year}{2012}) \bibinfo{pages}{170}, \bibinfo{doi}{\doi{10.1007/JHEP10(2012)170}}, \eprint{1208.1858}.

\bibtype{Article}%
\bibitem{Chen:2021bck}
\bibinfo{author}{H.-X. Chen}, \bibinfo{author}{W. Chen}, \bibinfo{author}{S.-L. Zhu}, \bibinfo{title}{{Two- and three-gluon glueballs of C=+}}, \bibinfo{journal}{Phys. Rev. D} \bibinfo{volume}{104} (\bibinfo{number}{9}) (\bibinfo{year}{2021}) \bibinfo{pages}{094050}, \bibinfo{doi}{\doi{10.1103/PhysRevD.104.094050}}, \eprint{2107.05271}.

\bibtype{Article}%
\bibitem{Szczepaniak:2003mr}
\bibinfo{author}{A.~P. Szczepaniak}, \bibinfo{author}{E.~S. Swanson}, \bibinfo{title}{{The Low lying glueball spectrum}}, \bibinfo{journal}{Phys. Lett. B} \bibinfo{volume}{577} (\bibinfo{year}{2003}) \bibinfo{pages}{61--66}, \bibinfo{doi}{\doi{10.1016/j.physletb.2003.10.008}}, \eprint{hep-ph/0308268}.

\bibtype{Article}%
\bibitem{Huber:2020ngt}
\bibinfo{author}{M.~Q. Huber}, \bibinfo{author}{C.~S. Fischer}, \bibinfo{author}{H. Sanchis-Alepuz}, \bibinfo{title}{{Spectrum of scalar and pseudoscalar glueballs from functional methods}}, \bibinfo{journal}{Eur. Phys. J. C} \bibinfo{volume}{80} (\bibinfo{number}{11}) (\bibinfo{year}{2020}) \bibinfo{pages}{1077}, \bibinfo{doi}{\doi{10.1140/epjc/s10052-020-08649-6}}, \eprint{2004.00415}.

\bibtype{Article}%
\bibitem{PANDA:2021ozp}
\bibinfo{author}{G. Barucca}, et al. (\bibinfo{collaboration}{PANDA}), \bibinfo{title}{{PANDA Phase One}}, \bibinfo{journal}{Eur. Phys. J. A} \bibinfo{volume}{57} (\bibinfo{number}{6}) (\bibinfo{year}{2021}) \bibinfo{pages}{184}, \bibinfo{doi}{\doi{10.1140/epja/s10050-021-00475-y}}, \eprint{2101.11877}.

\bibtype{Article}%
\bibitem{Jin:2021vct}
\bibinfo{author}{S. Jin}, \bibinfo{author}{X. Shen}, \bibinfo{title}{{Highlights of light meson spectroscopy at the BESIII experiment}}, \bibinfo{journal}{Natl. Sci. Rev.} \bibinfo{volume}{8} (\bibinfo{number}{11}) (\bibinfo{year}{2021}) \bibinfo{pages}{nwab198}, \bibinfo{doi}{\doi{10.1093/nsr/nwab198}}.

\bibtype{Article}%
\bibitem{Close:2002zu}
\bibinfo{author}{F.~E. Close}, \bibinfo{author}{N.s~A. Tornqvist}, \bibinfo{title}{{Scalar mesons above and below 1-GeV}}, \bibinfo{journal}{J. Phys. G} \bibinfo{volume}{28} (\bibinfo{year}{2002}) \bibinfo{pages}{R249--R267}, \bibinfo{doi}{\doi{10.1088/0954-3899/28/10/201}}, \eprint{hep-ph/0204205}.

\bibtype{Article}%
\bibitem{Oller:1998zr}
\bibinfo{author}{J.~A. Oller}, \bibinfo{author}{E. Oset}, \bibinfo{title}{{N/D description of two meson amplitudes and chiral symmetry}}, \bibinfo{journal}{Phys. Rev. D} \bibinfo{volume}{60} (\bibinfo{year}{1999}) \bibinfo{pages}{074023}, \bibinfo{doi}{\doi{10.1103/PhysRevD.60.074023}}, \eprint{hep-ph/9809337}.

\bibtype{Article}%
\bibitem{Amsler:2025wqz}
\bibinfo{author}{C. Amsler}, \bibinfo{title}{{Key Historical Experiments in Hadron Physics}}  (\bibinfo{year}{2025}), \eprint{2503.14689}.

\bibtype{Article}%
\bibitem{Lenske:2025idu}
\bibinfo{author}{Horst Lenske}, \bibinfo{author}{Igor Strakovsky}, \bibinfo{title}{{Hadron Production Processes}}  (\bibinfo{year}{2025}), \eprint{2507.21144}.

\bibtype{Article}%
\bibitem{BESIII:2023wfi}
\bibinfo{author}{M. Ablikim}, et al. (\bibinfo{collaboration}{BESIII}), \bibinfo{title}{{Determination of Spin-Parity Quantum Numbers of X(2370) as 0-+ from J/{\ensuremath{\psi}}{\textrightarrow}{\ensuremath{\gamma}}KS0KS0{\ensuremath{\eta}}'}}, \bibinfo{journal}{Phys. Rev. Lett.} \bibinfo{volume}{132} (\bibinfo{number}{18}) (\bibinfo{year}{2024}) \bibinfo{pages}{181901}, \bibinfo{doi}{\doi{10.1103/PhysRevLett.132.181901}}, \eprint{2312.05324}.

\bibtype{Article}%
\bibitem{Yukawa:1935xg}
\bibinfo{author}{H. Yukawa}, \bibinfo{title}{{On the Interaction of Elementary Particles I}}, \bibinfo{journal}{Proc. Phys. Math. Soc. Jap.} \bibinfo{volume}{17} (\bibinfo{year}{1935}) \bibinfo{pages}{48--57}, \bibinfo{doi}{\doi{10.1143/PTPS.1.1}}.

\bibtype{Article}%
\bibitem{Lattes:1947mw}
\bibinfo{author}{C.~M.~G. Lattes}, \bibinfo{author}{H. Muirhead}, \bibinfo{author}{G.~P.~S. Occhialini}, \bibinfo{author}{C.~F. Powell}, \bibinfo{title}{{PROCESSES INVOLVING CHARGED MESONS}}, \bibinfo{journal}{Nature} \bibinfo{volume}{159} (\bibinfo{year}{1947}) \bibinfo{pages}{694--697}, \bibinfo{doi}{\doi{10.1038/159694a0}}.

\bibtype{Article}%
\bibitem{Lattes:1947mx}
\bibinfo{author}{C.~M.~G. Lattes}, \bibinfo{author}{G.~P.~S. Occhialini}, \bibinfo{author}{C.~F. Powell}, \bibinfo{title}{{Observations on the Tracks of Slow Mesons in Photographic Emulsions. 1}}, \bibinfo{journal}{Nature} \bibinfo{volume}{160} (\bibinfo{year}{1947}) \bibinfo{pages}{453--456}, \bibinfo{doi}{\doi{10.1038/160453a0}}.

\bibtype{Article}%
\bibitem{Lattes:1947my}
\bibinfo{author}{C.~M.~G. Lattes}, \bibinfo{author}{G.~P.~S. Occhialini}, \bibinfo{author}{C.~F. Powell}, \bibinfo{title}{{Observations on the Tracks of Slow Mesons in Photographic Emulsions. 2}}, \bibinfo{journal}{Nature} \bibinfo{volume}{160} (\bibinfo{year}{1947}) \bibinfo{pages}{486--492}, \bibinfo{doi}{\doi{10.1038/160486a0}}.

\bibtype{Misc}%
\bibitem{Zee:2004}
\bibinfo{author}{A Zee}, \bibinfo{title}{Quantum Field Theory. Lecture 2 of 4} \bibinfo{year}{2014}, \bibinfo{note}{available on youtube (via aoflex)}, \bibinfo{url}{\urlprefix\url{https://www.youtube.com/watch?v=aypGTsLN0ck}}.

\bibtype{Article}%
\bibitem{Rochester:1947mi}
\bibinfo{author}{G.~D. Rochester}, \bibinfo{author}{C.~C. Butler}, \bibinfo{title}{{Evidence for the Existence of New Unstable Elementary Particles}}, \bibinfo{journal}{Nature} \bibinfo{volume}{160} (\bibinfo{year}{1947}) \bibinfo{pages}{855--857}, \bibinfo{doi}{\doi{10.1038/160855a0}}.

\bibtype{Article}%
\bibitem{Brown:1949mj}
\bibinfo{author}{R. Brown}, \bibinfo{author}{U. Camerini}, \bibinfo{author}{P.~H. Fowler}, \bibinfo{author}{H. Muirhead}, \bibinfo{author}{C.~F. Powell}, \bibinfo{author}{D.~M. Ritson}, \bibinfo{title}{{Observations With Electron Sensitive Plates Exposed to Cosmic Radiation}}, \bibinfo{journal}{Nature} \bibinfo{volume}{163} (\bibinfo{year}{1949}) \bibinfo{pages}{82}, \bibinfo{doi}{\doi{10.1038/163082a0}}.

\bibtype{Article}%
\bibitem{Pevsner:1961pa}
\bibinfo{author}{A. Pevsner}, et al., \bibinfo{title}{{Evidence for a Three Pion Resonance Near 550-{MeV}}}, \bibinfo{journal}{Phys. Rev. Lett.} \bibinfo{volume}{7} (\bibinfo{year}{1961}) \bibinfo{pages}{421--423}, \bibinfo{doi}{\doi{10.1103/PhysRevLett.7.421}}.

\bibtype{Article}%
\bibitem{Nambu:1960tm}
\bibinfo{author}{Y. Nambu}, \bibinfo{title}{{Quasiparticles and Gauge Invariance in the Theory of Superconductivity}}, \bibinfo{journal}{Phys. Rev.} \bibinfo{volume}{117} (\bibinfo{year}{1960}) \bibinfo{pages}{648--663}, \bibinfo{doi}{\doi{10.1103/PhysRev.117.648}}.

\bibtype{Article}%
\bibitem{Goldstone:1961eq}
\bibinfo{author}{J. Goldstone}, \bibinfo{title}{{Field Theories with Superconductor Solutions}}, \bibinfo{journal}{Nuovo Cim.} \bibinfo{volume}{19} (\bibinfo{year}{1961}) \bibinfo{pages}{154--164}, \bibinfo{doi}{\doi{10.1007/BF02812722}}.

\bibtype{Article}%
\bibitem{Goldstone:1962es}
\bibinfo{author}{J. Goldstone}, \bibinfo{author}{A. Salam}, \bibinfo{author}{S. Weinberg}, \bibinfo{title}{{Broken Symmetries}}, \bibinfo{journal}{Phys. Rev.} \bibinfo{volume}{127} (\bibinfo{year}{1962}) \bibinfo{pages}{965--970}, \bibinfo{doi}{\doi{10.1103/PhysRev.127.965}}.

\bibtype{Article}%
\bibitem{Nefediev:2025zkv}
\bibinfo{author}{Alexey Nefediev}, \bibinfo{title}{{Chiral symmetry and its breaking}}  (\bibinfo{year}{2025}), \eprint{2507.22502}.

\bibtype{Article}%
\bibitem{Gell-Mann:1968hlm}
\bibinfo{author}{M. Gell-Mann}, \bibinfo{author}{R.~J. Oakes}, \bibinfo{author}{B. Renner}, \bibinfo{title}{{Behavior of current divergences under SU(3) x SU(3)}}, \bibinfo{journal}{Phys. Rev.} \bibinfo{volume}{175} (\bibinfo{year}{1968}) \bibinfo{pages}{2195--2199}, \bibinfo{doi}{\doi{10.1103/PhysRev.175.2195}}.

\bibtype{Article}%
\bibitem{Glashow:1967rx}
\bibinfo{author}{S.~L. Glashow}, \bibinfo{author}{S. Weinberg}, \bibinfo{title}{{Breaking chiral symmetry}}, \bibinfo{journal}{Phys. Rev. Lett.} \bibinfo{volume}{20} (\bibinfo{year}{1968}) \bibinfo{pages}{224--227}, \bibinfo{doi}{\doi{10.1103/PhysRevLett.20.224}}.

\bibtype{Book}%
\bibitem{Weinberg:1996kr}
\bibinfo{author}{S. Weinberg}, \bibinfo{title}{{The quantum theory of fields. Vol. 2: Modern applications}}, \bibinfo{publisher}{Cambridge University Press} \bibinfo{year}{2013}, ISBN \bibinfo{isbn}{978-1-139-63247-8, 978-0-521-67054-8, 978-0-521-55002-4}, \bibinfo{doi}{\doi{10.1017/CBO9781139644174}}.

\bibtype{Article}%
\bibitem{Okubo:1961jc}
\bibinfo{author}{S. Okubo}, \bibinfo{title}{{Note on unitary symmetry in strong interactions}}, \bibinfo{journal}{Prog. Theor. Phys.} \bibinfo{volume}{27} (\bibinfo{year}{1962}) \bibinfo{pages}{949--966}, \bibinfo{doi}{\doi{10.1143/PTP.27.949}}.

\bibtype{Article}%
\bibitem{Klempt:2007cp}
\bibinfo{author}{E. Klempt}, \bibinfo{author}{A. Zaitsev}, \bibinfo{title}{{Glueballs, Hybrids, Multiquarks. Experimental facts versus QCD inspired concepts}}, \bibinfo{journal}{Phys. Rept.} \bibinfo{volume}{454} (\bibinfo{year}{2007}) \bibinfo{pages}{1--202}, \bibinfo{doi}{\doi{10.1016/j.physrep.2007.07.006}}, \eprint{0708.4016}.

\bibtype{Article}%
\bibitem{Weinberg:1978kz}
\bibinfo{author}{S. Weinberg}, \bibinfo{title}{{Phenomenological Lagrangians}}, \bibinfo{journal}{Physica A} \bibinfo{volume}{96} (\bibinfo{number}{1-2}) (\bibinfo{year}{1979}) \bibinfo{pages}{327--340}, \bibinfo{doi}{\doi{10.1016/0378-4371(79)90223-1}}.

\bibtype{Article}%
\bibitem{Gasser:1983yg}
\bibinfo{author}{J. Gasser}, \bibinfo{author}{H. Leutwyler}, \bibinfo{title}{{Chiral Perturbation Theory to One Loop}}, \bibinfo{journal}{Annals Phys.} \bibinfo{volume}{158} (\bibinfo{year}{1984}) \bibinfo{pages}{142}, \bibinfo{doi}{\doi{10.1016/0003-4916(84)90242-2}}.

\bibtype{Article}%
\bibitem{Gasser:1984gg}
\bibinfo{author}{J. Gasser}, \bibinfo{author}{H. Leutwyler}, \bibinfo{title}{{Chiral Perturbation Theory: Expansions in the Mass of the Strange Quark}}, \bibinfo{journal}{Nucl. Phys. B} \bibinfo{volume}{250} (\bibinfo{year}{1985}) \bibinfo{pages}{465--516}, \bibinfo{doi}{\doi{10.1016/0550-3213(85)90492-4}}.

\bibtype{Article}%
\bibitem{Meissner:2024ona}
\bibinfo{author}{Ulf-G. Mei{\ss}ner}, \bibinfo{title}{{Chiral perturbation theory}}  (\bibinfo{year}{2024}), \eprint{2410.21912}.

\bibtype{Article}%
\bibitem{Donoghue:1988ed}
\bibinfo{author}{J.~F. Donoghue}, \bibinfo{author}{C. Ramirez}, \bibinfo{author}{G. Valencia}, \bibinfo{title}{{The Spectrum of QCD and Chiral Lagrangians of the Strong and Weak Interactions}}, \bibinfo{journal}{Phys. Rev. D} \bibinfo{volume}{39} (\bibinfo{year}{1989}) \bibinfo{pages}{1947}, \bibinfo{doi}{\doi{10.1103/PhysRevD.39.1947}}.

\bibtype{Article}%
\bibitem{Ecker:1988te}
\bibinfo{author}{G. Ecker}, \bibinfo{author}{J. Gasser}, \bibinfo{author}{A. Pich}, \bibinfo{author}{E. de Rafael}, \bibinfo{title}{{The Role of Resonances in Chiral Perturbation Theory}}, \bibinfo{journal}{Nucl. Phys. B} \bibinfo{volume}{321} (\bibinfo{year}{1989}) \bibinfo{pages}{311--342}, \bibinfo{doi}{\doi{10.1016/0550-3213(89)90346-5}}.

\bibtype{Article}%
\bibitem{Dobado:1989qm}
\bibinfo{author}{A. Dobado}, \bibinfo{author}{M.~J. Herrero}, \bibinfo{author}{T.~N. Truong}, \bibinfo{title}{{Unitarized Chiral Perturbation Theory for Elastic Pion-Pion Scattering}}, \bibinfo{journal}{Phys. Lett. B} \bibinfo{volume}{235} (\bibinfo{year}{1990}) \bibinfo{pages}{134--140}, \bibinfo{doi}{\doi{10.1016/0370-2693(90)90109-J}}.

\bibtype{Article}%
\bibitem{Dobado:1996ps}
\bibinfo{author}{A. Dobado}, \bibinfo{author}{J.~R. Pelaez}, \bibinfo{title}{{The Inverse amplitude method in chiral perturbation theory}}, \bibinfo{journal}{Phys. Rev. D} \bibinfo{volume}{56} (\bibinfo{year}{1997}) \bibinfo{pages}{3057--3073}, \bibinfo{doi}{\doi{10.1103/PhysRevD.56.3057}}, \eprint{hep-ph/9604416}.

\bibtype{Article}%
\bibitem{Oller:1997ti}
\bibinfo{author}{J.~A. Oller}, \bibinfo{author}{E. Oset}, \bibinfo{title}{{Chiral symmetry amplitudes in the S wave isoscalar and isovector channels and the $\sigma$, f$_0$(980), a$_0$(980) scalar mesons}}, \bibinfo{journal}{Nucl. Phys. A} \bibinfo{volume}{620} (\bibinfo{year}{1997}) \bibinfo{pages}{438--456}, \bibinfo{doi}{\doi{10.1016/S0375-9474(97)00160-7}}, \bibinfo{note}{[Erratum: Nucl.Phys.A 652, 407--409 (1999)]}, \eprint{hep-ph/9702314}.

\bibtype{Article}%
\bibitem{Oller:1998hw}
\bibinfo{author}{J.~A. Oller}, \bibinfo{author}{E. Oset}, \bibinfo{author}{J.~R. Pelaez}, \bibinfo{title}{{Meson meson interaction in a nonperturbative chiral approach}}, \bibinfo{journal}{Phys. Rev. D} \bibinfo{volume}{59} (\bibinfo{year}{1999}) \bibinfo{pages}{074001}, \bibinfo{doi}{\doi{10.1103/PhysRevD.59.074001}}, \bibinfo{note}{[Erratum: Phys.Rev.D 60, 099906 (1999), Erratum: Phys.Rev.D 75, 099903 (2007)]}, \eprint{hep-ph/9804209}.

\bibtype{Article}%
\bibitem{Nieves:1998hp}
\bibinfo{author}{J. Nieves}, \bibinfo{author}{E. Ruiz~Arriola}, \bibinfo{title}{{Bethe-Salpeter approach for meson meson scattering in chiral perturbation theory}}, \bibinfo{journal}{Phys. Lett. B} \bibinfo{volume}{455} (\bibinfo{year}{1999}) \bibinfo{pages}{30--38}, \bibinfo{doi}{\doi{10.1016/S0370-2693(99)00461-X}}, \eprint{nucl-th/9807035}.

\bibtype{Article}%
\bibitem{Oller:2020guq}
\bibinfo{author}{J.~A. Oller}, \bibinfo{title}{{Unitarization Technics in Hadron Physics with Historical Remarks}}, \bibinfo{journal}{Symmetry} \bibinfo{volume}{12} (\bibinfo{number}{7}) (\bibinfo{year}{2020}) \bibinfo{pages}{1114}, \bibinfo{doi}{\doi{10.3390/sym12071114}}, \eprint{2005.14417}.

\bibtype{Article}%
\bibitem{Feldmann:1999uf}
\bibinfo{author}{T. Feldmann}, \bibinfo{title}{{Quark structure of pseudoscalar mesons}}, \bibinfo{journal}{Int. J. Mod. Phys. A} \bibinfo{volume}{15} (\bibinfo{year}{2000}) \bibinfo{pages}{159--207}, \bibinfo{doi}{\doi{10.1142/S0217751X00000082}}, \eprint{hep-ph/9907491}.

\bibtype{Article}%
\bibitem{Kroll:2005sd}
\bibinfo{author}{P. Kroll}, \bibinfo{title}{{Isospin symmetry breaking through pi0 - eta - eta-prime mixing}}, \bibinfo{journal}{Mod. Phys. Lett. A} \bibinfo{volume}{20} (\bibinfo{year}{2005}) \bibinfo{pages}{2667--2684}, \bibinfo{doi}{\doi{10.1142/S0217732305018633}}, \eprint{hep-ph/0509031}.

\bibtype{Article}%
\bibitem{Escribano:2005qq}
\bibinfo{author}{R. Escribano}, \bibinfo{author}{J.-M. Frere}, \bibinfo{title}{{Study of the eta - eta-prime system in the two mixing angle scheme}}, \bibinfo{journal}{JHEP} \bibinfo{volume}{06} (\bibinfo{year}{2005}) \bibinfo{pages}{029}, \bibinfo{doi}{\doi{10.1088/1126-6708/2005/06/029}}, \eprint{hep-ph/0501072}.

\bibtype{Article}%
\bibitem{Parganlija:2012fy}
\bibinfo{author}{D. Parganlija}, \bibinfo{author}{P. Kovacs}, \bibinfo{author}{G. Wolf}, \bibinfo{author}{F. Giacosa}, \bibinfo{author}{D.~H. Rischke}, \bibinfo{title}{{Meson vacuum phenomenology in a three-flavor linear sigma model with (axial-)vector mesons}}, \bibinfo{journal}{Phys. Rev. D} \bibinfo{volume}{87} (\bibinfo{number}{1}) (\bibinfo{year}{2013}) \bibinfo{pages}{014011}, \bibinfo{doi}{\doi{10.1103/PhysRevD.87.014011}}, \eprint{1208.0585}.

\bibtype{Article}%
\bibitem{Christ:2010dd}
\bibinfo{author}{N.~H. Christ}, \bibinfo{author}{C. Dawson}, \bibinfo{author}{T. Izubuchi}, \bibinfo{author}{C. Jung}, \bibinfo{author}{Q. Liu}, \bibinfo{author}{R.~D. Mawhinney}, \bibinfo{author}{C.~T. Sachrajda}, \bibinfo{author}{A. Soni}, \bibinfo{author}{R. Zhou}, \bibinfo{title}{{The $\eta$ and $\eta^\prime$ mesons from Lattice QCD}}, \bibinfo{journal}{Phys. Rev. Lett.} \bibinfo{volume}{105} (\bibinfo{year}{2010}) \bibinfo{pages}{241601}, \bibinfo{doi}{\doi{10.1103/PhysRevLett.105.241601}}, \eprint{1002.2999}.

\bibtype{Article}%
\bibitem{Escribano:2020jdy}
\bibinfo{author}{R. Escribano}, \bibinfo{author}{E. Royo}, \bibinfo{title}{{$\pi^0$-$\eta$-$\eta^{\prime}$ mixing from $V\!\rightarrow\!P\gamma$ and $P\!\rightarrow\!V\gamma$ decays}}, \bibinfo{journal}{Phys. Lett. B} \bibinfo{volume}{807} (\bibinfo{year}{2020}) \bibinfo{pages}{135534}, \bibinfo{doi}{\doi{10.1016/j.physletb.2020.135534}}, \eprint{2003.08379}.

\bibtype{Article}%
\bibitem{Ambrosino:2009sc}
\bibinfo{author}{F. Ambrosino}, et al., \bibinfo{title}{{A Global fit to determine the pseudoscalar mixing angle and the gluonium content of the eta-prime meson}}, \bibinfo{journal}{JHEP} \bibinfo{volume}{07} (\bibinfo{year}{2009}) \bibinfo{pages}{105}, \bibinfo{doi}{\doi{10.1088/1126-6708/2009/07/105}}, \eprint{0906.3819}.

\bibtype{Article}%
\bibitem{Wu:2011yx}
\bibinfo{author}{Jia-Jun Wu}, \bibinfo{author}{Xiao-Hai Liu}, \bibinfo{author}{Qiang Zhao}, \bibinfo{author}{Bing-Song Zou}, \bibinfo{title}{{The Puzzle of anomalously large isospin violations in $\eta(1405/1475)\to 3\pi$}}, \bibinfo{journal}{Phys. Rev. Lett.} \bibinfo{volume}{108} (\bibinfo{year}{2012}) \bibinfo{pages}{081803}, \bibinfo{doi}{\doi{10.1103/PhysRevLett.108.081803}}, \eprint{1108.3772}.

\bibtype{Article}%
\bibitem{Du:2019idk}
\bibinfo{author}{M.-C. Du}, \bibinfo{author}{Q. Zhao}, \bibinfo{title}{{Internal particle width effects on the triangle singularity mechanism in the study of the $\eta(1405)$ and $\eta(1475)$ puzzle}}, \bibinfo{journal}{Phys. Rev. D} \bibinfo{volume}{100} (\bibinfo{number}{3}) (\bibinfo{year}{2019}) \bibinfo{pages}{036005}, \bibinfo{doi}{\doi{10.1103/PhysRevD.100.036005}}, \eprint{1905.04207}.

\bibtype{Article}%
\bibitem{COMPASS:2025wkw}
\bibinfo{author}{G.~D. Alexeev}, et al. (\bibinfo{collaboration}{COMPASS}), \bibinfo{title}{{Spectroscopy of Strange Mesons and First Observation of a Strange Crypto-Exotic State with $J^P=0^-$}}  (\bibinfo{year}{2025}), \eprint{2504.09470}.

\bibtype{Article}%
\bibitem{BESIII:2022iwi}
\bibinfo{author}{M. Ablikim}, et al. (\bibinfo{collaboration}{BESIII}), \bibinfo{title}{{Partial wave analysis of J/{\ensuremath{\psi}}{\textrightarrow}{\ensuremath{\gamma}}{\ensuremath{\eta}}{\ensuremath{\eta}}'}}, \bibinfo{journal}{Phys. Rev. D} \bibinfo{volume}{106} (\bibinfo{number}{7}) (\bibinfo{year}{2022}) \bibinfo{pages}{072012}, \bibinfo{doi}{\doi{10.1103/PhysRevD.106.072012}}, \bibinfo{note}{[Erratum: Phys.Rev.D 107, 079901 (2023)]}, \eprint{2202.00623}.

\bibtype{Article}%
\bibitem{Johnson:1955zz}
\bibinfo{author}{M.~H. Johnson}, \bibinfo{author}{E. Teller}, \bibinfo{title}{{Classical Field Theory of Nuclear Forces}}, \bibinfo{journal}{Phys. Rev.} \bibinfo{volume}{98} (\bibinfo{year}{1955}) \bibinfo{pages}{783--787}, \bibinfo{doi}{\doi{10.1103/PhysRev.98.783}}.

\bibtype{Article}%
\bibitem{Schwinger:1957em}
\bibinfo{author}{J.~S. Schwinger}, \bibinfo{title}{{A Theory of the Fundamental Interactions}}, \bibinfo{journal}{Annals Phys.} \bibinfo{volume}{2} (\bibinfo{year}{1957}) \bibinfo{pages}{407--434}, \bibinfo{doi}{\doi{10.1016/0003-4916(57)90015-5}}.

\bibtype{Article}%
\bibitem{Nagasaki:1967bq}
\bibinfo{author}{M. Nagasaki}, \bibinfo{author}{T. Hirasawa}, \bibinfo{author}{M. Taketani}, \bibinfo{title}{{On the Analysis of nucleon-nucleon scattering with large momentum transfers}}, \bibinfo{journal}{Prog. Theor. Phys.} \bibinfo{volume}{38} (\bibinfo{year}{1967}) \bibinfo{pages}{514--516}, \bibinfo{doi}{\doi{10.1143/PTP.38.514}}.

\bibtype{Article}%
\bibitem{Machleidt:2000ge}
\bibinfo{author}{R. Machleidt}, \bibinfo{title}{{The High precision, charge dependent Bonn nucleon-nucleon potential (CD-Bonn)}}, \bibinfo{journal}{Phys. Rev. C} \bibinfo{volume}{63} (\bibinfo{year}{2001}) \bibinfo{pages}{024001}, \bibinfo{doi}{\doi{10.1103/PhysRevC.63.024001}}, \eprint{nucl-th/0006014}.

\bibtype{Article}%
\bibitem{Wiringa:1994wb}
\bibinfo{author}{R.~B. Wiringa}, \bibinfo{author}{V.~G.~J. Stoks}, \bibinfo{author}{R. Schiavilla}, \bibinfo{title}{{An Accurate nucleon-nucleon potential with charge independence breaking}}, \bibinfo{journal}{Phys. Rev. C} \bibinfo{volume}{51} (\bibinfo{year}{1995}) \bibinfo{pages}{38--51}, \bibinfo{doi}{\doi{10.1103/PhysRevC.51.38}}, \eprint{nucl-th/9408016}.

\bibtype{Inproceedings}%
\bibitem{Dalitz:1966fd}
\bibinfo{author}{R.~H. Dalitz}, \bibinfo{title}{{Resonant states and strong interactions}}, in: \bibinfo{booktitle}{{Oxford International Conference on Elementary Particles}} \bibinfo{year}{1966}, pp. \bibinfo{pages}{157--181}.

\bibtype{Article}%
\bibitem{Gell-Mann:1960mvl}
\bibinfo{author}{M. Gell-Mann}, \bibinfo{author}{M. Levy}, \bibinfo{title}{{The axial vector current in beta decay}}, \bibinfo{journal}{Nuovo Cim.} \bibinfo{volume}{16} (\bibinfo{year}{1960}) \bibinfo{pages}{705}, \bibinfo{doi}{\doi{10.1007/BF02859738}}.

\bibtype{Article}%
\bibitem{Nambu:1961fr}
\bibinfo{author}{Y. Nambu}, \bibinfo{author}{G. Jona-Lasinio}, \bibinfo{title}{{Dynamical model of elementary particles based on an analogy with superconductivity. II.}}, \bibinfo{journal}{Phys. Rev.} \bibinfo{volume}{124} (\bibinfo{year}{1961}) \bibinfo{pages}{246--254}, \bibinfo{doi}{\doi{10.1103/PhysRev.124.246}}.

\bibtype{Article}%
\bibitem{Nambu:1961tp}
\bibinfo{author}{Y. Nambu}, \bibinfo{author}{G. Jona-Lasinio}, \bibinfo{title}{{Dynamical Model of Elementary Particles Based on an Analogy with Superconductivity. 1.}}, \bibinfo{journal}{Phys. Rev.} \bibinfo{volume}{122} (\bibinfo{year}{1961}) \bibinfo{pages}{345--358}, \bibinfo{doi}{\doi{10.1103/PhysRev.122.345}}.

\bibtype{Article}%
\bibitem{Roy:1971tc}
\bibinfo{author}{S.~M. Roy}, \bibinfo{title}{{Exact integral equation for pion pion scattering involving only physical region partial waves}}, \bibinfo{journal}{Phys. Lett. B} \bibinfo{volume}{36} (\bibinfo{year}{1971}) \bibinfo{pages}{353--356}, \bibinfo{doi}{\doi{10.1016/0370-2693(71)90724-6}}.

\bibtype{Article}%
\bibitem{Ananthanarayan:2000ht}
\bibinfo{author}{B. Ananthanarayan}, \bibinfo{author}{G. Colangelo}, \bibinfo{author}{J. Gasser}, \bibinfo{author}{H. Leutwyler}, \bibinfo{title}{{Roy equation analysis of pi pi scattering}}, \bibinfo{journal}{Phys. Rept.} \bibinfo{volume}{353} (\bibinfo{year}{2001}) \bibinfo{pages}{207--279}, \bibinfo{doi}{\doi{10.1016/S0370-1573(01)00009-6}}, \eprint{hep-ph/0005297}.

\bibtype{Article}%
\bibitem{Colangelo:2001df}
\bibinfo{author}{G. Colangelo}, \bibinfo{author}{J. Gasser}, \bibinfo{author}{H. Leutwyler}, \bibinfo{title}{{$\pi \pi$ scattering}}, \bibinfo{journal}{Nucl. Phys. B} \bibinfo{volume}{603} (\bibinfo{year}{2001}) \bibinfo{pages}{125--179}, \bibinfo{doi}{\doi{10.1016/S0550-3213(01)00147-X}}, \eprint{hep-ph/0103088}.

\bibtype{Article}%
\bibitem{Caprini:2005zr}
\bibinfo{author}{I. Caprini}, \bibinfo{author}{G. Colangelo}, \bibinfo{author}{H. Leutwyler}, \bibinfo{title}{{Mass and width of the lowest resonance in QCD}}, \bibinfo{journal}{Phys. Rev. Lett.} \bibinfo{volume}{96} (\bibinfo{year}{2006}) \bibinfo{pages}{132001}, \bibinfo{doi}{\doi{10.1103/PhysRevLett.96.132001}}, \eprint{hep-ph/0512364}.

\bibtype{Article}%
\bibitem{Garcia-Martin:2011nna}
\bibinfo{author}{R. Garcia-Martin}, \bibinfo{author}{R. Kaminski}, \bibinfo{author}{J.~R. Pelaez}, \bibinfo{author}{J. Ruiz~de Elvira}, \bibinfo{title}{{Precise determination of the f0(600) and f0(980) pole parameters from a dispersive data analysis}}, \bibinfo{journal}{Phys. Rev. Lett.} \bibinfo{volume}{107} (\bibinfo{year}{2011}) \bibinfo{pages}{072001}, \bibinfo{doi}{\doi{10.1103/PhysRevLett.107.072001}}, \eprint{1107.1635}.

\bibtype{Article}%
\bibitem{Garcia-Martin:2011iqs}
\bibinfo{author}{R. Garcia-Martin}, \bibinfo{author}{R. Kaminski}, \bibinfo{author}{J.~R. Pelaez}, \bibinfo{author}{J. Ruiz~de Elvira}, \bibinfo{author}{F.~J. Yndurain}, \bibinfo{title}{{The Pion-pion scattering amplitude. IV: Improved analysis with once subtracted Roy-like equations up to 1100 MeV}}, \bibinfo{journal}{Phys. Rev. D} \bibinfo{volume}{83} (\bibinfo{year}{2011}) \bibinfo{pages}{074004}, \bibinfo{doi}{\doi{10.1103/PhysRevD.83.074004}}, \eprint{1102.2183}.

\bibtype{Article}%
\bibitem{Pelaez:2020gnd}
\bibinfo{author}{J.~R. Pel{\'a}ez}, \bibinfo{author}{A. Rodas}, \bibinfo{title}{{Dispersive {\ensuremath{\pi}}K{\textrightarrow}{\ensuremath{\pi}}K and {\ensuremath{\pi}}{\ensuremath{\pi}}{\textrightarrow}{\ensuremath{K \bar K}} amplitudes from scattering data, threshold parameters, and the lightest strange resonance {\ensuremath{\kappa}} or K0{\ensuremath{*}}(700)}}, \bibinfo{journal}{Phys. Rept.} \bibinfo{volume}{969} (\bibinfo{year}{2022}) \bibinfo{pages}{1--126}, \bibinfo{doi}{\doi{10.1016/j.physrep.2022.03.004}}, \eprint{2010.11222}.

\bibtype{Article}%
\bibitem{Steiner:1971ms}
\bibinfo{author}{F. Steiner}, \bibinfo{title}{{Partial wave crossing relations for meson-baryon scattering}}, \bibinfo{journal}{Fortsch. Phys.} \bibinfo{volume}{19} (\bibinfo{year}{1971}) \bibinfo{pages}{115--159}, \bibinfo{doi}{\doi{10.1002/prop.19710190302}}.

\bibtype{Article}%
\bibitem{Buettiker:2003pp}
\bibinfo{author}{P. Buettiker}, \bibinfo{author}{S. Descotes-Genon}, \bibinfo{author}{B. Moussallam}, \bibinfo{title}{{A new analysis of pi K scattering from Roy and Steiner type equations}}, \bibinfo{journal}{Eur. Phys. J. C} \bibinfo{volume}{33} (\bibinfo{year}{2004}) \bibinfo{pages}{409--432}, \bibinfo{doi}{\doi{10.1140/epjc/s2004-01591-1}}, \eprint{hep-ph/0310283}.

\bibtype{Article}%
\bibitem{Descotes-Genon:2006sdr}
\bibinfo{author}{S. Descotes-Genon}, \bibinfo{author}{B. Moussallam}, \bibinfo{title}{{The K*0 (800) scalar resonance from Roy-Steiner representations of pi K scattering}}, \bibinfo{journal}{Eur. Phys. J. C} \bibinfo{volume}{48} (\bibinfo{year}{2006}) \bibinfo{pages}{553}, \bibinfo{doi}{\doi{10.1140/epjc/s10052-006-0036-2}}, \eprint{hep-ph/0607133}.

\bibtype{Article}%
\bibitem{Pelaez:2020uiw}
\bibinfo{author}{J.~R. Pel{\'a}ez}, \bibinfo{author}{A. Rodas}, \bibinfo{title}{{Determination of the lightest strange resonance $K_0^*(700)$ or $\kappa$, from a dispersive data analysis}}, \bibinfo{journal}{Phys. Rev. Lett.} \bibinfo{volume}{124} (\bibinfo{number}{17}) (\bibinfo{year}{2020}) \bibinfo{pages}{172001}, \bibinfo{doi}{\doi{10.1103/PhysRevLett.124.172001}}, \eprint{2001.08153}.

\bibtype{Article}%
\bibitem{Danilkin:2020pak}
\bibinfo{author}{I. Danilkin}, \bibinfo{author}{O. Deineka}, \bibinfo{author}{M. Vanderhaeghen}, \bibinfo{title}{{Data-driven dispersive analysis of the {\ensuremath{\pi}}{\ensuremath{\pi}} and {\ensuremath{\pi}}K scattering}}, \bibinfo{journal}{Phys. Rev. D} \bibinfo{volume}{103} (\bibinfo{number}{11}) (\bibinfo{year}{2021}) \bibinfo{pages}{114023}, \bibinfo{doi}{\doi{10.1103/PhysRevD.103.114023}}, \eprint{2012.11636}.

\bibtype{Article}%
\bibitem{Briceno:2016mjc}
\bibinfo{author}{Raul~A. Briceno}, \bibinfo{author}{Jozef~J. Dudek}, \bibinfo{author}{Robert~G. Edwards}, \bibinfo{author}{David~J. Wilson}, \bibinfo{title}{{Isoscalar $\pi\pi$ scattering and the $\sigma$ meson resonance from QCD}}, \bibinfo{journal}{Phys. Rev. Lett.} \bibinfo{volume}{118} (\bibinfo{number}{2}) (\bibinfo{year}{2017}) \bibinfo{pages}{022002}, \bibinfo{doi}{\doi{10.1103/PhysRevLett.118.022002}}, \eprint{1607.05900}.

\bibtype{Article}%
\bibitem{Briceno:2017qmb}
\bibinfo{author}{Raul~A. Briceno}, \bibinfo{author}{Jozef~J. Dudek}, \bibinfo{author}{Robert~G. Edwards}, \bibinfo{author}{David~J. Wilson}, \bibinfo{title}{{Isoscalar $\pi\pi, K\overline{K}, \eta\eta$ scattering and the $\sigma, f_0, f_2$ mesons from QCD}}, \bibinfo{journal}{Phys. Rev. D} \bibinfo{volume}{97} (\bibinfo{number}{5}) (\bibinfo{year}{2018}) \bibinfo{pages}{054513}, \bibinfo{doi}{\doi{10.1103/PhysRevD.97.054513}}, \eprint{1708.06667}.

\bibtype{Article}%
\bibitem{Guo:2018zss}
\bibinfo{author}{Dehua Guo}, \bibinfo{author}{Andrei Alexandru}, \bibinfo{author}{Raquel Molina}, \bibinfo{author}{Maxim Mai}, \bibinfo{author}{Michael D{\"o}ring}, \bibinfo{title}{{Extraction of isoscalar $\pi\pi$ phase-shifts from lattice QCD}}, \bibinfo{journal}{Phys. Rev. D} \bibinfo{volume}{98} (\bibinfo{number}{1}) (\bibinfo{year}{2018}) \bibinfo{pages}{014507}, \bibinfo{doi}{\doi{10.1103/PhysRevD.98.014507}}, \eprint{1803.02897}.

\bibtype{Article}%
\bibitem{Rodas:2023gma}
\bibinfo{author}{Arkaitz Rodas}, \bibinfo{author}{Jozef~J. Dudek}, \bibinfo{author}{Robert~G. Edwards} (\bibinfo{collaboration}{Hadron Spectrum}), \bibinfo{title}{{Quark mass dependence of {\ensuremath{\pi}}{\ensuremath{\pi}} scattering in isospin 0, 1, and 2 from lattice QCD}}, \bibinfo{journal}{Phys. Rev. D} \bibinfo{volume}{108} (\bibinfo{number}{3}) (\bibinfo{year}{2023}) \bibinfo{pages}{034513}, \bibinfo{doi}{\doi{10.1103/PhysRevD.108.034513}}, \eprint{2303.10701}.

\bibtype{Article}%
\bibitem{Wilson:2014cna}
\bibinfo{author}{David~J. Wilson}, \bibinfo{author}{Jozef~J. Dudek}, \bibinfo{author}{Robert~G. Edwards}, \bibinfo{author}{Christopher~E. Thomas}, \bibinfo{title}{{Resonances in coupled $\pi K, \eta K$ scattering from lattice QCD}}, \bibinfo{journal}{Phys. Rev. D} \bibinfo{volume}{91} (\bibinfo{number}{5}) (\bibinfo{year}{2015}) \bibinfo{pages}{054008}, \bibinfo{doi}{\doi{10.1103/PhysRevD.91.054008}}, \eprint{1411.2004}.

\bibtype{Article}%
\bibitem{Dudek:2014qha}
\bibinfo{author}{Jozef~J. Dudek}, \bibinfo{author}{Robert~G. Edwards}, \bibinfo{author}{Christopher~E. Thomas}, \bibinfo{author}{David~J. Wilson} (\bibinfo{collaboration}{Hadron Spectrum}), \bibinfo{title}{{Resonances in coupled $\pi K -\eta K$ scattering from quantum chromodynamics}}, \bibinfo{journal}{Phys. Rev. Lett.} \bibinfo{volume}{113} (\bibinfo{number}{18}) (\bibinfo{year}{2014}) \bibinfo{pages}{182001}, \bibinfo{doi}{\doi{10.1103/PhysRevLett.113.182001}}, \eprint{1406.4158}.

\bibtype{Article}%
\bibitem{Cao:2023ntr}
\bibinfo{author}{Xiong-Hui Cao}, \bibinfo{author}{Qu-Zhi Li}, \bibinfo{author}{Zhi-Hui Guo}, \bibinfo{author}{Han-Qing Zheng}, \bibinfo{title}{{Roy equation analyses of {\ensuremath{\pi}}{\ensuremath{\pi}} scatterings at unphysical pion masses}}, \bibinfo{journal}{Phys. Rev. D} \bibinfo{volume}{108} (\bibinfo{number}{3}) (\bibinfo{year}{2023}) \bibinfo{pages}{034009}, \bibinfo{doi}{\doi{10.1103/PhysRevD.108.034009}}, \eprint{2303.02596}.

\bibtype{Article}%
\bibitem{Rodas:2023nec}
\bibinfo{author}{Arkaitz Rodas}, \bibinfo{author}{Jozef~J. Dudek}, \bibinfo{author}{Robert~G. Edwards} (\bibinfo{collaboration}{Hadron Spectrum}), \bibinfo{title}{{Determination of crossing-symmetric {\ensuremath{\pi}}{\ensuremath{\pi}} scattering amplitudes and the quark mass evolution of the {\ensuremath{\sigma}} constrained by lattice QCD}}, \bibinfo{journal}{Phys. Rev. D} \bibinfo{volume}{109} (\bibinfo{number}{3}) (\bibinfo{year}{2024}) \bibinfo{pages}{034513}, \bibinfo{doi}{\doi{10.1103/PhysRevD.109.034513}}, \eprint{2304.03762}.

\bibtype{Article}%
\bibitem{Cao:2024zuy}
\bibinfo{author}{Xiong-Hui Cao}, \bibinfo{author}{Feng-Kun Guo}, \bibinfo{author}{Zhi-Hui Guo}, \bibinfo{author}{Qu-Zhi Li}, \bibinfo{title}{{Rigorous Roy-Steiner equation analysis of {\ensuremath{\pi}}K scattering at unphysical quark masses}}, \bibinfo{journal}{Phys. Rev. D} \bibinfo{volume}{112} (\bibinfo{number}{3}) (\bibinfo{year}{2025}) \bibinfo{pages}{L031503}, \bibinfo{doi}{\doi{10.1103/38b9-m57d}}, \eprint{2412.03374}.

\bibtype{Article}%
\bibitem{Cao:2025hqm}
\bibinfo{author}{Xiong-Hui Cao}, \bibinfo{author}{Feng-Kun Guo}, \bibinfo{author}{Zhi-Hui Guo}, \bibinfo{author}{Qu-Zhi Li}, \bibinfo{title}{{Revisiting Roy-Steiner-equation analysis of pion-kaon scattering from lattice QCD data}}, \bibinfo{journal}{Phys. Rev. D} \bibinfo{volume}{112} (\bibinfo{number}{3}) (\bibinfo{year}{2025}) \bibinfo{pages}{034042}, \bibinfo{doi}{\doi{10.1103/hn8j-95vn}}, \eprint{2506.10619}.

\bibtype{Article}%
\bibitem{Braghin:2020yri}
\bibinfo{author}{Fabio~L. Braghin}, \bibinfo{title}{{Flavor-dependent U(3) Nambu{\textendash}Jona-Lasinio coupling constant}}, \bibinfo{journal}{Phys. Rev. D} \bibinfo{volume}{103} (\bibinfo{number}{9}) (\bibinfo{year}{2021}) \bibinfo{pages}{094028}, \bibinfo{doi}{\doi{10.1103/PhysRevD.103.094028}}, \eprint{2008.00346}.

\bibtype{Article}%
\bibitem{Braghin:2022uih}
\bibinfo{author}{Fabio~L. Braghin}, \bibinfo{title}{{Quark-antiquark states of the lightest scalar mesons within the Nambu-Jona-Lasinio model with flavor-dependent coupling constants}}, \bibinfo{journal}{J. Phys. G} \bibinfo{volume}{50} (\bibinfo{number}{9}) (\bibinfo{year}{2023}) \bibinfo{pages}{095101}, \bibinfo{doi}{\doi{10.1088/1361-6471/acdaea}}, \eprint{2212.06616}.

\bibtype{Article}%
\bibitem{Jaffe:1976ig}
\bibinfo{author}{R.~L. Jaffe}, \bibinfo{title}{{Multi-Quark Hadrons. 1. The Phenomenology of (2 Quark 2 anti-Quark) Mesons}}, \bibinfo{journal}{Phys. Rev. D} \bibinfo{volume}{15} (\bibinfo{year}{1977}) \bibinfo{pages}{267}, \bibinfo{doi}{\doi{10.1103/PhysRevD.15.267}}.

\bibtype{Article}%
\bibitem{vanBeveren:1986ea}
\bibinfo{author}{E. van Beveren}, \bibinfo{author}{T.~A. Rijken}, \bibinfo{author}{K. Metzger}, \bibinfo{author}{C. Dullemond}, \bibinfo{author}{G. Rupp}, \bibinfo{author}{J.~E. Ribeiro}, \bibinfo{title}{{A Low Lying Scalar Meson Nonet in a Unitarized Meson Model}}, \bibinfo{journal}{Z. Phys. C} \bibinfo{volume}{30} (\bibinfo{year}{1986}) \bibinfo{pages}{615--620}, \bibinfo{doi}{\doi{10.1007/BF01571811}}, \eprint{0710.4067}.

\bibtype{Article}%
\bibitem{vanBeveren:2006ua}
\bibinfo{author}{E. van Beveren}, \bibinfo{author}{D.~V. Bugg}, \bibinfo{author}{F. Kleefeld}, \bibinfo{author}{G. Rupp}, \bibinfo{title}{{The Nature of sigma, kappa, a(0)(980) and f(0)(980)}}, \bibinfo{journal}{Phys. Lett. B} \bibinfo{volume}{641} (\bibinfo{year}{2006}) \bibinfo{pages}{265--271}, \bibinfo{doi}{\doi{10.1016/j.physletb.2006.08.051}}, \eprint{hep-ph/0606022}.

\bibtype{Article}%
\bibitem{Black:1998zc}
\bibinfo{author}{D. Black}, \bibinfo{author}{A.~H. Fariborz}, \bibinfo{author}{F. Sannino}, \bibinfo{author}{J. Schechter}, \bibinfo{title}{{Evidence for a scalar kappa(900) resonance in pi K scattering}}, \bibinfo{journal}{Phys. Rev. D} \bibinfo{volume}{58} (\bibinfo{year}{1998}) \bibinfo{pages}{054012}, \bibinfo{doi}{\doi{10.1103/PhysRevD.58.054012}}, \eprint{hep-ph/9804273}.

\bibtype{Article}%
\bibitem{Black:1998wt}
\bibinfo{author}{D. Black}, \bibinfo{author}{A.~H. Fariborz}, \bibinfo{author}{F. Sannino}, \bibinfo{author}{J. Schechter}, \bibinfo{title}{{Putative light scalar nonet}}, \bibinfo{journal}{Phys. Rev. D} \bibinfo{volume}{59} (\bibinfo{year}{1999}) \bibinfo{pages}{074026}, \bibinfo{doi}{\doi{10.1103/PhysRevD.59.074026}}, \eprint{hep-ph/9808415}.

\bibtype{Article}%
\bibitem{Black:2000qq}
\bibinfo{author}{D. Black}, \bibinfo{author}{A.~H. Fariborz}, \bibinfo{author}{S. Moussa}, \bibinfo{author}{S. Nasri}, \bibinfo{author}{J. Schechter}, \bibinfo{title}{{Unitarized pseudoscalar meson scattering amplitudes in three flavor linear sigma models}}, \bibinfo{journal}{Phys. Rev. D} \bibinfo{volume}{64} (\bibinfo{year}{2001}) \bibinfo{pages}{014031}, \bibinfo{doi}{\doi{10.1103/PhysRevD.64.014031}}, \eprint{hep-ph/0012278}.

\bibtype{Article}%
\bibitem{Oller:2003vf}
\bibinfo{author}{Jose~A. Oller}, \bibinfo{title}{{The Mixing angle of the lightest scalar nonet}}, \bibinfo{journal}{Nucl. Phys. A} \bibinfo{volume}{727} (\bibinfo{year}{2003}) \bibinfo{pages}{353--369}, \bibinfo{doi}{\doi{10.1016/j.nuclphysa.2003.08.002}}, \eprint{hep-ph/0306031}.

\bibtype{Article}%
\bibitem{GomezNicola:2001as}
\bibinfo{author}{A. Gomez~Nicola}, \bibinfo{author}{J.~R. Pelaez}, \bibinfo{title}{{Meson meson scattering within one loop chiral perturbation theory and its unitarization}}, \bibinfo{journal}{Phys. Rev. D} \bibinfo{volume}{65} (\bibinfo{year}{2002}) \bibinfo{pages}{054009}, \bibinfo{doi}{\doi{10.1103/PhysRevD.65.054009}}, \eprint{hep-ph/0109056}.

\bibtype{Article}%
\bibitem{Pelaez:2004xp}
\bibinfo{author}{J.~R. Pelaez}, \bibinfo{title}{{Light scalars as tetraquarks or two-meson states from large N(c) and unitarized chiral perturbation theory}}, \bibinfo{journal}{Mod. Phys. Lett. A} \bibinfo{volume}{19} (\bibinfo{year}{2004}) \bibinfo{pages}{2879--2894}, \bibinfo{doi}{\doi{10.1142/S0217732304016160}}, \eprint{hep-ph/0411107}.

\bibtype{Article}%
\bibitem{Pelaez:2003dy}
\bibinfo{author}{J.~R. Pelaez}, \bibinfo{title}{{On the Nature of light scalar mesons from their large N(c) behavior}}, \bibinfo{journal}{Phys. Rev. Lett.} \bibinfo{volume}{92} (\bibinfo{year}{2004}) \bibinfo{pages}{102001}, \bibinfo{doi}{\doi{10.1103/PhysRevLett.92.102001}}, \eprint{hep-ph/0309292}.

\bibtype{Article}%
\bibitem{Pelaez:2006nj}
\bibinfo{author}{J.~R. Pelaez}, \bibinfo{author}{G. Rios}, \bibinfo{title}{{Nature of the f0(600) from its N(c) dependence at two loops in unitarized Chiral Perturbation Theory}}, \bibinfo{journal}{Phys. Rev. Lett.} \bibinfo{volume}{97} (\bibinfo{year}{2006}) \bibinfo{pages}{242002}, \bibinfo{doi}{\doi{10.1103/PhysRevLett.97.242002}}, \eprint{hep-ph/0610397}.

\bibtype{Article}%
\bibitem{tHooft:1973alw}
\bibinfo{author}{Gerard 't Hooft}, \bibinfo{title}{{A Planar Diagram Theory for Strong Interactions}}, \bibinfo{journal}{Nucl. Phys. B} \bibinfo{volume}{72} (\bibinfo{year}{1974}) \bibinfo{pages}{461}, \bibinfo{doi}{\doi{10.1016/0550-3213(74)90154-0}}.

\bibtype{Article}%
\bibitem{Witten:1980sp}
\bibinfo{author}{E. Witten}, \bibinfo{title}{{Large N Chiral Dynamics}}, \bibinfo{journal}{Annals Phys.} \bibinfo{volume}{128} (\bibinfo{year}{1980}) \bibinfo{pages}{363}, \bibinfo{doi}{\doi{10.1016/0003-4916(80)90325-5}}.

\bibtype{Article}%
\bibitem{Jaffe:2007id}
\bibinfo{author}{R.~L. Jaffe}, \bibinfo{title}{{Ordinary and extraordinary hadrons}}, \bibinfo{journal}{AIP Conf. Proc.} \bibinfo{volume}{964} (\bibinfo{number}{1}) (\bibinfo{year}{2007}) \bibinfo{pages}{1--13}, \bibinfo{doi}{\doi{10.1063/1.2823850}}, \eprint{hep-ph/0701038}.

\bibtype{Article}%
\bibitem{Weinberg:2013cfa}
\bibinfo{author}{S. Weinberg}, \bibinfo{title}{{Tetraquark Mesons in Large $N$ Quantum Chromodynamics}}, \bibinfo{journal}{Phys. Rev. Lett.} \bibinfo{volume}{110} (\bibinfo{year}{2013}) \bibinfo{pages}{261601}, \bibinfo{doi}{\doi{10.1103/PhysRevLett.110.261601}}, \eprint{1303.0342}.

\bibtype{Article}%
\bibitem{Knecht:2013yqa}
\bibinfo{author}{M. Knecht}, \bibinfo{author}{S. Peris}, \bibinfo{title}{{Narrow Tetraquarks at Large N}}, \bibinfo{journal}{Phys. Rev. D} \bibinfo{volume}{88} (\bibinfo{year}{2013}) \bibinfo{pages}{036016}, \bibinfo{doi}{\doi{10.1103/PhysRevD.88.036016}}, \eprint{1307.1273}.

\bibtype{Article}%
\bibitem{Harada:2003em}
\bibinfo{author}{M. Harada}, \bibinfo{author}{F. Sannino}, \bibinfo{author}{J. Schechter}, \bibinfo{title}{{Large N(c) and chiral dynamics}}, \bibinfo{journal}{Phys. Rev. D} \bibinfo{volume}{69} (\bibinfo{year}{2004}) \bibinfo{pages}{034005}, \bibinfo{doi}{\doi{10.1103/PhysRevD.69.034005}}, \eprint{hep-ph/0309206}.

\bibtype{Article}%
\bibitem{RuizdeElvira:2010cs}
\bibinfo{author}{J.~R. Pelaez}, \bibinfo{author}{M.~R. Pennington}, \bibinfo{author}{D.~J. Wilson}, \bibinfo{author}{J. Ruiz~de Elvira}, \bibinfo{title}{{Chiral Perturbation Theory, the ${1/N_c}$ expansion and Regge behaviour determine the structure of the lightest scalar meson}}, \bibinfo{journal}{Phys. Rev. D} \bibinfo{volume}{84} (\bibinfo{year}{2011}) \bibinfo{pages}{096006}, \bibinfo{doi}{\doi{10.1103/PhysRevD.84.096006}}, \eprint{1009.6204}.

\bibtype{Article}%
\bibitem{Maiani:2004uc}
\bibinfo{author}{L. Maiani}, \bibinfo{author}{F. Piccinini}, \bibinfo{author}{A.~D. Polosa}, \bibinfo{author}{V. Riquer}, \bibinfo{title}{{A New look at scalar mesons}}, \bibinfo{journal}{Phys. Rev. Lett.} \bibinfo{volume}{93} (\bibinfo{year}{2004}) \bibinfo{pages}{212002}, \bibinfo{doi}{\doi{10.1103/PhysRevLett.93.212002}}, \eprint{hep-ph/0407017}.

\bibtype{Article}%
\bibitem{tHooft:2008rus}
\bibinfo{author}{G. 't Hooft}, \bibinfo{author}{G. Isidori}, \bibinfo{author}{L. Maiani}, \bibinfo{author}{A.~D. Polosa}, \bibinfo{author}{V. Riquer}, \bibinfo{title}{{A Theory of Scalar Mesons}}, \bibinfo{journal}{Phys. Lett. B} \bibinfo{volume}{662} (\bibinfo{year}{2008}) \bibinfo{pages}{424--430}, \bibinfo{doi}{\doi{10.1016/j.physletb.2008.03.036}}, \eprint{0801.2288}.

\bibtype{Article}%
\bibitem{Heupel:2012ua}
\bibinfo{author}{Walter Heupel}, \bibinfo{author}{Gernot Eichmann}, \bibinfo{author}{Christian~S. Fischer}, \bibinfo{title}{{Tetraquark Bound States in a Bethe-Salpeter Approach}}, \bibinfo{journal}{Phys. Lett. B} \bibinfo{volume}{718} (\bibinfo{year}{2012}) \bibinfo{pages}{545--549}, \bibinfo{doi}{\doi{10.1016/j.physletb.2012.11.009}}, \eprint{1206.5129}.

\bibtype{Article}%
\bibitem{Eichmann:2015cra}
\bibinfo{author}{Gernot Eichmann}, \bibinfo{author}{Christian~S. Fischer}, \bibinfo{author}{Walter Heupel}, \bibinfo{title}{{The light scalar mesons as tetraquarks}}, \bibinfo{journal}{Phys. Lett. B} \bibinfo{volume}{753} (\bibinfo{year}{2016}) \bibinfo{pages}{282--287}, \bibinfo{doi}{\doi{10.1016/j.physletb.2015.12.036}}, \eprint{1508.07178}.

\bibtype{Article}%
\bibitem{Anisovich:2000kxa}
\bibinfo{author}{A.~V. Anisovich}, \bibinfo{author}{V.~V. Anisovich}, \bibinfo{author}{A.~V. Sarantsev}, \bibinfo{title}{{Systematics of q anti-q states in the (n, M**2) and (J, M**2) planes}}, \bibinfo{journal}{Phys. Rev. D} \bibinfo{volume}{62} (\bibinfo{year}{2000}) \bibinfo{pages}{051502}, \bibinfo{doi}{\doi{10.1103/PhysRevD.62.051502}}, \eprint{hep-ph/0003113}.

\bibtype{Article}%
\bibitem{Londergan:2013dza}
\bibinfo{author}{J.~T. Londergan}, \bibinfo{author}{J. Nebreda}, \bibinfo{author}{J.~R. Pelaez}, \bibinfo{author}{A. Szczepaniak}, \bibinfo{title}{{Identification of non-ordinary mesons from the dispersive connection between their poles and their Regge trajectories: The $f_0(500)$ resonance}}, \bibinfo{journal}{Phys. Lett. B} \bibinfo{volume}{729} (\bibinfo{year}{2014}) \bibinfo{pages}{9--14}, \bibinfo{doi}{\doi{10.1016/j.physletb.2013.12.061}}, \eprint{1311.7552}.

\bibtype{Article}%
\bibitem{Pelaez:2017sit}
\bibinfo{author}{J.~R. Pelaez}, \bibinfo{author}{A. Rodas}, \bibinfo{title}{{The non-ordinary Regge behavior of the $K^*_0(800)$ or $\kappa $ -meson versus the ordinary $K^*_0(1430)$}}, \bibinfo{journal}{Eur. Phys. J. C} \bibinfo{volume}{77} (\bibinfo{number}{6}) (\bibinfo{year}{2017}) \bibinfo{pages}{431}, \bibinfo{doi}{\doi{10.1140/epjc/s10052-017-4994-3}}, \eprint{1703.07661}.

\bibtype{Article}%
\bibitem{Carrasco:2015fva}
\bibinfo{author}{J.~A. Carrasco}, \bibinfo{author}{J. Nebreda}, \bibinfo{author}{J.~R. Pelaez}, \bibinfo{author}{A.~P. Szczepaniak}, \bibinfo{title}{{Dispersive calculation of complex Regge trajectories for the lightest $f_2$ resonances and the $K^*(892)$ }}, \bibinfo{journal}{Phys. Lett. B} \bibinfo{volume}{749} (\bibinfo{year}{2015}) \bibinfo{pages}{399--406}, \bibinfo{doi}{\doi{10.1016/j.physletb.2015.08.019}}, \eprint{1504.03248}.

\bibtype{Article}%
\bibitem{Amsler:1997up}
\bibinfo{author}{C. Amsler}, \bibinfo{title}{{Proton - anti-proton annihilation and meson spectroscopy with the crystal barrel}}, \bibinfo{journal}{Rev. Mod. Phys.} \bibinfo{volume}{70} (\bibinfo{year}{1998}) \bibinfo{pages}{1293--1340}, \bibinfo{doi}{\doi{10.1103/RevModPhys.70.1293}}, \eprint{hep-ex/9708025}.

\bibtype{Article}%
\bibitem{Amsler:1995td}
\bibinfo{author}{C. Amsler}, \bibinfo{author}{F.~E. Close}, \bibinfo{title}{{Is f0 (1500) a scalar glueball?}}, \bibinfo{journal}{Phys. Rev. D} \bibinfo{volume}{53} (\bibinfo{year}{1996}) \bibinfo{pages}{295--311}, \bibinfo{doi}{\doi{10.1103/PhysRevD.53.295}}, \eprint{hep-ph/9507326}.

\bibtype{Article}%
\bibitem{Amsler:1995tu}
\bibinfo{author}{C. Amsler}, \bibinfo{author}{F.~E. Close}, \bibinfo{title}{{Evidence for a scalar glueball}}, \bibinfo{journal}{Phys. Lett. B} \bibinfo{volume}{353} (\bibinfo{year}{1995}) \bibinfo{pages}{385--390}, \bibinfo{doi}{\doi{10.1016/0370-2693(95)00579-A}}, \eprint{hep-ph/9505219}.

\bibtype{Article}%
\bibitem{Etkin:1982se}
\bibinfo{author}{A. Etkin}, et al., \bibinfo{title}{{Evidence for Two New 0++ Mesons and a Possible Scalar Decuplet}}, \bibinfo{journal}{Phys. Rev. D} \bibinfo{volume}{25} (\bibinfo{year}{1982}) \bibinfo{pages}{2446}, \bibinfo{doi}{\doi{10.1103/PhysRevD.25.2446}}.

\bibtype{Article}%
\bibitem{Hyams:1973zf}
\bibinfo{author}{B. Hyams}, et al., \bibinfo{title}{{$\pi\pi$ Phase Shift Analysis from 600-MeV to 1900-MeV}}, \bibinfo{journal}{Nucl. Phys.} \bibinfo{volume}{B64} (\bibinfo{year}{1973}) \bibinfo{pages}{134--162}, \bibinfo{doi}{\doi{10.1016/0550-3213(73)90618-4}}.

\bibtype{Article}%
\bibitem{Hyams:1975mc}
\bibinfo{author}{B. Hyams}, et al., \bibinfo{title}{{A Study of All the pi pi Phase Shift Solutions in the Mass Region 1.0-GeV to 1.8-GeV from pi- p --> pi- pi+ n at 17.2-GeV}}, \bibinfo{journal}{Nucl. Phys.} \bibinfo{volume}{B100} (\bibinfo{year}{1975}) \bibinfo{pages}{205--224}, \bibinfo{doi}{\doi{10.1016/0550-3213(75)90616-1}}.

\bibtype{Article}%
\bibitem{Pelaez:2022qby}
\bibinfo{author}{J.~R. Pelaez}, \bibinfo{author}{A. Rodas}, \bibinfo{author}{Jacobo. Ruiz~de Elvira}, \bibinfo{title}{{f0(1370) Controversy from Dispersive Meson-Meson Scattering Data Analyses}}, \bibinfo{journal}{Phys. Rev. Lett.} \bibinfo{volume}{130} (\bibinfo{number}{5}) (\bibinfo{year}{2023}) \bibinfo{pages}{051902}, \bibinfo{doi}{\doi{10.1103/PhysRevLett.130.051902}}, \bibinfo{note}{[Erratum: Phys.Rev.Lett. 132, 239901 (2024)]}, \eprint{2206.14822}.

\bibtype{Article}%
\bibitem{Black:1999yz}
\bibinfo{author}{D. Black}, \bibinfo{author}{A.~H. Fariborz}, \bibinfo{author}{J. Schechter}, \bibinfo{title}{{Mechanism for a next-to-lowest lying scalar meson nonet}}, \bibinfo{journal}{Phys. Rev. D} \bibinfo{volume}{61} (\bibinfo{year}{2000}) \bibinfo{pages}{074001}, \bibinfo{doi}{\doi{10.1103/PhysRevD.61.074001}}, \eprint{hep-ph/9907516}.

\bibtype{Article}%
\bibitem{Brett:2019tzr}
\bibinfo{author}{R. Brett}, \bibinfo{author}{J. Bulava}, \bibinfo{author}{D. Darvish}, \bibinfo{author}{J. Fallica}, \bibinfo{author}{A. Hanlon}, \bibinfo{author}{B. H{\"o}rz}, \bibinfo{author}{C. Morningstar}, \bibinfo{title}{{Spectroscopy From The Lattice: The Scalar Glueball}}, \bibinfo{journal}{AIP Conf. Proc.} \bibinfo{volume}{2249} (\bibinfo{number}{1}) (\bibinfo{year}{2020}) \bibinfo{pages}{030032}, \bibinfo{doi}{\doi{10.1063/5.0008566}}, \eprint{1909.07306}.

\bibtype{Article}%
\bibitem{Morningstar:2024vjk}
\bibinfo{author}{C. Morningstar}, \bibinfo{title}{{Update on Glueballs}}, \bibinfo{journal}{PoS} \bibinfo{volume}{LATTICE2024} (\bibinfo{year}{2024}) \bibinfo{pages}{004}, \bibinfo{doi}{\doi{10.22323/1.466.0004}}, \eprint{2502.02547}.

\bibtype{Article}%
\bibitem{Molina:2008jw}
\bibinfo{author}{R. Molina}, \bibinfo{author}{D. Nicmorus}, \bibinfo{author}{E. Oset}, \bibinfo{title}{{The rho rho interaction in the hidden gauge formalism and the f(0)(1370) and f(2)(1270) resonances}}, \bibinfo{journal}{Phys. Rev. D} \bibinfo{volume}{78} (\bibinfo{year}{2008}) \bibinfo{pages}{114018}, \bibinfo{doi}{\doi{10.1103/PhysRevD.78.114018}}, \eprint{0809.2233}.

\bibtype{Article}%
\bibitem{Geng:2008gx}
\bibinfo{author}{L.~S. Geng}, \bibinfo{author}{E. Oset}, \bibinfo{title}{{Vector meson-vector meson interaction in a hidden gauge unitary approach}}, \bibinfo{journal}{Phys. Rev. D} \bibinfo{volume}{79} (\bibinfo{year}{2009}) \bibinfo{pages}{074009}, \bibinfo{doi}{\doi{10.1103/PhysRevD.79.074009}}, \eprint{0812.1199}.

\bibtype{Article}%
\bibitem{Albaladejo:2008qa}
\bibinfo{author}{M. Albaladejo}, \bibinfo{author}{J.~A. Oller}, \bibinfo{title}{{Identification of a Scalar Glueball}}, \bibinfo{journal}{Phys. Rev. Lett.} \bibinfo{volume}{101} (\bibinfo{year}{2008}) \bibinfo{pages}{252002}, \bibinfo{doi}{\doi{10.1103/PhysRevLett.101.252002}}, \eprint{0801.4929}.

\bibtype{Article}%
\bibitem{Chanowitz:2005du}
\bibinfo{author}{M. Chanowitz}, \bibinfo{title}{{Chiral suppression of scalar glueball decay}}, \bibinfo{journal}{Phys. Rev. Lett.} \bibinfo{volume}{95} (\bibinfo{year}{2005}) \bibinfo{pages}{172001}, \bibinfo{doi}{\doi{10.1103/PhysRevLett.95.172001}}, \eprint{hep-ph/0506125}.

\bibtype{Article}%
\bibitem{Chanowitz:2007ma}
\bibinfo{author}{M.~S. Chanowitz}, \bibinfo{title}{{Reply to `Comment on `Chiral suppression of scalar glueball decay''}}, \bibinfo{journal}{Phys. Rev. Lett.} \bibinfo{volume}{98} (\bibinfo{year}{2007}) \bibinfo{pages}{149104}, \bibinfo{doi}{\doi{10.1103/PhysRevLett.98.149104}}, \eprint{0704.1616}.

\bibtype{Article}%
\bibitem{Sarantsev:2021ein}
\bibinfo{author}{A.~V. Sarantsev}, \bibinfo{author}{I. Denisenko}, \bibinfo{author}{U. Thoma}, \bibinfo{author}{E. Klempt}, \bibinfo{title}{{Scalar isoscalar mesons and the scalar glueball from radiative $J/\psi$ decays}}, \bibinfo{journal}{Phys. Lett. B} \bibinfo{volume}{816} (\bibinfo{year}{2021}) \bibinfo{pages}{136227}, \bibinfo{doi}{\doi{10.1016/j.physletb.2021.136227}}, \eprint{2103.09680}.

\bibtype{Article}%
\bibitem{Klempt:2021wpg}
\bibinfo{author}{E. Klempt}, \bibinfo{author}{A.~V. Sarantsev}, \bibinfo{title}{{Singlet-octet-glueball mixing of scalar mesons}}, \bibinfo{journal}{Phys. Lett. B} \bibinfo{volume}{826} (\bibinfo{year}{2022}) \bibinfo{pages}{136906}, \bibinfo{doi}{\doi{10.1016/j.physletb.2022.136906}}, \eprint{2112.04348}.

\bibtype{Article}%
\bibitem{Gui:2012gx}
\bibinfo{author}{L.-C. Gui}, \bibinfo{author}{Y. Chen}, \bibinfo{author}{G. Li}, \bibinfo{author}{C. Liu}, \bibinfo{author}{Y.-B. Liu}, \bibinfo{author}{J.-P. Ma}, \bibinfo{author}{Y.-B. Yang}, \bibinfo{author}{J.-B. Zhang} (\bibinfo{collaboration}{CLQCD}), \bibinfo{title}{{Scalar Glueball in Radiative $J/\psi$ Decay on the Lattice}}, \bibinfo{journal}{Phys. Rev. Lett.} \bibinfo{volume}{110} (\bibinfo{number}{2}) (\bibinfo{year}{2013}) \bibinfo{pages}{021601}, \bibinfo{doi}{\doi{10.1103/PhysRevLett.110.021601}}, \eprint{1206.0125}.

\bibtype{Article}%
\bibitem{Novikov:1979uy}
\bibinfo{author}{V.~A. Novikov}, \bibinfo{author}{Mikhail~A. Shifman}, \bibinfo{author}{A.~I. Vainshtein}, \bibinfo{author}{Valentin~I. Zakharov}, \bibinfo{title}{{A Theory of the J/psi ---{\ensuremath{>}} eta (eta-prime) gamma Decays}}, \bibinfo{journal}{Nucl. Phys. B} \bibinfo{volume}{165} (\bibinfo{year}{1980}) \bibinfo{pages}{55--66}, \bibinfo{doi}{\doi{10.1016/0550-3213(80)90305-3}}.

\bibtype{Article}%
\bibitem{Gershtein:1983kc}
\bibinfo{author}{S.~S. Gershtein}, \bibinfo{author}{A.~K. Likhoded}, \bibinfo{author}{Yu.~D. Prokoshkin}, \bibinfo{title}{{G (1590) Meson and Possible Characteristic Features of a Glueball}}, \bibinfo{journal}{Z. Phys. C} \bibinfo{volume}{24} (\bibinfo{year}{1984}) \bibinfo{pages}{305}, \bibinfo{doi}{\doi{10.1007/BF01410369}}.

\bibtype{Article}%
\bibitem{Ball:1995zv}
\bibinfo{author}{Patricia Ball}, \bibinfo{author}{J.~M. Frere}, \bibinfo{author}{M. Tytgat}, \bibinfo{title}{{Phenomenological evidence for the gluon content of eta and eta-prime}}, \bibinfo{journal}{Phys. Lett. B} \bibinfo{volume}{365} (\bibinfo{year}{1996}) \bibinfo{pages}{367--376}, \bibinfo{doi}{\doi{10.1016/0370-2693(95)01287-7}}, \eprint{hep-ph/9508359}.

\bibtype{Article}%
\bibitem{Frere:2015xxa}
\bibinfo{author}{Jean-Marie Fr{\`e}re}, \bibinfo{author}{Julian Heeck}, \bibinfo{title}{{Scalar glueballs: Constraints from the decays into $\eta$ or $\eta'$}}, \bibinfo{journal}{Phys. Rev. D} \bibinfo{volume}{92} (\bibinfo{number}{11}) (\bibinfo{year}{2015}) \bibinfo{pages}{114035}, \bibinfo{doi}{\doi{10.1103/PhysRevD.92.114035}}, \eprint{1506.04766}.

\bibtype{Article}%
\bibitem{Rodas:2021tyb}
\bibinfo{author}{A. Rodas}, \bibinfo{author}{A. Pilloni}, \bibinfo{author}{M. Albaladejo}, \bibinfo{author}{C. Fernandez-Ramirez}, \bibinfo{author}{V. Mathieu}, \bibinfo{author}{A.~P. Szczepaniak} (\bibinfo{collaboration}{Joint Physics Analysis Center}), \bibinfo{title}{{Scalar and tensor resonances in $J/\psi $ radiative decays}}, \bibinfo{journal}{Eur. Phys. J. C} \bibinfo{volume}{82} (\bibinfo{number}{1}) (\bibinfo{year}{2022}) \bibinfo{pages}{80}, \bibinfo{doi}{\doi{10.1140/epjc/s10052-022-10014-8}}, \eprint{2110.00027}.

\bibtype{Article}%
\bibitem{BESIII:2015rug}
\bibinfo{author}{M. Ablikim}, et al. (\bibinfo{collaboration}{BESIII}), \bibinfo{title}{{Amplitude analysis of the $\pi^{0}\pi^{0}$~system produced in radiative $J/\psi$~decays}}, \bibinfo{journal}{Phys. Rev. D} \bibinfo{volume}{92} (\bibinfo{number}{5}) (\bibinfo{year}{2015}) \bibinfo{pages}{052003}, \bibinfo{doi}{\doi{10.1103/PhysRevD.92.052003}}, \bibinfo{note}{[Erratum: Phys.Rev.D 93, 039906 (2016)]}, \eprint{1506.00546}.

\bibtype{Article}%
\bibitem{BESIII:2018ubj}
\bibinfo{author}{M. Ablikim}, et al. (\bibinfo{collaboration}{BESIII}), \bibinfo{title}{{Amplitude analysis of the $K_{S}K_{S}$ system produced in radiative $J/\psi$ decays}}, \bibinfo{journal}{Phys. Rev. D} \bibinfo{volume}{98} (\bibinfo{number}{7}) (\bibinfo{year}{2018}) \bibinfo{pages}{072003}, \bibinfo{doi}{\doi{10.1103/PhysRevD.98.072003}}, \eprint{1808.06946}.

\bibtype{Article}%
\bibitem{BaBar:2021fkz}
\bibinfo{author}{J.~P. Lees}, et al. (\bibinfo{collaboration}{BaBar}), \bibinfo{title}{{Light meson spectroscopy from Dalitz plot analyses of $\eta_c$ decays to $\eta' K^+ K^-$, $\eta' \pi^+ \pi^-$, and $\eta \pi^+ \pi^-$ produced in two-photon interactions}}, \bibinfo{journal}{Phys. Rev. D} \bibinfo{volume}{104} (\bibinfo{number}{7}) (\bibinfo{year}{2021}) \bibinfo{pages}{072002}, \bibinfo{doi}{\doi{10.1103/PhysRevD.104.072002}}, \eprint{2106.05157}.

\bibtype{Article}%
\bibitem{BESIII:2022npc}
\bibinfo{author}{M. Ablikim}, et al. (\bibinfo{collaboration}{BESIII}), \bibinfo{title}{{Observation of an a0-like State with Mass of 1.817~GeV in the Study of Ds+{\textrightarrow}KS0K+{\ensuremath{\pi}}0 Decays}}, \bibinfo{journal}{Phys. Rev. Lett.} \bibinfo{volume}{129} (\bibinfo{number}{18}) (\bibinfo{year}{2022}) \bibinfo{pages}{182001}, \bibinfo{doi}{\doi{10.1103/PhysRevLett.129.182001}}, \eprint{2204.09614}.

\bibtype{Article}%
\bibitem{Zhu:2022wzk}
\bibinfo{author}{X. Zhu}, \bibinfo{author}{D.-M. Li}, \bibinfo{author}{E. Wang}, \bibinfo{author}{L.-S. Geng}, \bibinfo{author}{J.-J. Xie}, \bibinfo{title}{{Theoretical study of the process Ds+{\textrightarrow}{\ensuremath{\pi}}+KS0KS0 and the isovector partner of f0(1710)}}, \bibinfo{journal}{Phys. Rev. D} \bibinfo{volume}{105} (\bibinfo{number}{11}) (\bibinfo{year}{2022}) \bibinfo{pages}{116010}, \bibinfo{doi}{\doi{10.1103/PhysRevD.105.116010}}, \eprint{2204.09384}.

\bibtype{Article}%
\bibitem{Oset:2023hyt}
\bibinfo{author}{Eulogio Oset}, \bibinfo{author}{Lian-Rong Dai}, \bibinfo{author}{Li-Sheng Geng}, \bibinfo{title}{{Repercussion of the a0(1710) [a0(1817)] resonance and future developments}}, \bibinfo{journal}{Sci. Bull.} \bibinfo{volume}{68} (\bibinfo{year}{2023}) \bibinfo{pages}{243--246}, \bibinfo{doi}{\doi{10.1016/j.scib.2023.01.011}}, \eprint{2301.08532}.

\bibtype{Article}%
\bibitem{Nambu:1957wzj}
\bibinfo{author}{Y. Nambu}, \bibinfo{title}{{Possible existence of a heavy neutral meson}}, \bibinfo{journal}{Phys. Rev.} \bibinfo{volume}{106} (\bibinfo{year}{1957}) \bibinfo{pages}{1366--1367}, \bibinfo{doi}{\doi{10.1103/PhysRev.106.1366}}.

\bibtype{Article}%
\bibitem{Maglich:1961rtx}
\bibinfo{author}{B. Maglich}, \bibinfo{author}{L.~W. Alvarez}, \bibinfo{author}{A.~H. Rosenfeld}, \bibinfo{author}{M.~L. Stevenson}, \bibinfo{title}{{Evidence for a $T=0$ Three Pion Resonance}}, \bibinfo{journal}{Phys. Rev. Lett.} \bibinfo{volume}{7} (\bibinfo{year}{1961}) \bibinfo{pages}{178--182}, \bibinfo{doi}{\doi{10.1103/PhysRevLett.7.178}}.

\bibtype{Article}%
\bibitem{Frazer:1959gy}
\bibinfo{author}{W.~R. Frazer}, \bibinfo{author}{J.~R. Fulco}, \bibinfo{title}{{Effect of a pion pion scattering resonance on nucleon structure}}, \bibinfo{journal}{Phys. Rev. Lett.} \bibinfo{volume}{2} (\bibinfo{year}{1959}) \bibinfo{pages}{365}, \bibinfo{doi}{\doi{10.1103/PhysRevLett.2.365}}.

\bibtype{Article}%
\bibitem{Breit1960}
\bibinfo{author}{G. Breit}, \bibinfo{title}{{The Nucleon-Nucleon Spin-Orbit Potential}}, \bibinfo{journal}{Proc. Nat. Acad. Sci. USA} \bibinfo{volume}{46} (\bibinfo{year}{1960}) \bibinfo{pages}{746--756}.

\bibtype{Article}%
\bibitem{Bowcock1960}
\bibinfo{author}{J.~E. Bowcock}, \bibinfo{author}{W.~N. Cottingham}, \bibinfo{author}{D. Lurie}, \bibinfo{title}{{Effect of a pion-pion scattering resonance on nucleon structure}}, \bibinfo{journal}{Nuovo Cimento} \bibinfo{volume}{16} (\bibinfo{year}{1960}) \bibinfo{pages}{918--938}.

\bibtype{Article}%
\bibitem{Sakurai:1960ju}
\bibinfo{author}{J.~J. Sakurai}, \bibinfo{title}{{Theory of strong interactions}}, \bibinfo{journal}{Annals Phys.} \bibinfo{volume}{11} (\bibinfo{year}{1960}) \bibinfo{pages}{1--48}, \bibinfo{doi}{\doi{10.1016/0003-4916(60)90126-3}}.

\bibtype{Article}%
\bibitem{Bergia:1961zz}
\bibinfo{author}{S. Bergia}, \bibinfo{author}{A. Stanghellini}, \bibinfo{author}{S. Fubini}, \bibinfo{author}{C. Villi}, \bibinfo{title}{{Electromagnetic Form Factors of the Nucleon and Pion-Pion Interaction}}, \bibinfo{journal}{Phys. Rev. Lett.} \bibinfo{volume}{6} (\bibinfo{year}{1961}) \bibinfo{pages}{367--371}, \bibinfo{doi}{\doi{10.1103/PhysRevLett.6.367}}.

\bibtype{Article}%
\bibitem{Derado1960}
\bibinfo{author}{I. Derado}, \bibinfo{title}{{Experimental evidence for the pion-pion interaction at 1 GeV}}, \bibinfo{journal}{Nuovo Cimento} \bibinfo{volume}{15} (\bibinfo{year}{1960}) \bibinfo{pages}{853--5}.

\bibtype{Article}%
\bibitem{Stonehill:1961zz}
\bibinfo{author}{D. Stonehill}, \bibinfo{author}{C. Baltay}, \bibinfo{author}{H. Courant}, \bibinfo{author}{W. Fickinger}, \bibinfo{author}{E.~C. Fowler}, \bibinfo{author}{H. Kraybill}, \bibinfo{author}{J. Sandweiss}, \bibinfo{author}{J. Sanford}, \bibinfo{author}{H. Taft}, \bibinfo{title}{{Pion-Pion Interaction in Pion Production by pi+-p Collisions}}, \bibinfo{journal}{Phys. Rev. Lett.} \bibinfo{volume}{6} (\bibinfo{year}{1961}) \bibinfo{pages}{624--625}, \bibinfo{doi}{\doi{10.1103/PhysRevLett.6.624}}.

\bibtype{Article}%
\bibitem{Erwin:1961ny}
\bibinfo{author}{A.~R. Erwin}, \bibinfo{author}{R. March}, \bibinfo{author}{W.~D. Walker}, \bibinfo{author}{E. West}, \bibinfo{title}{{Evidence for a $\pi \pi$ Resonance in the I = 1, $J=1$ State}}, \bibinfo{journal}{Phys. Rev. Lett.} \bibinfo{volume}{6} (\bibinfo{year}{1961}) \bibinfo{pages}{628--630}, \bibinfo{doi}{\doi{10.1103/PhysRevLett.6.628}}.

\bibtype{Article}%
\bibitem{BESIII:2022wxz}
\bibinfo{author}{M. Ablikim}, et al. (\bibinfo{collaboration}{BESIII}), \bibinfo{title}{{Measurement of $e^{+}e^{-} \to K^{+}K^{-}\pi^{0}$ cross section and observation of a resonant structure}}, \bibinfo{journal}{JHEP} \bibinfo{volume}{07} (\bibinfo{year}{2022}) \bibinfo{pages}{045}, \bibinfo{doi}{\doi{10.1007/JHEP07(2022)045}}, \eprint{2202.06447}.

\bibtype{Article}%
\bibitem{BESIII:2023sbq}
\bibinfo{author}{M. Ablikim}, et al. (\bibinfo{collaboration}{BESIII}), \bibinfo{title}{{Measurement of the cross sections for e+e-{\textrightarrow}{\ensuremath{\eta}}{\ensuremath{\pi}}+{\ensuremath{\pi}}- at center-of-mass energies between 2.00 and 3.08~GeV}}, \bibinfo{journal}{Phys. Rev. D} \bibinfo{volume}{108} (\bibinfo{number}{11}) (\bibinfo{year}{2023}) \bibinfo{pages}{L111101}, \bibinfo{doi}{\doi{10.1103/PhysRevD.108.L111101}}, \eprint{2310.10452}.

\bibtype{Article}%
\bibitem{Sakurai:1972wk}
\bibinfo{author}{J.~J. Sakurai}, \bibinfo{author}{D. Schildknecht}, \bibinfo{title}{{Generalized vector dominance and inelastic electron - proton scattering}}, \bibinfo{journal}{Phys. Lett. B} \bibinfo{volume}{40} (\bibinfo{year}{1972}) \bibinfo{pages}{121--126}, \bibinfo{doi}{\doi{10.1016/0370-2693(72)90300-0}}.

\bibtype{Article}%
\bibitem{Schildknecht:2005xr}
\bibinfo{author}{Dieter Schildknecht}, \bibinfo{title}{{Vector meson dominance}}, \bibinfo{journal}{Acta Phys. Polon. B} \bibinfo{volume}{37} (\bibinfo{year}{2006}) \bibinfo{pages}{595--608}, \eprint{hep-ph/0511090}.

\bibtype{Article}%
\bibitem{Bando:1987br}
\bibinfo{author}{Masako Bando}, \bibinfo{author}{Taichiro Kugo}, \bibinfo{author}{Koichi Yamawaki}, \bibinfo{title}{{Nonlinear Realization and Hidden Local Symmetries}}, \bibinfo{journal}{Phys. Rept.} \bibinfo{volume}{164} (\bibinfo{year}{1988}) \bibinfo{pages}{217--314}, \bibinfo{doi}{\doi{10.1016/0370-1573(88)90019-1}}.

\bibtype{Article}%
\bibitem{Gardner:1997ie}
\bibinfo{author}{S. Gardner}, \bibinfo{author}{H.~B. O'Connell}, \bibinfo{title}{{$\rho - \omega$ mixing and the pion form-factor in the timelike region}}, \bibinfo{journal}{Phys. Rev. D} \bibinfo{volume}{57} (\bibinfo{year}{1998}) \bibinfo{pages}{2716--2726}, \bibinfo{doi}{\doi{10.1103/PhysRevD.57.2716}}, \bibinfo{note}{[Erratum: Phys.Rev.D 62, 019903 (2000)]}, \eprint{hep-ph/9707385}.

\bibtype{Article}%
\bibitem{Barnes:1996ff}
\bibinfo{author}{Ted Barnes}, \bibinfo{author}{F.~E. Close}, \bibinfo{author}{P.~R. Page}, \bibinfo{author}{E.~S. Swanson}, \bibinfo{title}{{Higher quarkonia}}, \bibinfo{journal}{Phys. Rev. D} \bibinfo{volume}{55} (\bibinfo{year}{1997}) \bibinfo{pages}{4157--4188}, \bibinfo{doi}{\doi{10.1103/PhysRevD.55.4157}}, \eprint{hep-ph/9609339}.

\bibtype{Article}%
\bibitem{Close:1997dj}
\bibinfo{author}{Frank~E. Close}, \bibinfo{author}{Philip~R. Page}, \bibinfo{title}{{How to distinguish hybrids from radial quarkonia}}, \bibinfo{journal}{Phys. Rev. D} \bibinfo{volume}{56} (\bibinfo{year}{1997}) \bibinfo{pages}{1584--1588}, \bibinfo{doi}{\doi{10.1103/PhysRevD.56.1584}}, \eprint{hep-ph/9701425}.

\bibtype{Article}%
\bibitem{Ke:2018evd}
\bibinfo{author}{Hong-Wei Ke}, \bibinfo{author}{Xue-Qian Li}, \bibinfo{title}{{Study of the strong decays of $\phi(2170)$ and the future charm-tau factory}}, \bibinfo{journal}{Phys. Rev. D} \bibinfo{volume}{99} (\bibinfo{number}{3}) (\bibinfo{year}{2019}) \bibinfo{pages}{036014}, \bibinfo{doi}{\doi{10.1103/PhysRevD.99.036014}}, \eprint{1810.07912}.

\bibtype{Article}%
\bibitem{Agaev:2019coa}
\bibinfo{author}{S.~S. Agaev}, \bibinfo{author}{K. Azizi}, \bibinfo{author}{H. Sundu}, \bibinfo{title}{{Nature of the vector resonance $Y(2175)$}}, \bibinfo{journal}{Phys. Rev. D} \bibinfo{volume}{101} (\bibinfo{number}{7}) (\bibinfo{year}{2020}) \bibinfo{pages}{074012}, \bibinfo{doi}{\doi{10.1103/PhysRevD.101.074012}}, \eprint{1911.09743}.

\bibtype{Article}%
\bibitem{Yan:2023vbh}
\bibinfo{author}{Mao-Jun Yan}, \bibinfo{author}{Jorgivan~M. Dias}, \bibinfo{author}{Adolfo Guevara}, \bibinfo{author}{Feng-Kun Guo}, \bibinfo{author}{Bing-Song Zou}, \bibinfo{title}{{On the {\ensuremath{\eta}}$_{1}$(1855), {\ensuremath{\pi}}$_{1}$(1400) and {\ensuremath{\pi}}$_{1}$(1600) as Dynamically Generated States and Their SU(3) Partners}}, \bibinfo{journal}{Universe} \bibinfo{volume}{9} (\bibinfo{number}{2}) (\bibinfo{year}{2023}) \bibinfo{pages}{109}, \bibinfo{doi}{\doi{10.3390/universe9020109}}, \eprint{2301.04432}.

\bibtype{Article}%
\bibitem{BESIII:2022riz}
\bibinfo{author}{M. Ablikim}, et al. (\bibinfo{collaboration}{BESIII}), \bibinfo{title}{{Observation of an Isoscalar Resonance with Exotic JPC=1-+ Quantum Numbers in J/{\ensuremath{\psi}}{\textrightarrow}{\ensuremath{\gamma}}{\ensuremath{\eta}}{\ensuremath{\eta}}'}}, \bibinfo{journal}{Phys. Rev. Lett.} \bibinfo{volume}{129} (\bibinfo{number}{19}) (\bibinfo{year}{2022}) \bibinfo{pages}{192002}, \bibinfo{doi}{\doi{10.1103/PhysRevLett.129.192002}}, \bibinfo{note}{[Erratum: Phys.Rev.Lett. 130, 159901 (2023)]}, \eprint{2202.00621}.

\bibtype{Article}%
\bibitem{COMPASS:2018uzl}
\bibinfo{author}{M. Aghasyan}, et al. (\bibinfo{collaboration}{COMPASS}), \bibinfo{title}{{Light isovector resonances in $\pi^- p \to \pi^-\pi^-\pi^+ p$ at 190 GeV/${\it c}$}}, \bibinfo{journal}{Phys. Rev. D} \bibinfo{volume}{98} (\bibinfo{number}{9}) (\bibinfo{year}{2018}) \bibinfo{pages}{092003}, \bibinfo{doi}{\doi{10.1103/PhysRevD.98.092003}}, \eprint{1802.05913}.

\bibtype{Article}%
\bibitem{COMPASS:2021ogp}
\bibinfo{author}{M.~G. Alexeev}, et al. (\bibinfo{collaboration}{COMPASS}), \bibinfo{title}{{Exotic meson $\pi_1(1600)$ with $J^{PC} = 1^{-+}$ and its decay into $\rho(770)\pi$}}, \bibinfo{journal}{Phys. Rev. D} \bibinfo{volume}{105} (\bibinfo{number}{1}) (\bibinfo{year}{2022}) \bibinfo{pages}{012005}, \bibinfo{doi}{\doi{10.1103/PhysRevD.105.012005}}, \eprint{2108.01744}.

\bibtype{Article}%
\bibitem{JPAC:2018zyd}
\bibinfo{author}{A. Rodas}, et al. (\bibinfo{collaboration}{JPAC}), \bibinfo{title}{{Determination of the pole position of the lightest hybrid meson candidate}}, \bibinfo{journal}{Phys. Rev. Lett.} \bibinfo{volume}{122} (\bibinfo{number}{4}) (\bibinfo{year}{2019}) \bibinfo{pages}{042002}, \bibinfo{doi}{\doi{10.1103/PhysRevLett.122.042002}}, \eprint{1810.04171}.

\bibtype{Article}%
\bibitem{COMPASS:2014vkj}
\bibinfo{author}{C. Adolph}, et al. (\bibinfo{collaboration}{COMPASS}), \bibinfo{title}{{Odd and even partial waves of $\eta\pi^-$ and $\eta'\pi^-$ in $\pi^-p\to\eta^{(\prime)}\pi^-p$ at $191\,\textrm{GeV}/c$}}, \bibinfo{journal}{Phys. Lett. B} \bibinfo{volume}{740} (\bibinfo{year}{2015}) \bibinfo{pages}{303--311}, \bibinfo{doi}{\doi{10.1016/j.physletb.2014.11.058}}, \bibinfo{note}{[Erratum: Phys.Lett.B 811, 135913 (2020)]}, \eprint{1408.4286}.

\bibtype{Article}%
\bibitem{Kopf:2020yoa}
\bibinfo{author}{B. Kopf}, \bibinfo{author}{M. Albrecht}, \bibinfo{author}{H. Koch}, \bibinfo{author}{M. K{\"u}{\ss}ner}, \bibinfo{author}{J. Pychy}, \bibinfo{author}{X. Qin}, \bibinfo{author}{U. Wiedner}, \bibinfo{title}{{Investigation of the lightest hybrid meson candidate with a coupled-channel analysis of ${{\bar{p}}p}$-, $\pi ^- p$- and ${\pi \pi }$-Data}}, \bibinfo{journal}{Eur. Phys. J. C} \bibinfo{volume}{81} (\bibinfo{number}{12}) (\bibinfo{year}{2021}) \bibinfo{pages}{1056}, \bibinfo{doi}{\doi{10.1140/epjc/s10052-021-09821-2}}, \eprint{2008.11566}.

\bibtype{Article}%
\bibitem{Frere:1988ac}
\bibinfo{author}{J.~M. Frere}, \bibinfo{author}{S. Titard}, \bibinfo{title}{{A NEW LOOK AT EXOTIC DECAYS rho-tilde (1-+, I = 1) ---{\ensuremath{>}} eta-prime pi versus rho pi}}, \bibinfo{journal}{Phys. Lett. B} \bibinfo{volume}{214} (\bibinfo{year}{1988}) \bibinfo{pages}{463--466}, \bibinfo{doi}{\doi{10.1016/0370-2693(88)91395-0}}.

\bibtype{Article}%
\bibitem{Frere:2025mpf}
\bibinfo{author}{Jean-Marie Frere}, \bibinfo{title}{{Prospects for new glueballs and exotics searches}}  (\bibinfo{year}{2025}), \eprint{2504.21611}.

\bibtype{Article}%
\bibitem{Chen:2022qpd}
\bibinfo{author}{Hua-Xing Chen}, \bibinfo{author}{Niu Su}, \bibinfo{author}{Shi-Lin Zhu}, \bibinfo{title}{{QCD Axial Anomaly Enhances the {\ensuremath{\eta}}{\ensuremath{\eta}}' Decay of the Hybrid Candidate {\ensuremath{\eta}} $_{1}$(1855)}}, \bibinfo{journal}{Chin. Phys. Lett.} \bibinfo{volume}{39} (\bibinfo{number}{5}) (\bibinfo{year}{2022}) \bibinfo{pages}{051201}, \bibinfo{doi}{\doi{10.1088/0256-307X/39/5/051201}}, \eprint{2202.04918}.

\bibtype{Article}%
\bibitem{Cheng:2011pb}
\bibinfo{author}{H.-Y. Cheng}, \bibinfo{title}{{Revisiting Axial-Vector Meson Mixing}}, \bibinfo{journal}{Phys. Lett. B} \bibinfo{volume}{707} (\bibinfo{year}{2012}) \bibinfo{pages}{116--120}, \bibinfo{doi}{\doi{10.1016/j.physletb.2011.12.013}}, \eprint{1110.2249}.

\bibtype{Article}%
\bibitem{Divotgey:2013jba}
\bibinfo{author}{F. Divotgey}, \bibinfo{author}{L. Olbrich}, \bibinfo{author}{F. Giacosa}, \bibinfo{title}{{Phenomenology of axial-vector and pseudovector mesons: decays and mixing in the kaonic sector}}, \bibinfo{journal}{Eur. Phys. J. A} \bibinfo{volume}{49} (\bibinfo{year}{2013}) \bibinfo{pages}{135}, \bibinfo{doi}{\doi{10.1140/epja/i2013-13135-3}}, \eprint{1306.1193}.

\bibtype{Article}%
\bibitem{Geng:2006yb}
\bibinfo{author}{L.~S. Geng}, \bibinfo{author}{E. Oset}, \bibinfo{author}{L. Roca}, \bibinfo{author}{J.~A. Oller}, \bibinfo{title}{{Clues for the existence of two K(1)(1270) resonances}}, \bibinfo{journal}{Phys. Rev. D} \bibinfo{volume}{75} (\bibinfo{year}{2007}) \bibinfo{pages}{014017}, \bibinfo{doi}{\doi{10.1103/PhysRevD.75.014017}}, \eprint{hep-ph/0610217}.

\bibtype{Article}%
\bibitem{Debastiani:2016xgg}
\bibinfo{author}{V.~R. Debastiani}, \bibinfo{author}{F. Aceti}, \bibinfo{author}{Wei-Hong Liang}, \bibinfo{author}{E. Oset}, \bibinfo{title}{{Revising the $f_1(1420)$ resonance}}, \bibinfo{journal}{Phys. Rev. D} \bibinfo{volume}{95} (\bibinfo{number}{3}) (\bibinfo{year}{2017}) \bibinfo{pages}{034015}, \bibinfo{doi}{\doi{10.1103/PhysRevD.95.034015}}, \eprint{1611.05383}.

\bibtype{Article}%
\bibitem{Jafarzade:2022uqo}
\bibinfo{author}{S. Jafarzade}, \bibinfo{author}{A. Vereijken}, \bibinfo{author}{M. Piotrowska}, \bibinfo{author}{F. Giacosa}, \bibinfo{title}{{From well-known tensor mesons to yet unknown axial-tensor mesons}}, \bibinfo{journal}{Phys. Rev. D} \bibinfo{volume}{106} (\bibinfo{number}{3}) (\bibinfo{year}{2022}) \bibinfo{pages}{036008}, \bibinfo{doi}{\doi{10.1103/PhysRevD.106.036008}}, \eprint{2203.16585}.

\bibtype{Article}%
\bibitem{MartinezTorres:2020hus}
\bibinfo{author}{A. Martinez~Torres}, \bibinfo{author}{K.~P. Khemchandani}, \bibinfo{author}{L. Roca}, \bibinfo{author}{E. Oset}, \bibinfo{title}{{Few-body systems consisting of mesons}}, \bibinfo{journal}{Few Body Syst.} \bibinfo{volume}{61} (\bibinfo{number}{4}) (\bibinfo{year}{2020}) \bibinfo{pages}{35}, \bibinfo{doi}{\doi{10.1007/s00601-020-01568-y}}, \eprint{2005.14357}.

\bibtype{Article}%
\bibitem{Vereijken:2023jor}
\bibinfo{author}{A. Vereijken}, \bibinfo{author}{S. Jafarzade}, \bibinfo{author}{M. Piotrowska}, \bibinfo{author}{F. Giacosa}, \bibinfo{title}{{Is f2(1950) the tensor glueball?}}, \bibinfo{journal}{Phys. Rev. D} \bibinfo{volume}{108} (\bibinfo{number}{1}) (\bibinfo{year}{2023}) \bibinfo{pages}{014023}, \bibinfo{doi}{\doi{10.1103/PhysRevD.108.014023}}, \eprint{2304.05225}.

\bibtype{Article}%
\bibitem{BESIII:2016qzq}
\bibinfo{author}{M. Ablikim}, et al. (\bibinfo{collaboration}{BESIII}), \bibinfo{title}{{Observation of pseudoscalar and tensor resonances in $J/\psi\to \gamma \phi \phi$}}, \bibinfo{journal}{Phys. Rev. D} \bibinfo{volume}{93} (\bibinfo{number}{11}) (\bibinfo{year}{2016}) \bibinfo{pages}{112011}, \bibinfo{doi}{\doi{10.1103/PhysRevD.93.112011}}, \eprint{1602.01523}.

\bibtype{Article}%
\bibitem{Yang:2013xba}
\bibinfo{author}{Yi-Bo Yang}, \bibinfo{author}{Long-Cheng Gui}, \bibinfo{author}{Ying Chen}, \bibinfo{author}{Chuan Liu}, \bibinfo{author}{Yu-Bin Liu}, \bibinfo{author}{Jian-Ping Ma}, \bibinfo{author}{Jian-Bo Zhang} (\bibinfo{collaboration}{CLQCD}), \bibinfo{title}{{Lattice Study of Radiative J/{\ensuremath{\psi}} Decay to a Tensor Glueball}}, \bibinfo{journal}{Phys. Rev. Lett.} \bibinfo{volume}{111} (\bibinfo{number}{9}) (\bibinfo{year}{2013}) \bibinfo{pages}{091601}, \bibinfo{doi}{\doi{10.1103/PhysRevLett.111.091601}}, \eprint{1304.3807}.

\bibtype{Article}%
\bibitem{Etkin:1987rj}
\bibinfo{author}{A. Etkin}, et al., \bibinfo{title}{{Increased Statistics and Observation of the $g(T$), $g(T$)-prime, and $g(T$)-prime-prime 2++ Resonances in the Glueball Enhanced Channel $\pi^- p \to \phi \phi n$}}, \bibinfo{journal}{Phys. Lett. B} \bibinfo{volume}{201} (\bibinfo{year}{1988}) \bibinfo{pages}{568--572}, \bibinfo{doi}{\doi{10.1016/0370-2693(88)90619-3}}.

\bibtype{Article}%
\bibitem{Booth:1985kr}
\bibinfo{author}{P.~S.~L. Booth}, et al., \bibinfo{title}{{Angular Correlations in the $\phi \phi$ System and Evidence for Hadronic $\eta_c$ Production}}, \bibinfo{journal}{Nucl. Phys. B} \bibinfo{volume}{273} (\bibinfo{year}{1986}) \bibinfo{pages}{689--702}, \bibinfo{doi}{\doi{10.1016/0550-3213(86)90385-8}}.

\bibtype{Article}%
\bibitem{WA102:1998nzq}
\bibinfo{author}{D Barberis}, et al. (\bibinfo{collaboration}{WA102}), \bibinfo{title}{{A Study of the centrally produced phi phi system in p p interactions at 450-GeV/c}}, \bibinfo{journal}{Phys. Lett. B} \bibinfo{volume}{432} (\bibinfo{year}{1998}) \bibinfo{pages}{436--442}, \bibinfo{doi}{\doi{10.1016/S0370-2693(98)00661-3}}, \eprint{hep-ex/9805018}.

\bibtype{Article}%
\bibitem{Klempt:2022qjf}
\bibinfo{author}{E. Klempt}, \bibinfo{author}{K.~V. Nikonov}, \bibinfo{author}{A.~V. Sarantsev}, \bibinfo{author}{I. Denisenko}, \bibinfo{title}{{Search for the tensor glueball}}, \bibinfo{journal}{Phys. Lett. B} \bibinfo{volume}{830} (\bibinfo{year}{2022}) \bibinfo{pages}{137171}, \bibinfo{doi}{\doi{10.1016/j.physletb.2022.137171}}, \eprint{2205.07239}.

\bibtype{Article}%
\bibitem{Koenigstein:2016tjw}
\bibinfo{author}{A. Koenigstein}, \bibinfo{author}{F. Giacosa}, \bibinfo{title}{{Phenomenology of pseudotensor mesons and the pseudotensor glueball}}, \bibinfo{journal}{Eur. Phys. J. A} \bibinfo{volume}{52} (\bibinfo{number}{12}) (\bibinfo{year}{2016}) \bibinfo{pages}{356}, \bibinfo{doi}{\doi{10.1140/epja/i2016-16356-x}}, \eprint{1608.08777}.

\bibtype{Article}%
\bibitem{Aston:1993qc}
\bibinfo{author}{D. Aston}, et al., \bibinfo{title}{{Evidence for two J(P) = 2- strange meson states in the K(2) (1770) region}}, \bibinfo{journal}{Phys. Lett. B} \bibinfo{volume}{308} (\bibinfo{year}{1993}) \bibinfo{pages}{186--192}, \bibinfo{doi}{\doi{10.1016/0370-2693(93)90620-W}}.

\bibtype{Article}%
\bibitem{ACCMOR:1981yww}
\bibinfo{author}{C. Daum}, et al. (\bibinfo{collaboration}{ACCMOR}), \bibinfo{title}{{Diffractive Production of Strange Mesons at 63-{GeV}}}, \bibinfo{journal}{Nucl. Phys. B} \bibinfo{volume}{187} (\bibinfo{year}{1981}) \bibinfo{pages}{1--41}, \bibinfo{doi}{\doi{10.1016/0550-3213(81)90114-0}}.

\bibtype{Article}%
\bibitem{LHCb:2016axx}
\bibinfo{author}{R. Aaij}, et al. (\bibinfo{collaboration}{LHCb}), \bibinfo{title}{{Observation of $J/\psi\phi$ structures consistent with exotic states from amplitude analysis of $B^+\to J/\psi \phi K^+$ decays}}, \bibinfo{journal}{Phys. Rev. Lett.} \bibinfo{volume}{118} (\bibinfo{number}{2}) (\bibinfo{year}{2017}) \bibinfo{pages}{022003}, \bibinfo{doi}{\doi{10.1103/PhysRevLett.118.022003}}, \eprint{1606.07895}.

\bibtype{Article}%
\bibitem{CrystalBarrel:1996bnu}
\bibinfo{author}{J. Adomeit}, et al. (\bibinfo{collaboration}{Crystal Barrel}), \bibinfo{title}{{Evidence for two isospin zero J(PC) = 2-+ mesons at 1645-MeV and 1875-MeV}}, \bibinfo{journal}{Z. Phys. C} \bibinfo{volume}{71} (\bibinfo{year}{1996}) \bibinfo{pages}{227--238}, \bibinfo{doi}{\doi{10.1007/s002880050166}}.

\bibtype{Article}%
\bibitem{Jafarzade:2021vhh}
\bibinfo{author}{S. Jafarzade}, \bibinfo{author}{A. Koenigstein}, \bibinfo{author}{F. Giacosa}, \bibinfo{title}{{Phenomenology of $J^{PC}$ = $3^{--}$ tensor mesons}}, \bibinfo{journal}{Phys. Rev. D} \bibinfo{volume}{103} (\bibinfo{number}{9}) (\bibinfo{year}{2021}) \bibinfo{pages}{096027}, \bibinfo{doi}{\doi{10.1103/PhysRevD.103.096027}}, \eprint{2101.03195}.

\bibtype{Article}%
\bibitem{Roca:2010tf}
\bibinfo{author}{L. Roca}, \bibinfo{author}{E. Oset}, \bibinfo{title}{{A description of the f2(1270), rho3(1690), f4(2050), rho5(2350) and f6(2510) resonances as multi-rho(770) states}}, \bibinfo{journal}{Phys. Rev. D} \bibinfo{volume}{82} (\bibinfo{year}{2010}) \bibinfo{pages}{054013}, \bibinfo{doi}{\doi{10.1103/PhysRevD.82.054013}}, \eprint{1005.0283}.

\bibtype{Article}%
\bibitem{Yamagata-Sekihara:2010muv}
\bibinfo{author}{J. Yamagata-Sekihara}, \bibinfo{author}{L. Roca}, \bibinfo{author}{E. Oset}, \bibinfo{title}{{On the nature of the $K^*_2(1430)$, $K^*_3(1780)$, $K^*_4(2045)$, $K^*_5(2380)$ and $K^*6$ as $K^*$ - multi-$\rho$ states}}, \bibinfo{journal}{Phys. Rev. D} \bibinfo{volume}{82} (\bibinfo{year}{2010}) \bibinfo{pages}{094017}, \bibinfo{doi}{\doi{10.1103/PhysRevD.82.094017}}, \bibinfo{note}{[Erratum: Phys.Rev.D 85, 119905 (2012)]}, \eprint{1010.0525}.

\bibtype{Article}%
\bibitem{Regge:1959mz}
\bibinfo{author}{T. Regge}, \bibinfo{title}{{Introduction to complex orbital momenta}}, \bibinfo{journal}{Nuovo Cim.} \bibinfo{volume}{14} (\bibinfo{year}{1959}) \bibinfo{pages}{951}, \bibinfo{doi}{\doi{10.1007/BF02728177}}.

\bibtype{Book}%
\bibitem{Collins:1977jy}
\bibinfo{author}{P.~D.~B. Collins}, \bibinfo{title}{{An Introduction to Regge Theory and High Energy Physics}}, \bibinfo{publisher}{Cambridge University Press} \bibinfo{year}{1977}, ISBN \bibinfo{isbn}{978-1-009-40326-9, 978-1-009-40329-0, 978-1-009-40328-3, 978-0-521-11035-8}, \bibinfo{doi}{\doi{10.1017/9781009403269}}.

\bibtype{Book}%
\bibitem{Gribov:2003nw}
\bibinfo{author}{V.~N. Gribov}, \bibinfo{title}{{The theory of complex angular momenta: Gribov lectures on theoretical physics}}, Cambridge Monographs on Mathematical Physics, \bibinfo{publisher}{Cambridge University Press} \bibinfo{year}{2007}, ISBN \bibinfo{isbn}{978-0-521-03703-7, 978-0-521-81834-6, 978-0-511-05504-1}, \bibinfo{doi}{\doi{10.1017/CBO9780511534959}}.

\bibtype{Article}%
\bibitem{Chew:1962eu}
\bibinfo{author}{G.~F. Chew}, \bibinfo{author}{S.~C. Frautschi}, \bibinfo{title}{{Regge Trajectories and the Principle of Maximum Strength for Strong Interactions}}, \bibinfo{journal}{Phys. Rev. Lett.} \bibinfo{volume}{8} (\bibinfo{year}{1962}) \bibinfo{pages}{41--44}, \bibinfo{doi}{\doi{10.1103/PhysRevLett.8.41}}.

\bibtype{Article}%
\bibitem{Ebert:2009ub}
\bibinfo{author}{D. Ebert}, \bibinfo{author}{R.~N. Faustov}, \bibinfo{author}{V.~O. Galkin}, \bibinfo{title}{{Mass spectra and Regge trajectories of light mesons in the relativistic quark model}}, \bibinfo{journal}{Phys. Rev. D} \bibinfo{volume}{79} (\bibinfo{year}{2009}) \bibinfo{pages}{114029}, \bibinfo{doi}{\doi{10.1103/PhysRevD.79.114029}}, \eprint{0903.5183}.

\bibtype{Misc}%
\bibitem{E12-06-102}
\bibinfo{author}{C. Meyer}, et al., \bibinfo{title}{{E12-06-102:Mapping the Spectrum of Light Quark Mesons and Gluonic Excitations with Linearly Polarized Photons}} \bibinfo{year}{2006}, \bibinfo{url}{\urlprefix\url{https://www.jlab.org/exp_prog/proposals/06/PR12-06-102.pdf}}.

\bibtype{Article}%
\bibitem{GlueX:2014hxq}
\bibinfo{author}{M. Dugger}, et al. (\bibinfo{collaboration}{GlueX}), \bibinfo{title}{{A study of decays to strange final states with GlueX in Hall D using components of the BaBar DIRC}}, \bibinfo{journal}{arXiv:1408.0215}  (\bibinfo{year}{2014}).

\bibtype{Article}%
\bibitem{Adhikari:2020cvz}
\bibinfo{author}{S. Adhikari}, et al. (\bibinfo{collaboration}{GlueX}), \bibinfo{title}{{The GLUEX beamline and detector}}, \bibinfo{journal}{Nucl. Instrum. Meth. A} \bibinfo{volume}{987} (\bibinfo{year}{2021}) \bibinfo{pages}{164807}, \bibinfo{doi}{\doi{10.1016/j.nima.2020.164807}}, \eprint{2005.14272}.

\bibtype{Misc}%
\bibitem{E12-11-005}
\bibinfo{author}{M. Battaglieri}, \bibinfo{author}{R.~De Vita}, \bibinfo{author}{C. Salgado}, \bibinfo{author}{S. Stepanyan}, \bibinfo{author}{D. Watts}, \bibinfo{author}{D. Weygand}, \bibinfo{title}{{E12-11-005:Meson Spectroscopy with low $Q^2$ electron scattering in CLAS12 }} \bibinfo{year}{2011}, \bibinfo{url}{\urlprefix\url{https://www.jlab.org/exp_prog/proposals/11/PR12-11-005.pdf}}.

\bibtype{Article}%
\bibitem{Arrington:2021alx}
\bibinfo{author}{J. Arrington}, et al., \bibinfo{title}{{Physics with CEBAF at 12 GeV and future opportunities}}, \bibinfo{journal}{Prog. Part. Nucl. Phys.} \bibinfo{volume}{127} (\bibinfo{year}{2022}) \bibinfo{pages}{103985}, \bibinfo{doi}{\doi{10.1016/j.ppnp.2022.103985}}, \eprint{2112.00060}.

\bibtype{Article}%
\bibitem{BESIII:2020nme}
\bibinfo{author}{M. Ablikim}, et al. (\bibinfo{collaboration}{BESIII}), \bibinfo{title}{{Future Physics Programme of BESIII}}, \bibinfo{journal}{Chin. Phys. C} \bibinfo{volume}{44} (\bibinfo{number}{4}) (\bibinfo{year}{2020}) \bibinfo{pages}{040001}, \bibinfo{doi}{\doi{10.1088/1674-1137/44/4/040001}}, \eprint{1912.05983}.

\bibtype{Article}%
\bibitem{Battaglieri:2014gca}
\bibinfo{author}{M. Battaglieri}, et al., \bibinfo{title}{{Analysis Tools for Next-Generation Hadron Spectroscopy Experiments}}, \bibinfo{journal}{Acta Phys. Polon. B} \bibinfo{volume}{46} (\bibinfo{year}{2015}) \bibinfo{pages}{257}, \bibinfo{doi}{\doi{10.5506/APhysPolB.46.257}}, \eprint{1412.6393}.

\bibtype{Article}%
\bibitem{JPAC:2021rxu}
\bibinfo{author}{Miguel Albaladejo}, et al. (\bibinfo{collaboration}{JPAC}), \bibinfo{title}{{Novel approaches in hadron spectroscopy}}, \bibinfo{journal}{Prog. Part. Nucl. Phys.} \bibinfo{volume}{127} (\bibinfo{year}{2022}) \bibinfo{pages}{103981}, \bibinfo{doi}{\doi{10.1016/j.ppnp.2022.103981}}, \eprint{2112.13436}.

\bibtype{Article}%
\bibitem{Sombillo:2020ccg}
\bibinfo{author}{Denny Lane~B. Sombillo}, \bibinfo{author}{Yoichi Ikeda}, \bibinfo{author}{Toru Sato}, \bibinfo{author}{Atsushi Hosaka}, \bibinfo{title}{{Classifying the pole of an amplitude using a deep neural network}}, \bibinfo{journal}{Phys. Rev. D} \bibinfo{volume}{102} (\bibinfo{number}{1}) (\bibinfo{year}{2020}) \bibinfo{pages}{016024}, \bibinfo{doi}{\doi{10.1103/PhysRevD.102.016024}}, \eprint{2003.10770}.

\bibtype{Article}%
\bibitem{Sombillo:2021rxv}
\bibinfo{author}{Denny Lane~B. Sombillo}, \bibinfo{author}{Yoichi Ikeda}, \bibinfo{author}{Toru Sato}, \bibinfo{author}{Atsushi Hosaka}, \bibinfo{title}{{Model independent analysis of coupled-channel scattering: A deep learning approach}}, \bibinfo{journal}{Phys. Rev. D} \bibinfo{volume}{104} (\bibinfo{number}{3}) (\bibinfo{year}{2021}) \bibinfo{pages}{036001}, \bibinfo{doi}{\doi{10.1103/PhysRevD.104.036001}}, \eprint{2105.04898}.

\bibtype{Article}%
\bibitem{Ng:2021ibr}
\bibinfo{author}{L. Ng}, \bibinfo{author}{L. Bibrzycki}, \bibinfo{author}{J. Nys}, \bibinfo{author}{C. Fernandez-Ramirez}, \bibinfo{author}{A. Pilloni}, \bibinfo{author}{V. Mathieu}, \bibinfo{author}{A.~J. Rasmusson}, \bibinfo{author}{A.~P. Szczepaniak} (\bibinfo{collaboration}{Joint Physics Analysis Center, JPAC}), \bibinfo{title}{{Deep learning exotic hadrons}}, \bibinfo{journal}{Phys. Rev. D} \bibinfo{volume}{105} (\bibinfo{number}{9}) (\bibinfo{year}{2022}) \bibinfo{pages}{L091501}, \bibinfo{doi}{\doi{10.1103/PhysRevD.105.L091501}}, \eprint{2110.13742}.

\bibtype{Article}%
\bibitem{Malekhosseini:2024eot}
\bibinfo{author}{M. Malekhosseini}, \bibinfo{author}{S. Rostami}, \bibinfo{author}{A.~R. Olamaei}, \bibinfo{author}{R. Ostovar}, \bibinfo{author}{K. Azizi}, \bibinfo{title}{{Meson mass and width: Deep learning approach}}, \bibinfo{journal}{Phys. Rev. D} \bibinfo{volume}{110} (\bibinfo{number}{5}) (\bibinfo{year}{2024}) \bibinfo{pages}{054011}, \bibinfo{doi}{\doi{10.1103/PhysRevD.110.054011}}, \eprint{2404.00448}.

\end{thebibliography*}

%%%%%%%%%%%%%%
\end{document}